\documentclass[12pt]{article}
\usepackage{makeidx}
\usepackage[english]{babel}
\usepackage{hyperref}
\usepackage{amsmath}
\usepackage{amsfonts}
\usepackage{amssymb}

\begin{document}

\title{Non-Abelian gauge theories with composite fields in the background
field method}
\author{\textsc{P.Yu. Moshin},$^{a}$\thanks{%
e-mail: moshin@phys.tsu.ru} \textsc{A.A. Reshetnyak,}$^{a,b,c}$\thanks{%
e-mail: reshet@tspu.edu.ru} \textsc{R.A. Castro}$^{d}$\thanks{%
e-mail: rialcap@usp.br} \\
\textit{${}$}$^{a}${\normalsize National Research Tomsk State University,
634050, Tomsk, Russia}\\
$^{b}${\normalsize Tomsk State Pedagogical University, 634041, Tomsk, Russia}%
\\
$^{c}${\normalsize National Research Tomsk Polytechnic University, 634050,
Tomsk, Russia}\\
$^{d}${\normalsize University of S\~{a}o Paulo, CEP 05508-090, S\~{a}o
Paulo, Brazil}}
\date{}
\maketitle

\begin{abstract}
Non-Abelian gauge theories with composite fields are examined in the
background field method. Generating functionals of Green's functions for a
Yang--Mills theory with composite and background fields are introduced,
including the generating functional of vertex Green's functions (effective
action). The corresponding Ward identities are obtained, and the issue of
gauge dependence is investigated. A gauge variation of the effective action
is found in terms of a nilpotent operator depending on the composite and
background fields. On-shell independence from the choice of gauge fixing for
the effective action is established. In the study of the Ward identities and
gauge dependence, finite field-dependent BRST transformations with a
background field are introduced and employed on a systematic basis. On the
one hand, this involves the consideration of (modified) Ward identities with
a field-dependent anticommuting parameter, also depending on a non-trivial
background. On the other hand, the issue of gauge dependence is studied with
reference to a finite variation of the gauge Fermion. The concept of a joint
introduction of composite and background fields to non-Abelian gauge
theories is exemplified by the Gribov--Zwanziger theory, including the case
of a local BRST-invariant horizon, and also by the Volovich--Katanaev model
of two-dimensional gravity with dynamical torsion.
\end{abstract}

\section{Introduction}

\setcounter{equation}{0}\renewcommand{\theequation}{1.\arabic{equation}}

The use of background \cite{DeW, AFS,Abbott} and composite \cite{17,18}
fields has gained considerable attention in quantum field theory. {%
Composite fields are normally introduced as }$m${-degree polynomials }%
$\sigma ^{m}\left( \phi \right) ${\ in the quantum fields }$\phi ^{A}$%
{\ of a theory. The composite fields are associated with their
respective sources }$L_{m}${\ in a way the quantum fields are coupled
to sources }$J_{A}$.{\ A less trivial example is the
Gribov--Zwanziger model \cite{Zwanziger}, which involves a quantum theory of
Yang--Mills fields }$A_{\mu }${\ with a non-local composite field
entering through the so-called Gribov horizon functional }$H\left( A\right) $%
{\ applied to cure the problem of residual gauge invariance \cite%
{Gribov}. In a quantum theory of Yang--Mills fields with a classical action }%
$S_{0}\left( A\right) ${, one can generally introduce some background
fields }$B_{\mu }${\ entering through }$S_{0}\left( A+B\right) $%
{, as well as through certain gauge-fixing conditions. Path
integration is then carried out with respect to }$A_{\mu }${, so that
the resulting generating functional }$Z${\ of Green's functions
retains a dependence on }$B_{\mu }${, namely, }$Z=Z\left( B,J\right) $%
. The background field method \cite{DeW, AFS,Abbott} reformulates the
quantization of Yang--Mills theories under {the background gauge
conditions} \cite{Abbott, Abbott:1981ke, Weinberg} in a manner which
provides an {effective action} $\Gamma _{\mathrm{eff}}\left( B\right)
$ invariant under the gauge transformations of {the background fields}
$B_{\mu }$ and reproduces physical results with {essential
simplifications} in calculating the Feynman diagrams, which allows one to
study a wide range of quantum properties in gauge theories \cite%
{'tH,K-SZ,GvanNW,CMacL,IO,GS,Ven,Gr,BC,FPQ}; see also \cite%
{BQ,Barv,FT,BLT-YM,BFMc} for recent developments.

{In this connection, the issue of a gauge-invariant effective action
in the Gribov--Zwanziger model \cite{Zwanziger,Gribov} has so far remained
unsolved, which generally calls for an introduction of a background into a
quantum theory involving composite fields. A suitable framework for dealing
with the Gribov--Zwanziger theory is given by the concept of soft BRST
symmetry \cite{llr1,Reshetnyak1,Reshetnyak,LL2,Reshetnyak2} and the local
composite operator technique \cite{Capri0, Capri2} on arbitrary backgrounds
\cite{Canfora}. }The interest in composite fields {is also due to the
fact} that the effective action for composite fields (see \cite{18} for a
review) introduced in \cite{17} has found diverse applications to quantum
field models such as \cite{ExamComp,HockMossT}, including the early
Universe, the inflationary Universe, the Standard Model and SUSY theories
\cite{DSCB,20,GT,21}. It seems to be of particular interest in this regard
to {apply the techniques \cite%
{llr1,Reshetnyak1,Reshetnyak,LL2,Reshetnyak2,Capri0, Capri2,Canfora} to the
investigation of} BRST-invariant renormalizability in Yang--Mills theories,
which includes the $N=1$ SUSY formulations \cite{Stepanyantz,
Shifman:2015doa, Capri1} and the functional renormalization group \cite
{FRG1, FRG2, FRG3, FRG4, FRG5}.

{Let us emphasize the following. Composite fields in a quantum
Yang--Mils theory without a residual gauge symmetry have not been introduced
so far. A consistent quantization using the method of path integrals on
arbitrary backgrounds in the Gribov--Zwanziger model, involving not only a
non-local horizon, but also its localized description, has not yet been
proposed. This has been a direct obstacle to the construction of a
gauge-invariant effective action in non-Abelian gauge theories without a
residual gauge symmetry, both with and without composite fields on arbitrary
backgrounds of classical Yang--Mills fields. The subject of the present
paper is an introduction of the background field method in a way consistent
with the formalism of composite fields.}

In this paper, we address the issue of quantum non-Abelian gauge theories
including both composite and background fields with a systematic treatment
based on Yang--Mills theories quantized using the Faddeev--Popov method \cite%
{FP}. A combined treatment of Yang--Mills fields $A_{\mu }$ with composite
and background ones calls for a joint introduction of these ingredients on a
systematic basis. We suggest to use the symmetry principle as such a
systematic guideline. Thus, suppose that a generating functional $Z\left(
J,L\right) $ of Green's functions with composite fields is given, depending
on the sources $J_{A}$ for the usual fields $\phi ^{A}$, and also on the
sources $L_{m}$ for the composite fields $\sigma ^{m}\left( \phi \right) $.
One may ask how some background fields $B_{\mu }$ can be introduced in such
a way as to produce an extended functional $Z\left( B,J,L\right) $ which
reflects the symmetries inherent in $Z\left( J,L\right) $. Suppose, on the
other hand, that a generating functional $Z\left( B,J\right) $ of Green's
functions in the background field method is given, also featuring some
symmetries, and then one may ask how some composite fields $\sigma
^{m}\left( \phi ,B\right) $ with sources $L_{m}$ can be introduced for the
resulting $Z\left( B,J,L\right) $ to inherit the original symmetries. These
two approaches prove to be equivalent in the sense outlined in the following
preliminary exposition.

In the first approach, one is given a generating functional $Z\left(
J,L\right) $,
\begin{equation}
Z\left( J,L\right) =\int d\phi \exp \left\{ \frac{i}{\hbar }\left[ S_{%
\mathrm{FP}}\left( \phi \right) +J_{A}\phi ^{A}+L_{m}\sigma ^{m}\left( \phi
\right) \right] \right\} ,  \label{approach1}
\end{equation}%
corresponding to the Faddeev--Popov action $S_{\mathrm{FP}}\left( \phi
\right) $ of a Yang--Mills theory with composite fields $\sigma ^{m}\left(
\phi \right) $. Then a background field $B_{\mu }$ can be introduced by
\emph{localizing} the inherent global symmetry of $Z\left( J,L\right) $
under $SU(N)$ transformations (\textquotedblleft
rotations\textquotedblright\ for $J_{A}$ and tensor transformations for $%
L_{m}$) in such a way that $Z\left( B,J,L\right) $ defined as\footnote{%
More specifically, the replacement $\partial _{\mu }\rightarrow D_{\mu
}\left( B\right) $ is described by the relations (\ref{rule}), (\ref{F(A+B)}%
) of Section \ref{sec2}. This rule is unambiguous for local fields $\sigma
^{m}$ without higher derivatives, $\sigma ^{m}=\sigma ^{m}\left( \phi
,\partial \phi \right) $. For more details, see Section \ref{sec2}.}%
\begin{equation}
Z\left( B,J,L\right) =\left. Z\left( J,L\right) \right\vert _{\partial _{\mu
}\rightarrow D_{\mu }\left( B\right) }  \label{intro1}
\end{equation}%
turns out to be invariant under local $SU(N)$ transformations of the sources
$J_{A}$, $L_{m}$, accompanied by gauge transformations of the field $B_{\mu
} $ with an associated covariant derivative $D_{\mu }\left( B\right) $.
Besides, the original $S_{\mathrm{FP}}\left( \phi \right) $ becomes modified
to the Faddeev--Popov action $S_{\mathrm{FP}}\left( \phi ,B\right) $ of the
background field method and related to $S_{\mathrm{FP}}\left( \phi \right) $
by so-called \emph{background} and \emph{quantum} transformations of this
method (see Section \ref{sec2}).

In the second approach, one is given a generating functional $Z\left(
B,J\right) $ constructed using the background field method for Yang--Mills
theories,%
\begin{equation}
Z\left( B,J\right) =\int d\phi \exp \left\{ \frac{i}{\hbar }\left[ S_{%
\mathrm{FP}}\left( \phi ,B\right) +J_{A}\phi ^{A}\right] \right\} ,
\label{approach2}
\end{equation}%
which implies that $S_{\mathrm{FP}}\left( \phi ,B\right) $ is given by%
\begin{equation*}
S_{\mathrm{FP}}\left( \phi ,B\right) =\left. S_{\mathrm{FP}}\left( \phi
\right) \right\vert _{\partial _{\mu }\rightarrow D_{\mu }\left( B\right) }\
.
\end{equation*}%
One can then introduce some composite fields $\sigma ^{m}\left( \phi
,B\right) $ on condition that the resulting generating functional%
\begin{equation}
Z\left( B,J,L\right) =\int d\phi \exp \left\{ \frac{i}{\hbar }\left[ S_{%
\mathrm{FP}}\left( \phi ,B\right) +J_{A}\phi ^{A}+L_{m}\sigma ^{m}\left(
\phi ,B\right) \right] \right\}  \label{intro2a}
\end{equation}%
inherit the symmetry of $Z\left( B,J\right) $ under local $SU(N)$ rotations
of the sources $J_{A}$ accompanied by gauge transformations of the
background field $B_{\mu }$ with the covariant derivative $D_{\mu }\left(
B\right) $. This symmetry condition for $Z\left( B,J,L\right) $ is met by a
local $SU(N)$ tensor transformation law imposed on $\sigma ^{m}\left( \phi
,B\right) $, which is provided by $B_{\mu }$ entering into the composite
fields $\sigma ^{m}\left( \phi ,B\right) $ via the covariant derivatives $%
D_{\mu }\left( B\right) $ and implies%
\begin{equation}
\sigma ^{m}\left( \phi ,B\right) =\left. \sigma ^{m}\left( \phi \right)
\right\vert _{\partial _{\mu }\rightarrow D_{\mu }\left( B\right) }
\label{intro2b}
\end{equation}%
for certain $\sigma ^{m}\left( \phi \right) $, which brings us back to the
first approach.

In the main part of the present work, we choose the first approach as a
starting point of our systematic treatment assuming the composite fields to
be\ \emph{local}, and then, in the remaining part, we show how the first and
second approaches can be extended beyond the given assumptions by
considering some examples. The principal research issues to be addressed are
as follows:

\begin{enumerate}
\item Introduction of generating functionals of Green's functions with
composite and background fields in Yang--Mills theories; investigation of
the related symmetry properties;

\item Extension of finite field-dependent BRST (FD BRST) transformations
\cite{Reshetnyak,LL1} to the case of background field dependence;

\item Investigation of the Ward identities and gauge dependence for the
above generating functionals on a basis of finite FD BRST transformations;

\item Introduction of background and composite fields field into the
Gribov--Zwanziger theory \cite{Zwanziger,Gribov};

\item Introduction of composite fields into a quantized Volovich--Katanaev
model \cite{LM}.
\end{enumerate}

The paper is organized as follows. In Section \ref{sec2}, we introduce a
generating functional of Green's functions with composite and background
fields in Yang--Mills theories. Section \ref{sec3} is devoted to the
corresponding Ward identities and the properties of the generating
functional of vertex Green's functions (effective action). Thus, the
effective action $\Gamma _{\mathrm{eff}}\left( B,\Sigma \right) $, depending
on the background field $B_{\mu }$ and a set of tensor auxiliary fields $%
\Sigma ^{m}$ associated with $\sigma ^{m}$, is found to exhibit a gauge
symmetry under the gauge transformations of $B_{\mu }$ along with the local $%
SU(N)$ transformations of $\Sigma ^{m}$. The study of the Ward identities
systematically utilizes the concept of finite FD BRST transformations \cite%
{Reshetnyak,LL1}, first suggested\footnote{%
See \cite{BLTfin} for a field-antifield BV formalism and \cite{MR1, MR2,
SMMR, UMR, N34} for extended $N\geq 2$ BRST symmetries.} in \cite{JM,
Upadhyay3} and now depending also on the background field $B_{\mu }$. In
Section \ref{sec4}, we use the finite FD BRST transformations and the
related (modified) Ward identities to analyze the dependence of the
generating functionals of Green's functions upon a choice of gauge-fixing.
In doing so, we evaluate a finite gauge variation of the effective action in
terms of a nilpotent operator depending on the composite and background
fields, and also determine the conditions of on-shell gauge-independence.
Loop expansion properties and a one-loop effective action with composite and
background fields are examined in Section \ref{sec45}. In Section \ref{sec5}%
, we consider an example of the Gribov--Zwanziger theory \cite{Zwanziger},
which is a quantum Yang--Mills theory including the Gribov horizon \cite%
{Gribov}, using a local and a non-local BRST-invariant representations, in
terms of composite fields. The quantum theory \cite{Zwanziger} is then
extended by introducing a background field, which modifies our first
approach (\ref{approach1}), (\ref{intro1}) beyond the case of local
composite fields, along with a study of the gauge-independence problem. In
Section \ref{sec6}, we consider an example of the two-dimensional gravity
with dynamical torsion by Volovich and Katanaev \cite{VK}, quantized
according to the background field method in \cite{LM} and featuring a
gauge-invariant background effective action.\ As a modification of our
second approach (\ref{approach2}), (\ref{intro2a}), (\ref{intro2b}), when
extended beyond the Yang--Mills case, the quantized two-dimensional gravity
\cite{LM} is generalized to the presence of composite fields, and the
corresponding effective action with composite and background fields is found
to be gauge-invariant, in a way similar to the Yang--Mills case. Section \ref%
{sec7} presents a summary of our results. Appendices \ref{A}, \ref{B}, \ref%
{C} support the consideration of the respective Yang--Mills,
Gribov--Zwanziger and Volovich--Katanaev models.

We use DeWitt's condensed notation \cite{DeWcond}. The Grassmann parity and
ghost number of a quantity $F$ are denoted by $\epsilon \left( F\right) $, $%
\mathrm{gh}\left( F\right) $, respectively, and $\left[ F,G\right\} $ stands
for the supercommutator of any quantities $F$, $G$ with a definite Grassmann
parity, $\left[ F,G\right\} =FG-\left( -1\right) ^{\epsilon \left( F\right)
\epsilon \left( G\right) }GF$.\ Unless specifically indicated by an arrow,
derivatives with respect to fields and sources are understood as left-hand
ones.

\section{Generating Functional of Green's Functions\label{sec2}}

\setcounter{equation}{0}\renewcommand{\theequation}{2.\arabic{equation}}

Consider a generating functional $Z\left( J,L\right) $ corresponding to the
Faddeev--Popov action $S_{\mathrm{FP}}\left( \phi \right) $ of a Yang--Mills
theory with local composite fields,%
\begin{equation}
Z\left( J,L\right) =\int d\phi \exp \left\{ \frac{i}{\hbar }\left[ S_{%
\mathrm{FP}}\left( \phi \right) +J_{A}\phi ^{A}+L_{m}\sigma ^{m}\left( \phi
\right) \right] \right\} ,  \label{Z(J,L)}
\end{equation}%
where $L_{m}$ are sources to the composite fields $\sigma ^{m}\left( \phi
\right) $,%
\begin{equation}
\sigma ^{m}\left( \phi \right) =\sum_{n=2}\frac{1}{n!}\Lambda _{A_{1}\ldots
A_{n}}^{m}\phi ^{A_{n}}\ldots \phi ^{A_{1}}\ ,  \label{compfield}
\end{equation}%
and $J_{A}$ are sources to the fields $\phi ^{A}=\left( A^{i},b^{\alpha },%
\bar{c}^{\alpha },c^{\alpha }\right) $ composed by gauge fields $A^{i}$,
(anti)ghost fields $\bar{c}^{\alpha }$, $c^{\alpha }$, and
Nakanishi--Lautrup fields $b^{\alpha }$, with the following distribution of
Grassmann parity and ghost number:%
\begin{equation*}
\begin{tabular}{|c|c|c|c|c|}
\hline
& $\phi ^{A}$ & $J_{A}$ & $\sigma ^{m}$ & $L_{m}$ \\ \hline
$\epsilon $ & $\left( 0,0,1,1\right) $ & $\epsilon \left( \phi ^{A}\right) $
& $\epsilon \left( \sigma ^{m}\right) $ & $\epsilon \left( \sigma
^{m}\right) $ \\ \hline
$\mathrm{gh}$ & $\left( 0,0,-1,1\right) $ & $-\mathrm{gh}\left( \phi
^{A}\right) $ & $\mathrm{gh}\left( \sigma ^{m}\right) $ & $-\mathrm{gh}%
\left( \sigma ^{m}\right) $ \\ \hline
\end{tabular}%
\end{equation*}%
The Faddeev--Popov action $S_{\mathrm{FP}}\left( \phi \right) $,
\begin{equation*}
S_{\mathrm{FP}}\left( \phi \right) =S_{0}\left( A\right) +\Psi \left( \phi
\right) \overleftarrow{s},
\end{equation*}%
is given in terms of a gauge-invariant classical action $S_{0}\left(
A\right) $, invariant, $\delta S_{0}\left( A\right) =0$, under infinitesimal
gauge transformations $\delta A^{i}=R_{\alpha }^{i}\left( A\right) \xi
^{\alpha }$ with a closed algebra of gauge generators $R_{\alpha }^{i}\left(
A\right) $,
\begin{equation*}
R_{\alpha \,j}^{i}\left( A\right) R_{\beta }^{j}\left( A\right) -R_{\beta
\,j}^{i}\left( A\right) R_{\alpha }^{j}\left( A\right) =F_{\alpha \beta
}^{\gamma }R_{\gamma }^{i}\left( A\right) ,\ \ \ F_{\alpha \beta }^{\gamma }=%
\mathrm{const},\ \ \ R_{\alpha ,\,j}^{i}\equiv R_{\alpha }^{i}\frac{%
\overleftarrow{\delta }}{\delta A^{j}}\ ,
\end{equation*}%
and a nilpotent Slavnov variation $\overleftarrow{s}$ applied to a gauge
Fermion $\Psi \left( \phi \right) $, $\epsilon \left( \Psi \right) =-\mathrm{%
gh}\left( \Psi \right) =1$,%
\begin{equation}
S_{\mathrm{FP}}\left( \phi \right) =S_{0}\left( A\right) +\Psi \left( \phi
\right) \overleftarrow{s},\ \ \ \Psi \left( \phi \right) =\bar{c}^{\alpha
}\chi _{\alpha }\left( \phi \right) ,\ \overleftarrow{s}\ ^{2}=0,
\label{S_FP}
\end{equation}%
where%
\begin{equation*}
\phi ^{A}\overleftarrow{s}=\left( R_{\alpha }^{i}\left( A\right) c^{\alpha
},0,b^{\alpha },1/2F_{\beta \gamma }^{\alpha }c^{\gamma }c^{\beta }\right) .
\end{equation*}%
For the explicit field content%
\begin{eqnarray*}
i &=&(x,p,\mu ),\ \ \ \alpha =\left( x,p\right) ,\ \ \ \mu =0,\ldots ,D-1,\
\ \ p=1,\ldots ,N^{2}-1, \\
\phi ^{A} &=&\left( A^{p|\mu },b^{p},\bar{c}^{p},c^{p}\right) ,\ \ \ \left(
A^{\mu },b,\bar{c},c\right) \equiv T^{p}\left( A^{p|\mu },b^{p},\bar{c}%
^{p},c^{p}\right) ,\ \ \ \left[ T^{p},T^{q}\right] =f^{pqr}T^{r},
\end{eqnarray*}%
the field variations $\phi ^{A}\overleftarrow{s}$ have the form%
\begin{eqnarray}
\left( A_{\mu },b,\bar{c},c\right) \overleftarrow{s} &=&\left( \left[ D_{\mu
}\left( A\right) ,c\right] ,0,b,g/2\left[ c,c\right] _{+}\right) ,  \notag \\
\left( A_{\mu }^{p},b^{p},\bar{c}^{p},c^{p}\right) \overleftarrow{s}
&=&\left( D_{\mu }^{pq}\left( A\right)
c^{q},0,b^{p},g/2f^{pqr}c^{q}c^{r}\right) ,  \label{BRST}
\end{eqnarray}%
where%
\begin{equation*}
D_{\mu }\left( A\right) \equiv \partial _{\mu }+gA_{\mu }\ ,\ \ \ D_{\mu
}^{pq}\left( A\right) =\delta ^{pq}\partial _{\mu }+gf^{prq}A_{\mu }^{r}\ .
\end{equation*}%
The classical action $S_{0}\left( A\right) $ has the form (in the adjoint
representation with Hermitian $T^{p}$),%
\begin{eqnarray}
S_{0}\left( A\right) &=&\frac{1}{2g^{2}}\int d^{D}x\ \mathrm{Tr}\left(
F_{\mu \nu }F^{\mu \nu }\right) =\frac{1}{4}\int d^{D}x\ F_{\mu \nu
}^{p}F^{p|\mu \nu },\ \ \ \mathrm{Tr}\left( T^{p}T^{q}\right) =\frac{1}{2}%
\delta ^{pq},  \label{S_0} \\
F_{\mu \nu } &\equiv &\left[ D_{\mu }\left( A\right) ,D_{\nu }\left(
A\right) \right] ,\ \ \ F_{\mu \nu }^{p}=\partial _{\mu }A_{\nu
}^{p}-\partial _{\nu }A_{\mu }^{p}+gf^{prs}A_{\mu }^{r}A_{\nu }^{s}\ ,
\notag
\end{eqnarray}%
and the gauge Fermion $\Psi \left( \phi \right) =\bar{c}^{\alpha }\chi
_{\alpha }\left( \phi \right) $ with gauge-fixing functions $\chi _{\alpha
}\left( \phi \right) =\chi ^{p}\left( \phi \left( x\right) \right) $ reads%
\begin{equation}
\Psi \left( \phi \right) =\int d^{D}x\ \bar{c}^{p}\chi ^{p}\left( \phi
\right) =2\int d^{D}x\ \mathrm{Tr}\left[ \bar{c}\chi \left( \phi \right) %
\right] ,\ \ \ \chi \left( \phi \right) =T^{p}\chi ^{p}\left( \phi \right) .
\label{Psi}
\end{equation}

The Faddeev--Popov action $S_{\mathrm{FP}}\left( \phi \right) $ is invariant
under two kinds of global transformations: BRST transformations \cite%
{BRST1,BRST2}, $\delta _{\lambda }\phi ^{A}=\phi ^{A}\overleftarrow{s}%
\lambda $, with an anticommuting parameter $\lambda $, $\epsilon \left(
\lambda \right) =\mathrm{gh}\left( \lambda \right) =1$, and $SU(N)$
rotations (finite $\phi ^{A}\overset{U}{\rightarrow }\phi ^{\prime A}$ and
infinitesimal $\delta _{\varsigma }\phi ^{A}$)\ with even parameters $%
\varsigma ^{p}$,%
\begin{eqnarray}
\left( A_{\mu },b,\bar{c},c\right) \overset{U}{\rightarrow }\left( A_{\mu
},b,\bar{c},c\right) ^{\prime } &=&U\left( A_{\mu },b,\bar{c},c\right)
U^{-1},\ \ \ U=\exp \left( -gT^{p}\varsigma ^{p}\right) ,\ \ \ \varsigma
^{p}=\mathrm{const},  \label{consistent} \\
\delta _{\varsigma }\left( A_{\mu }^{p},b^{p},\bar{c}^{p},c^{p}\right)
&=&gf^{prq}\left( A_{\mu }^{r},b^{r},\bar{c}^{r},c^{r}\right) \varsigma ^{q},
\notag
\end{eqnarray}%
or, in a tensor form, via the adjoint representation with a matrix $%
M^{pq}\left( \varsigma \right) $,%
\begin{equation*}
\left( A_{\mu }^{p},b^{p},\bar{c}^{p},c^{p}\right) ^{\prime }=M^{pq}\left(
\varsigma \right) \left( A_{\mu }^{q},b^{q},\bar{c}^{q},c^{q}\right) ,\ \ \
M^{pq}\left( \varsigma \right) =\delta ^{pq}+gf^{pqr}\varsigma ^{r}+O\left(
\varsigma ^{2}\right) .
\end{equation*}%
The classical action $S_{0}\left( A\right) $ in $S_{\mathrm{FP}}\left( \phi
\right) =S_{0}\left( A\right) +\Psi \left( \phi \right) \overleftarrow{s}$
is invariant under $A_{\mu }\overset{U}{\rightarrow }A_{\mu }^{\prime }$ as
a particular case ($\xi ^{p}(x)=\mathrm{const}$) of invariance under the
finite form $A_{\mu }\overset{V}{\rightarrow }A_{\mu }^{\prime }$ of gauge
transformations%
\begin{equation}
A_{\mu }^{\prime }=VA_{\mu }V^{-1}+g^{-1}V\left( \partial _{\mu
}V^{-1}\right) ,\ \ \ D_{\mu }\left( A^{\prime }\right) =VD_{\mu }\left(
A\right) V^{-1},\ \ \ V=\exp \left( -gT^{p}\xi ^{p}\right) \ ,\ \ \ \xi
^{p}=\xi ^{p}\left( x\right) ,  \label{gauge_A}
\end{equation}%
whereas the invariance of $\Psi \left( \phi \right) \overleftarrow{s}$ under
$\phi ^{A}\overset{U}{\rightarrow }\phi ^{\prime A}$ is implied by the
explicit form of $\overleftarrow{s}$ and the fact that the gauge functions $%
\chi ^{p}\left( \phi \right) $ are local and constructed\footnote{%
Note that $F^{pq}=f^{prq}F^{r}$, $f^{pqr}F^{q}G^{r}=\left[ F,G\right\} ^{p}$%
, $\partial _{\mu }F^{p}=\left[ \partial _{\mu },F\right] ^{p}$, $%
F^{p}G^{p}\thicksim \mathrm{Tr}\left( FG\right) $, $%
F_{1}^{pp_{1}}F_{2}^{p_{1}p_{2}}\cdots F_{n}^{p_{n-2}p}=\mathrm{Tr}\left(
F_{1}\cdots F_{n}\right) $ for any quantities $F^{p}$ and $G^{p}$ carrying
the index~$p$, with $F^{\prime }=UFU^{-1}$, $G^{\prime }=UGU^{-1}$. Given
this, the explicit form (\ref{BRST}) of the field variations $\phi ^{A}%
\overleftarrow{s}$ implies the property $\left( \chi \overleftarrow{s}%
\right) ^{\prime }=U\left( \chi \overleftarrow{s}\right) U^{-1}$.} from the
fields $\phi ^{A}$, structure constants $f^{pqr}$ and derivatives $\partial
_{\mu }$, for instance, in Landau and Feynman gauges,%
\begin{equation}
\chi _{\mathrm{L}}^{p}\left( \phi \right) =\partial ^{\mu }A_{\mu }^{p}\ ,\
\ \ \chi _{\mathrm{F}}^{p}\left( \phi \right) =b^{p}+\partial ^{\mu }A_{\mu
}^{p}\ .  \label{LF_gauges}
\end{equation}%
so that, in particular, $\chi ^{p}\left( \phi \right) $ transform as $SU(N)$
vectors, with $\Psi \left( \phi \right) $ being invariant under $\phi ^{A}%
\overset{U}{\rightarrow }\phi ^{\prime A}$,%
\begin{equation*}
\chi \left( \phi ^{\prime }\right) =U\chi \left( \phi \right) U^{-1},\ \ \
\delta \chi ^{p}\left( \phi \right) =gf^{prq}\chi ^{r}\left( \phi \right)
\varsigma ^{q}.
\end{equation*}

By similar reasonings, one can see that local composite fields $\sigma
^{m}\left( \phi \right) $, also constructed from the fields $\phi ^{A}$,
structure constants $f^{pqr}$ and derivatives $\partial _{\mu }$,%
\begin{equation*}
\sigma ^{m}\left( \phi \right) =\sigma ^{p_{1}\cdots p_{k}|\mu _{1}\cdots
\mu _{l}}\left( \phi (x)\right) ,\ \ \ m=\left( x,p_{1}\cdots p_{k},\mu
_{1}\cdots \mu _{l}\right) ,
\end{equation*}%
in the path integral (\ref{Z(J,L)}) for $Z\left( J,L\right) $ transform
under $\phi ^{A}\overset{U}{\rightarrow }\phi ^{\prime A}$ as tensors with
respect to the indices $p_{1},\ldots ,p_{k}$, namely,%
\begin{eqnarray}
\sigma ^{\prime \ p_{1}\cdots p_{k}|\mu _{1}\cdots \mu _{l}}
&=&M^{p_{1}q_{1}}\cdots M^{p_{k}q_{k}}\sigma ^{q_{1}\cdots q_{k}|\mu
_{1}\cdots \mu _{l}},  \notag \\
\delta _{\varsigma }\sigma ^{p_{1}\cdots p_{k}|\mu _{1}\cdots \mu _{l}}
&=&g\sum_{r_{s}\in \left\{ r_{1},\ldots ,r_{k}\right\}
}f^{p_{s}r_{s}q}\sigma ^{p_{1}\cdots r_{s}\cdots p_{k}|\mu _{1}\cdots \mu
_{l}}\varsigma ^{q}\equiv gf^{\left\{ p\right\} \hat{r}q}\sigma
^{p_{1}\cdots \hat{r}\cdots p_{k}|\mu _{1}\cdots \mu _{l}}\varsigma ^{q},
\label{shorthand}
\end{eqnarray}%
which generalizes the transformation of a vector in the index $p$. As a
consequence, the exponential in the path integral (\ref{Z(J,L)}) for $%
Z\left( J,L\right) $ is invariant under $\phi ^{A}\overset{U}{\rightarrow }%
\phi ^{\prime A}$, accompanied by global transformations of the sources $%
J_{A}$, $L_{m}$,%
\begin{equation}
\left( J_{A},L_{m}\right) \overset{U}{\rightarrow }\left( J_{A},L_{m}\right)
^{\prime },\ \ \ J_{A}=(J_{\left( A\right) \mu }^{p},J_{\left( b\right)
}^{p},J_{\left( \bar{c}\right) }^{p},J_{\left( c\right) }^{p}),\ \ \
L_{m}=L_{\mu _{1}\cdots \mu _{j}}^{p_{1}\cdots p_{i}}\ ,
\label{global_sources}
\end{equation}%
in a tensor and infinitesimal form:%
\begin{equation}
\begin{array}{ll}
(J_{\left( A\right) }^{p|\mu },J_{\left( b\right) }^{p},J_{\left( \bar{c}%
\right) }^{p},J_{\left( c\right) }^{p})^{\prime }=M^{pq}(J_{\left( A\right)
}^{q|\mu },J_{\left( b\right) }^{q},J_{\left( \bar{c}\right) }^{q},J_{\left(
c\right) }^{q}), & L_{\mu _{1}\cdots \mu _{j}}^{\prime \ p_{1}\cdots
p_{i}}=M^{p_{1}q_{1}}\cdots M^{p_{i}q_{i}}L_{\mu _{1}\cdots \mu
_{j}}^{q_{1}\cdots q_{i}}\ , \\
\delta _{\varsigma }(J_{\left( A\right) }^{p|\mu },J_{\left( b\right)
}^{p},J_{\left( \bar{c}\right) }^{p},J_{\left( c\right)
}^{p})=gf^{prq}(J_{\left( A\right) }^{r|\mu },J_{\left( b\right)
}^{r},J_{\left( \bar{c}\right) }^{r},J_{\left( c\right) }^{r})\varsigma ^{q},
& \delta _{\varsigma }L_{\mu _{1}\cdots \mu _{j}}^{p_{1}\cdots
p_{i}}=gf^{\left\{ p\right\} \hat{r}q}L_{\mu _{1}\cdots \mu
_{j}}^{p_{1}\cdots \hat{r}\cdots p_{i}}\varsigma ^{q},%
\end{array}
\label{tensor&infinite}
\end{equation}%
which provides the invariance of the source term $J_{A}\phi ^{A}+L_{m}\sigma
^{m}\left( \phi \right) $.

Let us introduce an additional gauge field $B_{\mu }=$ $B_{\mu }^{p}T^{p}$
which transforms as in (\ref{gauge_A}),%
\begin{equation}
B_{\mu }\overset{V}{\rightarrow }B_{\mu }^{\prime }=VB_{\mu
}V^{-1}+g^{-1}V\left( \partial _{\mu }V^{-1}\right) ,  \label{gauge_B}
\end{equation}%
with the inherent property%
\begin{equation}
D_{\mu }\left( B^{\prime }\right) =VD_{\mu }\left( B\right) V^{-1}\ ,\ \ \
D_{\mu }\left( B\right) \equiv \partial _{\mu }+gB_{\mu }\ ,
\label{property}
\end{equation}%
and subject the exponential in the path integral (\ref{Z(J,L)}) to the
following modification:%
\begin{equation}
\left. \exp \left\{ \frac{i}{\hbar }\left[ S_{\mathrm{FP}}\left( \phi
\right) +J\phi +L\sigma \left( \phi \right) \right] \right\} \right\vert
_{\partial _{\mu }\rightarrow D_{\mu }\left( B\right) }\equiv \exp \left\{
\frac{i}{\hbar }\left[ S_{\mathrm{FP}}\left( \phi ,B\right) +J\phi +L\sigma
\left( \phi ,B\right) \right] \right\} ,  \label{integrand}
\end{equation}%
where the replacement $\partial _{\mu }\rightarrow D_{\mu }\left( B\right) $
is understood as%
\begin{equation}
\partial _{\mu }\rightarrow D_{\mu }\left( B\right) :\ \ \ \left[ \partial
_{\mu },\bullet \right] \rightarrow \left[ D_{\mu }\left( B\right) ,\bullet %
\right] \Rightarrow \left[ D_{\mu }\left( A\right) ,\bullet \right]
\rightarrow \left[ D_{\mu }\left( A+B\right) ,\bullet \right] ,  \label{rule}
\end{equation}%
in particular,%
\begin{equation}
F_{\mu \nu }\left( A\right) \rightarrow \left[ D_{\mu }\left( A+B\right)
,D_{\nu }\left( A+B\right) \right] =F_{\mu \nu }\left( A+B\right) .
\label{F(A+B)}
\end{equation}%
The quantum action $S_{\mathrm{FP}}\left( \phi \right) $ in (\ref{integrand}%
) is then replaced by the background action $S_{\mathrm{FP}}\left( \phi
,B\right) $ of the form (\ref{S_FP(B)}), (\ref{givenby}), while as regards
the replacement $\sigma ^{m}\left( \phi \right) \rightarrow \sigma
^{m}\left( \phi ,B\right) $ one should notice the following. Namely, in the
case of local composite fields without higher derivatives, $\sigma
^{m}=\sigma ^{m}\left( \phi ,\partial \phi \right) $, where $\partial _{\mu
} $ enter only via the structures $\left[ \partial _{\mu },\phi \right] $\
and $\left[ D_{\mu }\left( A\right) ,\phi \right] $ in a matrix form, the
introduction of $B_{\mu }$ according to (\ref{rule}) is unambiguous. In the
case of higher derivatives, $\sigma ^{m}=\sigma ^{m}\left( \phi ,\partial
\phi ,\ldots ,\partial \cdots \partial \phi \right) $, the introduction of $%
B_{\mu }$ is not unique, since, prior to including the background field to $%
\partial _{\mu _{1}}\cdots \partial _{\mu _{n}}\phi $, these structures can
be modified by adding terms with a difference of cross derivatives. Such
extra terms are zero in the absence of a background; however, they are
non-vanishing (and arbitrary) in the presence of a background: %
\begin{equation*}
\lbrack \partial _{\mu },\partial _{\nu }]=0,\ \ [D_{\mu }(B),D_{\nu
}(B)]\neq 0.
\end{equation*}
At the same time, the general considerations below remain valid irrespective
of a particular representation of $\sigma ^{m}\left( \phi ,B\right) $
according to (\ref{rule}).

Due to the transformation property (\ref{property}) of the derivative $%
D_{\mu }\left( B\right) $, the generating functional $Z\left( B,J,L\right) $
modified by the field $B_{\mu }$ according to (\ref{integrand}), (\ref{rule}%
),%
\begin{equation}
Z\left( B,J,L\right) =\int d\phi \exp \left\{ \frac{i}{\hbar }\left[ S_{%
\mathrm{FP}}\left( \phi ,B\right) +J_{A}\phi ^{A}+L_{m}\sigma ^{m}\left(
\phi ,B\right) \right] \right\} ,  \label{Z_mod}
\end{equation}%
is invariant under a set of local transformations,%
\begin{eqnarray}
\delta _{\xi }B_{\mu }^{p} &=&D_{\mu }^{pq}\left( B\right) \xi ^{q},  \notag
\\
\delta _{\xi }(J_{\left( A\right) }^{p|\mu },J_{\left( b\right)
}^{p},J_{\left( \bar{c}\right) }^{p},J_{\left( c\right) }^{p})
&=&gf^{prq}(J_{\left( A\right) }^{r|\mu },J_{\left( b\right) }^{r},J_{\left(
\bar{c}\right) }^{r},J_{\left( c\right) }^{r})\xi ^{q},  \label{set_local} \\
\delta _{\xi }L_{\mu _{1}\cdots \mu _{l}}^{p_{1}\cdots p_{k}} &=&gf^{\left\{
p\right\} \hat{r}q}L_{\mu _{1}\cdots \mu _{l}}^{p_{1}\cdots \hat{r}\cdots
p_{k}}\xi ^{q},  \notag
\end{eqnarray}%
given by the gauge transformations (\ref{gauge_B}) of the field $B_{\mu }$
combined with a localized form $U\left( \varsigma \right) \rightarrow
V\left( \xi \right) $ of the transformations (\ref{global_sources}), (\ref%
{tensor&infinite}) for the sources $J_{A}$, $L_{m}$ with infinitesimal
parameters $\xi ^{p}$,%
\begin{equation*}
\left( B_{\mu },J_{A},L_{m}\right) \overset{V}{\rightarrow }\left( B_{\mu
},J_{A},L_{m}\right) ^{\prime }.
\end{equation*}%
The invariance property $Z\left( B^{\prime },J^{\prime },L^{\prime }\right)
=Z\left( B,J,L\right) $ can be established by applying to the transformed
path integral $Z\left( B^{\prime },J^{\prime },L^{\prime }\right) $ a
compensating change of the integration variables:%
\begin{equation*}
\delta _{\xi }\left( A_{\mu }^{p},b^{p},\bar{c}^{p},c^{p}\right)
=gf^{prq}\left( A_{\mu }^{r},b^{r},\bar{c}^{r},c^{r}\right) \xi ^{q},\ \ \
\left( A_{\mu },b,\bar{c},c\right) \overset{V}{\rightarrow }\left( A_{\mu
},b,\bar{c},c\right) ^{\prime },
\end{equation*}%
whose Jacobian equals to unity in view of the antisymmetry of the structure
constants. The invariance of $Z\left( B,J,L\right) $ can be recast in the
form%
\begin{eqnarray}
&&\ \int d^{D}x\ \left\{ \left[ D_{\mu }^{pq}\left( B\right) \xi ^{q}\right]
\frac{\overrightarrow{\delta }}{\delta B_{\mu }^{p}}+g\xi ^{q}f^{\left\{
p\right\} \hat{r}q}L_{\mu _{1}\cdots \mu _{l}}^{p_{1}\cdots \hat{r}\cdots
p_{k}}\frac{\overrightarrow{\delta }}{\delta L_{\mu _{1}\cdots \mu
_{l}}^{p_{1}\cdots p_{k}}}\right.  \notag \\
&&\ +\left. g\xi ^{q}f^{prq}\left( J_{\left( A\right) }^{r|\mu }\frac{%
\overrightarrow{\delta }}{\delta J_{\left( A\right) }^{p|\mu }}+J_{\left(
b\right) }^{r}\frac{\overrightarrow{\delta }}{\delta J_{\left( b\right) }^{p}%
}+J_{\left( \bar{c}\right) }^{r}\frac{\overrightarrow{\delta }}{\delta
J_{\left( \bar{c}\right) }^{p}}+J_{\left( c\right) }^{r}\frac{%
\overrightarrow{\delta }}{\delta J_{\left( c\right) }^{p}}\right) \right\}
Z\left( B,J,L\right) =0.  \label{local_Z}
\end{eqnarray}

To interpret the generating functional $Z\left( B,J,L\right) $ in (\ref%
{Z_mod}), notice that the modified Faddeev--Popov action $S_{\mathrm{FP}%
}\left( \phi ,B\right) $ constructed by the rule (\ref{integrand}), (\ref%
{rule}), (\ref{F(A+B)}) is invariant under the finite local transformations%
\begin{equation}
\left( A_{\mu },b,\bar{c},c\right) \overset{V}{\rightarrow }V\left( A_{\mu
},b,\bar{c},c\right) V^{-1}\ ,\ \ \ B_{\mu }\overset{V}{\rightarrow }VB_{\mu
}V^{-1}+g^{-1}V\partial _{\mu }V^{-1}  \label{A,B,b,c}
\end{equation}%
and takes the form%
\begin{equation}
S_{\mathrm{FP}}\left( \phi ,B\right) =S_{0}\left( A+B\right) +\Psi \left(
\phi ,B\right) \overleftarrow{s}_{\mathrm{q}}\ ,  \label{S_FP(B)}
\end{equation}%
where%
\begin{eqnarray}
S_{0}\left( A+B\right) &=&\left. S_{0}\left( A\right) \right\vert _{\left[
\partial _{\mu },\bullet \right] \rightarrow \left[ D_{\mu }\left( B\right)
,\bullet \right] }=\left. S_{0}\left( A\right) \right\vert _{D_{\mu }\left(
A\right) \rightarrow D_{\mu }\left( A+B\right) }\ ,  \notag \\
\Psi \left( \phi ,B\right) &=&\left. \Psi \left( \phi \right) \right\vert _{
\left[ \partial _{\mu },\bullet \right] \rightarrow \left[ D_{\mu }\left(
B\right) ,\bullet \right] },\ \ \ \overleftarrow{s}_{\mathrm{q}}=\left.
\overleftarrow{s}\right\vert _{D_{\mu }\left( A\right) \rightarrow D_{\mu
}\left( A+B\right) }\ ,  \label{givenby}
\end{eqnarray}%
namely,%
\begin{eqnarray}
\left( A_{\mu },b,\bar{c},c\right) \overleftarrow{s}_{\mathrm{q}} &=&\left( %
\left[ D_{\mu }\left( A+B\right) ,c\right] ,0,b,g/2\left[ c,c\right]
_{+}\right) \ ,  \notag \\
\left( A_{\mu }^{p},b^{p},\bar{c}^{p},c^{p}\right) \overleftarrow{s}_{%
\mathrm{q}} &=&\left( D_{\mu }^{pq}\left( A+B\right)
c^{q},0,b^{p},g/2f^{pqr}c^{q}c^{r}\right) \ .  \label{mod_Slavnov}
\end{eqnarray}%
For instance, the Landau and Feynman gauges (\ref{LF_gauges}) are modified
to the background gauges%
\begin{equation}
\chi _{\mathrm{L}}^{p}\left( \phi ,B\right) =D_{\mu }^{pq}\left( B\right)
A^{q|\mu }\ ,\ \ \ \chi _{\mathrm{F}}^{p}\left( \phi ,B\right) =b^{p}+D_{\mu
}^{pq}\left( B\right) A^{q|\mu }.  \label{background_gauges}
\end{equation}

In the background field method, a quantum action $S_{\mathrm{FP}}\left( \phi
,B\right) $ constructed according to (\ref{S_FP(B)}), (\ref{mod_Slavnov}) is
known as the Faddeev--Popov action with a background field $B_{\mu }$. The
quantum action $S_{\mathrm{FP}}\left( \phi ,B\right) $ is invariant under
global transformations of $\phi ^{A}$, with a nilpotent generator $%
\overleftarrow{s}_{\mathrm{q}}$ and an anticommuting parameter $\lambda $:%
\begin{equation}
\delta S_{\mathrm{FP}}\left( \phi ,B\right) =0,\ \ \ \delta \phi ^{A}=\phi
^{A}\overleftarrow{s}_{\mathrm{q}}\lambda ,\ \ \ \overleftarrow{s}_{\mathrm{q%
}}\overleftarrow{s}_{\mathrm{q}}=0,\ \ \ \epsilon \left( \lambda \right) =%
\mathrm{gh}\left( \lambda \right) =1.  \label{BRST_mod}
\end{equation}%
At the infinitesimal level, the local transformations (\ref{A,B,b,c}) for
the fields $A_{\mu }$, $B_{\mu }$ are known as \emph{background
transformations}, and the transformations of $A_{\mu }$, $B_{\mu }$
corresponding to the modified Slavnov variation $\overleftarrow{s}_{\mathrm{q%
}}$ in (\ref{S_FP(B)}), (\ref{mod_Slavnov}), (\ref{BRST_mod}) are known as
\emph{quantum transformations},%
\begin{equation*}
\begin{array}{r}
\mathrm{background:} \\
\mathrm{quantum:}%
\end{array}%
\begin{array}{ll}
\delta _{\mathrm{b}}A_{\mu }=g\left[ A_{\mu },T^{p}\xi ^{p}\right] , &
\delta _{\mathrm{b}}B_{\mu }=\left[ D_{\mu }\left( B\right) ,T^{p}\xi ^{p}%
\right] , \\
\delta _{\mathrm{q}}A_{\mu }=\left[ D_{\mu }\left( A+B\right) ,T^{p}\xi ^{p}%
\right] , & \delta _{\mathrm{q}}B_{\mu }=0,%
\end{array}%
\end{equation*}%
whereas the classical action $S_{0}\left( A+B\right) $ is left invariant by
both of these types of transformations. In this connection, the family of
background gauges $\chi ^{p}\left( \phi ,B\right) =\tilde{\chi}^{p}\left(
A,B\right) +\left( \alpha /2\right) b^{p}$, parameterized by $\alpha \not=0$
and defined according to (\ref{givenby}),%
\begin{equation*}
\chi ^{p}\left( A,B\right) =\left. \chi ^{p}\left( A\right) \right\vert _{
\left[ \partial _{\mu },\bullet \right] \rightarrow \left[ D_{\mu }\left(
B\right) ,\bullet \right] }\ ,
\end{equation*}%
with the Nakanishi--Lautrup fields $b^{p}$ integrated out of (\ref{Z_mod})
by the shift $b^{p}\rightarrow b^{p}+\alpha ^{-1}\tilde{\chi}$ at the
vanishing sources, $J=L=0$, reduces the vacuum functional $Z\left( B\right) $
to the form (for future convenience, we denote $A\equiv Q$),%
\begin{eqnarray}
Z(B) &=&\int dQ\;d\overline{c}\;dc\;\exp \left\{ \frac{i}{\hbar }\left[
S_{0}(Q+B)+S_{\mathrm{gf}}(Q,B)+S_{\mathrm{gh}}(Q,B;\overline{c},c)\right]
\right\} ,  \notag \\
S_{\mathrm{gf}}\left( Q,B\right) &=&-\frac{1}{2\alpha }\int d^{D}x\ \tilde{%
\chi}^{p}\tilde{\chi}^{p},\ \ \ S_{\mathrm{gh}}\left( Q,B\right) =\int
d^{D}x\ \bar{c}^{p}\left. \delta _{\mathrm{q}}\tilde{\chi}^{p}\right\vert
_{\xi \rightarrow c}\ ,  \label{rule2}
\end{eqnarray}%
where the gauge-fixing term $S_{\mathrm{gf}}=S_{\mathrm{gf}}\left(
Q,B\right) $ is invariant under the background transformations, $\delta _{%
\mathrm{b}}S_{\mathrm{gf}}=0$, due to $\delta _{\mathrm{b}}\tilde{\chi}%
^{p}=gf^{prq}\tilde{\chi}^{r}\xi ^{q}$, which may be employed to define the
quantum action in background gauges $\tilde{\chi}^{p}\left( Q,B\right) $
depending on the quantum $Q$ and background $B$ fields with the associated
background and quantum transformations (see also \cite{Abbott}),%
\begin{equation}
\begin{array}{ll}
\delta _{\mathrm{b}}B_{\mu }^{p}=D_{\mu }^{pq}\left( B\right) \xi ^{q}, &
\delta _{\mathrm{b}}Q_{\mu }^{p}=gf^{prq}Q_{\mu }^{r}\xi ^{q}, \\
\delta _{\mathrm{q}}B_{\mu }^{p}=0, & \delta _{\mathrm{q}}Q_{\mu
}^{p}=D_{\mu }^{pq}\left( Q+B\right) \xi ^{q}.%
\end{array}
\label{bq}
\end{equation}%
By construction, the quantum action and the integrand of $Z(B)$\ in (\ref%
{rule2}) are invariant under the residual local transformations (\ref%
{A,B,b,c}), namely,
\begin{equation}
\left( Q_{\mu },\bar{c},c\right) \overset{V}{\rightarrow }V\left( Q_{\mu },%
\bar{c},c\right) V^{-1}\ ,\ \ \ B_{\mu }\overset{V}{\rightarrow }VB_{\mu
}V^{-1}+g^{-1}V\partial _{\mu }V^{-1},  \label{Q,B,c}
\end{equation}%
which translates infinitesimally, $\delta \left( B_{\mu },Q_{\mu },\bar{c}%
,c\right) $, to the background transformations $\delta _{\mathrm{b}}$\
accompanied by some compensating transformations of the ghost fields:%
\begin{equation*}
\delta \left( B_{\mu }^{p},\ Q_{\mu }^{p},\ \bar{c}^{p},\ c^{p}\right)
=(\delta _{\mathrm{b}}B_{\mu }^{p},\ \delta _{\mathrm{b}}Q_{\mu }^{p},\
gf^{prq}\bar{c}^{r}\xi ^{q},\ gf^{prq}c^{r}\xi ^{q}).
\end{equation*}%
Given this, we interpret $Z\left( B,J,L\right) $ defined according to (\ref%
{Z(J,L)}), (\ref{integrand}), (\ref{rule}), (\ref{Z_mod}) as a generating
functional of Green's functions for Yang--Mills theories with composite
fields in the background field method, or as a generating functional of
Green's functions with composite and background fields for such theories. As
we shall see below, this interpretation provides the existence of a
corresponding gauge-invariant effective action with composite and background
fields.

\section{Ward Identities, Effective Action\label{sec3}}

\setcounter{equation}{0}\renewcommand{\theequation}{3.\arabic{equation}}

Let us now present the corresponding generating functionals of connected and
vertex Green's functions with composite and background fields and examine
their properties. To this end, we first introduce an extended generating
functional $Z\left( B,J,L,\phi ^{\ast }\right) $,%
\begin{eqnarray}
Z\left( B,J,L,\phi ^{\ast }\right) &=&\int d\phi \exp \left\{ \frac{i}{\hbar
}\left[ S_{\mathrm{ext}}\left( \phi ,\phi ^{\ast },B\right) +J_{A}\phi
^{A}+L_{m}\sigma ^{m}\left( \phi ,B\right) \right] \right\}  \notag \\
&\equiv &\int \mathcal{I}_{\phi ,\phi ^{\ast },B}^{\Psi }\exp \left\{ \frac{i%
}{\hbar }\left[ J_{A}\phi ^{A}+L_{m}\sigma ^{m}\left( \phi ,B\right) \right]
\right\} ,  \label{Z_ext}
\end{eqnarray}%
with an extended quantum action $S_{\mathrm{ext}}\left( \phi ,\phi ^{\ast
},B\right) $ given by%
\begin{eqnarray*}
S_{\mathrm{ext}}\left( \phi ,\phi ^{\ast },B\right) &=&S_{\mathrm{FP}}\left(
\phi ,B\right) +\phi _{A}^{\ast }(\phi ^{A}\overleftarrow{s}_{\mathrm{q}}),
\\
\phi _{A}^{\ast }(\phi ^{A}\overleftarrow{s}_{\mathrm{q}}) &=&\int
d^{D}x\left( A_{\mu }^{\ast p}D^{pq|\mu }\left( A+B\right) c^{q}+\bar{c}%
^{\ast p}b^{p}+\left( g/2\right) f^{pqr}c^{\ast p}c^{q}c^{r}\right) , \\
&=&2\int d^{D}x\ \mathrm{Tr}\left( A_{\mu }^{\ast }\left[ D^{\mu }\left(
A+B\right) ,c\right] +\bar{c}^{\ast }b+\left( g/2\right) c^{\ast }\left[ c,c%
\right] _{+}\right) ,
\end{eqnarray*}%
where $\phi _{A}^{\ast }$, $\epsilon \left( \phi _{A}^{\ast }\right)
=\epsilon \left( \phi ^{A}\right) +1$, $\mathrm{gh}\left( \phi _{A}^{\ast
}\right) =-\mathrm{gh}\left( \phi ^{A}\right) $, is a set of antifields
introduced as sources to the variations $\phi ^{A}\overleftarrow{s}_{\mathrm{%
q}}$,%
\begin{equation*}
\phi _{A}^{\ast }=\left( A_{\mu }^{\ast p},b^{\ast p},\bar{c}^{\ast
p},c^{\ast p}\right) ,\ \ \ \left( A_{\mu }^{\ast },b^{\ast },\bar{c}^{\ast
},c^{\ast }\right) \equiv T^{p}\left( A_{\mu }^{\ast p},b^{\ast p},\bar{c}%
^{\ast p},c^{\ast p}\right) .
\end{equation*}

Due to the invariance property (\ref{BRST_mod}) of $S_{\mathrm{FP}}\left(
\phi ,B\right) $, the extended quantum action $S_{\mathrm{ext}}\left( \phi
,\phi ^{\ast },B\right) $ satisfies the identity%
\begin{equation*}
S_{\mathrm{ext}}\left( \phi ,\phi ^{\ast },B\right) \overleftarrow{s}_{%
\mathrm{q}}=0,
\end{equation*}%
which can be recast in the form of a master equation,%
\begin{equation*}
\left( S_{\mathrm{ext}},S_{\mathrm{ext}}\right) =0,\ \ \ \left( F,G\right)
\equiv F\frac{\overleftarrow{\delta }}{\delta \phi ^{A}}\frac{%
\overrightarrow{\delta }}{\delta \phi _{A}^{\ast }}G-\left( -1\right)
^{\left( \epsilon \left( F\right) +1\right) \left( \epsilon \left( G\right)
+1\right) }G\frac{\overleftarrow{\delta }}{\delta \phi ^{A}}\frac{%
\overrightarrow{\delta }}{\delta \phi _{A}^{\ast }}F,
\end{equation*}%
or, equivalently,%
\begin{equation*}
\left( -1\right) ^{\epsilon \left( \phi ^{A}\right) }\frac{\overrightarrow{%
\delta }}{\delta \phi ^{A}}\frac{\overrightarrow{\delta }}{\delta \phi
_{A}^{\ast }}\exp \left[ \left( i/\hbar \right) S_{\mathrm{ext}}\right] =0,
\end{equation*}%
which holds due to the complete antisymmetry of the structure constants $%
f^{pqr}$.

Let us make in the integrand (\ref{Z_ext}) a finite FD BRST transformation
(see Appendix~\ref{A.I} for details) with a generator $\overleftarrow{s}_{%
\mathrm{q}}$ given by (\ref{mod_Slavnov}) and a Grassmann-odd functional
parameter $\lambda (\phi ,B)$,%
\begin{equation}
\phi ^{A}\rightarrow \phi ^{\prime A}=\phi ^{A}+\phi ^{A}\overleftarrow{s}_{%
\mathrm{q}}\lambda (\phi ,B),  \label{FFDBRST}
\end{equation}%
where $\lambda (\phi ,B)$ is related to a finite change \cite{Reshetnyak,LL1}
of the gauge fermion $\Delta \Psi \left( \phi ,B\right) $ depending also on
the background field. For a finite constant $\lambda $, the following
invariance property holds true:%
\begin{equation}
\mathcal{I}_{\phi +\phi \overleftarrow{s}_{\mathrm{q}}\lambda ,\phi ^{\ast
},B}^{\Psi }=\mathcal{I}_{\phi ,\phi ^{\ast },B}^{\Psi }\ ,\ \ \ \lambda
\frac{\overleftarrow{\delta }}{\delta \phi }=\lambda \frac{\overleftarrow{%
\delta }}{\delta B}=0,  \label{brstinv}
\end{equation}%
whereas the FD parameter $\lambda (\phi ,B)$ in the Jacobian $\mathrm{Sdet}%
\left\Vert {\delta \phi ^{\prime }}/{\delta \phi }\right\Vert =\left[
1+\lambda (\phi ,B)\overleftarrow{s}_{q}\right] ^{-1}$ of (\ref{JacobianB})
for the change of variables (\ref{FFDBRST}), given the choice of $\lambda
(\phi ,B)=\lambda (\phi ,B|\Delta \Psi )$ in the form%
\begin{eqnarray}
\lambda (\phi ,B\left\vert \Delta \Psi \right. ) &=&\Delta \Psi (\phi
,B)\left\{ \left[ \Delta \Psi (\phi ,B)\right] \overleftarrow{s}_{\mathrm{q}%
}\right\} ^{-1}\left( \exp \left\{ -\frac{i}{\hbar }\left[ {\Delta }\Psi
(\phi ,B)\right] \overleftarrow{s}_{\mathrm{q}}\right\} -1\right)  \notag \\
&=&-\frac{i}{\hbar }\Delta \Psi (\phi ,B)+{o}(\Delta \Psi ),  \label{lamDPsi}
\end{eqnarray}%
implies the independence of the extended vacuum functional, $Z_{\Psi }\left(
B,\phi ^{\ast }\right) =Z\left( B,0,0,\phi ^{\ast }\right) $, from a finite
variation of an admissible gauge condition, $\Psi (\phi ,B)\rightarrow $ $%
\Psi (\phi ,B)+\Delta \Psi (\phi ,B)$, namely,
\begin{equation}
Z_{\Psi }\left( B,\phi ^{\ast }\right) =Z_{\Psi +\Delta \Psi }\left( B,\phi
^{\ast }\right) \Longleftrightarrow \mathcal{I}_{\phi +\phi \overleftarrow{s}%
_{\mathrm{q}}\lambda ,\phi ^{\ast },B}^{\Psi }=\mathcal{I}_{\phi ,\phi
^{\ast },B}^{\Psi +\Delta \Psi }\ .  \label{gindvacbcomp}
\end{equation}%
The latter property, once finite FD BRST transformations are applied to the
integrand of the generating functional (\ref{Z_ext}), leads to a \emph{%
modified Ward identity}, suggested for the first time within the BV
formalism in \cite{BLTfin}, now with respect to a functional $Z\left(
B,J,L,\phi ^{\ast }\right) $ extended by antifields and a background field:
\begin{equation}
\left\langle \left[ 1+\frac{i}{\hbar }J_{A}\left( \phi ^{A}\overleftarrow{s}%
_{\mathrm{q}}\right) \lambda (\phi ,B)\right] \left[ 1+\lambda (\phi ,B)%
\overleftarrow{s}_{\mathrm{q}}\right] ^{-1}\right\rangle _{\Psi ,B,J,L,\phi
^{\ast }}=1\ .  \label{smWIclalg}
\end{equation}%
Here, the notation $\langle \mathcal{D}\rangle _{\Psi ,B,J,L,\phi ^{\ast }}$%
, with a certain functional $\mathcal{D}\left( \phi ,\phi ^{\ast },B\right) $%
, implies a source-dependent expectation value corresponding to a
gauge-fixing functional $\Psi (\phi ,B)$,%
\begin{equation}
\left\langle \mathcal{D}\right\rangle _{\Psi ,B,J,L,\phi ^{\ast
}}=Z^{-1}(B,J,L,\phi ^{\ast })\int d\phi \ \mathcal{D}\left( \phi ,\phi
^{\ast },B\right) \exp \left\{ \frac{i}{\hbar }\left[ S_{\mathrm{ext}}\left(
\phi ,\phi ^{\ast },B\right) +J_{A}\phi ^{A}+L_{m}\sigma ^{m}\left( \phi
,B\right) \right] \right\} ,  \label{aexv}
\end{equation}%
with the normalization $\left\langle 1\right\rangle _{\cdots }=1$, where the
dots stand for $\Psi ,B,J,L,\phi ^{\ast }$\ as in (\ref{smWIclalg}). Using
the familiar rules $\langle \phi ^{A}\rangle _{\cdots }=Z^{-1}{\frac{\hbar }{%
i}\overrightarrow{\delta }/\delta J_{A}}Z$ and $\langle \phi ^{A}%
\overleftarrow{s}_{\mathrm{q}}\rangle _{\cdots }=Z^{-1}{\frac{\hbar }{i}%
\delta /\delta \phi _{A}^{\ast }}Z$, one presents the modified identity (\ref%
{smWIclalg}) in the form%
\begin{equation}
\left\{ \hat{\omega}\lambda \left( \frac{\hbar \overrightarrow{\delta }}{%
i\delta J},B\right) +\left[ \sum_{n=1}(-1)^{n}\left( \lambda \overleftarrow{s%
}_{\mathrm{q}}\right) ^{n}\left( \frac{\hbar \overrightarrow{\delta }}{%
i\delta J},B\right) \right] \left[ 1+\hat{\omega}\lambda \left( \frac{\hbar
\overrightarrow{\delta }}{i\delta J},B\right) \right] \right\} Z=0,
\label{mWIclalg}
\end{equation}%
with a nilpotent Grassmann-odd operator $\hat{\omega}$,%
\begin{equation}
\hat{\omega}=\left[ J_{A}+L_{m}\sigma _{,A}^{m}\left( \frac{\hbar }{i}\frac{%
\overrightarrow{\delta }}{\delta J},B\right) \right] \frac{\overrightarrow{%
\delta }}{\delta \phi _{A}^{\ast }},\ \ \ \hat{\omega}^{2}=0.  \label{omega}
\end{equation}%
In deriving the $\lambda $-dependent identity (\ref{mWIclalg}), we have used
the expansion $(1+x)^{-1}$ = $1+\sum_{n>0}(-1)^{n}x^{n}$ with $x=\lambda
\overleftarrow{s}_{\mathrm{q}}$. Notice that, instead of the monomial $%
\left( \lambda \overleftarrow{s}_{\mathrm{q}}\right) ^{n}$, we can
equivalently apply $[(\hbar /i)\lambda _{,A}\overrightarrow{\delta }/\delta
\phi _{A}^{\ast }]^{n}$ with $\lambda _{,A}=\lambda \overleftarrow{\delta }%
/\delta \phi ^{A}$ under the sign of functional integral, which leads to
another representation of the identity (\ref{mWIclalg}),
\begin{equation}
\left\{ \hat{\omega}\lambda \left( \frac{\hbar \overrightarrow{\delta }}{%
i\delta J},B\right) +\left( \sum_{n=1}(-1)^{n}\left[ \lambda _{,A}\left(
\frac{\hbar \overrightarrow{\delta }}{i\delta J},B\right) \frac{\hbar
\overrightarrow{\delta }}{i\delta \phi _{A}^{\ast }}\right] ^{n}\right) %
\left[ 1+\hat{\omega}\lambda \left( \frac{\hbar \overrightarrow{\delta }}{%
i\delta J},B\right) \right] \right\} Z=0.  \label{mWIclalg1}
\end{equation}%
For an infinitesimal FD parameter $\lambda $, the identity (\ref{mWIclalg1})
acquires the form
\begin{equation}
\left[ \hat{\omega}\lambda \left( \frac{\hbar \overrightarrow{\delta }}{%
i\delta J},B\right) -\frac{\hbar }{i}\lambda _{,A}\left( \frac{\hbar
\overrightarrow{\delta }}{i\delta J},B\right) \frac{\overrightarrow{\delta }%
}{\delta \phi _{A}^{\ast }}\right] Z=0.  \label{mWIclalglin}
\end{equation}%
For a constant $\lambda $, namely, $\lambda \overleftarrow{s}_{\mathrm{q}}=0$%
, the relation (\ref{mWIclalg}) contains the usual Ward identity \cite{6},
depending parametrically on a background field:
\begin{equation}
\hat{\omega}Z\left( B,J,L,\phi ^{\ast }\right) =0.  \label{Ward_Z}
\end{equation}%
The generating functional $W\left( B,J,L,\phi ^{\ast }\right) $ of connected
Green's functions, $W=\left( \hbar /i\right) \ln Z$, satisfies a related
modified Ward identity,%
\begin{eqnarray}
&&\hat{\Omega}\langle \lambda (B)\rangle W+\left[ \sum_{n=1}(-1)^{n}\langle
\left( \lambda (B)\overleftarrow{s}_{\mathrm{q}}\rangle \right) ^{n}\right] %
\left[ 1+\hat{\Omega}\langle \lambda \left( B\right) \rangle \right] W=0,
\label{modWard_W} \\
&&\ \hat{\Omega}=\left[ J_{A}+L_{m}\sigma _{,A}^{m}\left( \frac{\hbar }{i}%
\frac{\overrightarrow{\delta }}{\delta J}+\frac{\overrightarrow{\delta }W}{%
\delta J},B\right) \right] \frac{\overrightarrow{\delta }}{\delta \phi
_{A}^{\ast }},\ \ \langle \lambda (B)\rangle \equiv \lambda \left( \frac{%
\hbar }{i}\frac{\overrightarrow{\delta }}{\delta J}+\frac{\overrightarrow{%
\delta }W}{\delta J},B\right) ,  \label{Omega}
\end{eqnarray}%
deduced by a unitary transformation of the operator $\hat{\omega}$ in (\ref%
{omega}),
\begin{equation}
\hat{\Omega}=\hat{U}^{-1}\hat{\omega}\hat{U},\ \ \ \hat{U}=\exp \left(
i/\hbar W\right) .  \label{unittransf}
\end{equation}%
Once again, an infinitesimal FD $\lambda $ reduces the identity (\ref%
{modWard_W}), or, equivalently, (\ref{mWIclalg1}), to the form%
\begin{equation}
\left[ \hat{\Omega}\langle \lambda (B)\rangle -\frac{\hbar }{i}\langle
\lambda _{,A}(B)\rangle \frac{\overrightarrow{\delta }}{\delta \phi
_{A}^{\ast }}\right] W=0.  \label{modWIclalglin}
\end{equation}%
In turn, for a constant $\lambda $ the relation (\ref{modWard_W}) contains
the usual Ward identity \cite{6}
\begin{equation}
\hat{\Omega}W=\left[ J_{A}+L_{m}\sigma _{,A}^{m}\left( \frac{\hbar }{i}\frac{%
\overrightarrow{\delta }}{\delta J}+\frac{\overrightarrow{\delta }W}{\delta J%
},B\right) \right] \frac{\overrightarrow{\delta }}{\delta \phi _{A}^{\ast }}%
W=0.  \label{Ward_W}
\end{equation}%
As we introduce a generating functional $\Gamma \left( B,\phi ,\Sigma ,\phi
^{\ast }\right) $ of vertex Green's functions with composite fields (on a
background)\ by using a double Legendre transformation \cite{6},
\begin{equation}
\Gamma \left( B,\phi ,\Sigma ,\phi ^{\ast }\right) =W\left( B,J,L,\phi
^{\ast }\right) -J_{A}\phi ^{A}-L_{m}\left[ \sigma ^{m}\left( \phi ,B\right)
+\Sigma ^{m}\right] \ ,  \label{Gamma}
\end{equation}%
where%
\begin{equation*}
\phi ^{A}=\frac{\overrightarrow{\delta }W}{\delta J_{A}},\ \Sigma ^{m}=\frac{%
\overrightarrow{\delta }W}{\delta L_{m}}-\sigma ^{m}\left( \frac{%
\overrightarrow{\delta }W}{\delta J},B\right) ,\ -J_{A}=\Gamma \frac{%
\overleftarrow{\delta }}{\delta \phi ^{A}}+L_{m}\sigma _{,A}^{m}(\phi ,B),\
-L_{m}=\Gamma \frac{\overleftarrow{\delta }}{\delta \Sigma ^{m}}\ ,
\end{equation*}%
the modified Ward identity (\ref{modWard_W}) acquires the form (see Appendix~%
\ref{A.II} for details)
\begin{eqnarray}
&&\hat{\omega}_{\Gamma }\langle \langle \lambda (B)\rangle \rangle \Gamma
+\left\{ \sum_{n=1}(-1)^{n}\left[ \langle \langle \lambda (B)\overleftarrow{s%
}_{\mathrm{q}}\rangle \rangle \right] ^{n}\right\} \left[ 1+\hat{\omega}%
_{\Gamma }\langle \langle \lambda \left( B\right) \rangle \rangle \right]
\Gamma =0,  \label{modWard_Gamma_1} \\
&&\hat{\omega}_{\Gamma }=(\Gamma ,\bullet )+\left[ \sigma _{,A}^{m}(\hat{\phi%
},B)-\sigma _{,A}^{m}(\phi ,B)\right] \frac{\overrightarrow{\delta }\Gamma }{%
\delta \phi _{A}^{\ast }}\frac{\overrightarrow{\delta }}{\delta \Sigma ^{m}}
\notag \\
&&\ \ -\frac{i}{\hbar }\left( \left[ \Gamma \frac{\overleftarrow{\delta }}{%
\delta \Sigma ^{m}}\left( \sigma _{,C}^{m}(\hat{\phi},B)\frac{%
\overrightarrow{\delta }\Gamma }{\delta \phi _{C}^{\ast }}\right) ,\Phi ^{%
\mathsf{a}}\right\} \frac{\overrightarrow{\delta }}{\delta \Phi ^{\mathsf{a}}%
}+\left[ \Gamma \frac{\overleftarrow{\delta }}{\delta \Sigma ^{m}}\left(
\sigma _{,C}^{m}(\hat{\phi},B)\frac{\overrightarrow{\delta }\Gamma }{\delta
\phi _{C}^{\ast }}\right) ,\,\sigma ^{n}({\phi },B)\right\} \frac{%
\overrightarrow{\delta }}{\delta \Sigma ^{n}}\right)  \notag \\
&&\ \ +\frac{i}{\hbar }(-1)^{\epsilon (\sigma ^{n})+\epsilon (\phi
^{D})}\sigma _{,D}^{n}({\phi },B)\left[ \Gamma \frac{\overleftarrow{\delta }%
}{\delta \Sigma ^{m}}\left( \sigma _{,C}^{m}(\hat{\phi},B)\frac{%
\overrightarrow{\delta }\Gamma }{\delta \phi _{C}^{\ast }}\right) ,\phi
^{D}\right\} \frac{\overrightarrow{\delta }}{\delta \Sigma ^{n}}  \notag \\
&&\ \ +(-1)^{\epsilon (\sigma ^{m})+\epsilon (\phi ^{D})\epsilon (\phi
^{A})} \left[ \sigma _{,D}^{m}({\phi },B),\Gamma \frac{\overleftarrow{\delta
}}{\delta \Sigma ^{n}}\sigma _{,A}^{n}(\hat{\phi},B)\right\} \left(
G^{\prime \prime -1}\right) ^{A\mathsf{a}}\left( \frac{\overrightarrow{%
\delta }}{\delta \Phi ^{\mathsf{a}}}\frac{\overrightarrow{\delta }\Gamma }{%
\delta \phi _{D}^{\ast }}\right) \frac{\overrightarrow{\delta }}{\delta
\Sigma ^{m}},  \label{omegaG}
\end{eqnarray}%
where $\langle \langle \lambda (B)\rangle \rangle \equiv \lambda (\hat{\phi}%
,B)$, and the following notation is used:
\begin{eqnarray}
\hat{\phi}{}^{A} &\equiv &\phi ^{A}+i\hbar \left( G^{\prime \prime
-1}\right) ^{A\mathsf{a}}\overrightarrow{\delta }/\delta \Phi ^{\mathsf{a}}\
,\ \ \ \left( G^{\prime \prime }\right) _{\mathsf{ab}}\equiv \overrightarrow{%
\delta }F_{\mathsf{b}}/\delta \Phi ^{\mathsf{a}}\ ,  \label{phi_hat} \\
\Phi ^{\mathsf{a}} &=&\left( \phi ^{A},\Sigma ^{m}\right) ,\ \ \ F_{\mathsf{a%
}}=\left( \Gamma _{,A}-\Gamma _{,n}\sigma _{,A}^{n}\left( \phi ,B\right)
,\Gamma _{,m}\right) ,\ \ \ \Gamma _{,\mathsf{a}}\equiv \Gamma
\overleftarrow{\delta }/\delta \Phi ^{\mathsf{a}}\ .  \notag
\end{eqnarray}%
For a constant infinitesimal parameter $\lambda $, the modified Ward
identity (\ref{modWard_Gamma_1}) is reduced to the usual one \cite{6},
however, now with a background included,%
\begin{equation}
\frac{1}{2}\left( \Gamma ,\Gamma \right) =-\Gamma \frac{\overleftarrow{%
\delta }}{\delta \Sigma ^{m}}\left[ \sigma _{,A}^{m}(\hat{\phi},B)-\sigma
_{,A}^{m}(\phi ,B)\right] \frac{\overrightarrow{\delta }}{\delta \phi
_{A}^{\ast }}\Gamma .  \label{Ward_Gamma_1}
\end{equation}

The extended generating functional $Z\left( B,J,L,\phi ^{\ast }\right) $ of
Green's functions (\ref{Z_ext}) and the related functional $W\left(
B,J,L,\phi ^{\ast }\right) $ exhibit an invariance under the local
transformations (\ref{set_local}) accompanied by the following
transformations of the antifields:%
\begin{equation}
\delta _{\xi }\left( A_{\mu }^{\ast p},b^{\ast p},\bar{c}^{\ast p},c^{\ast
p}\right) =gf^{prq}\left( A_{\mu }^{\ast r},b^{\ast r},\bar{c}^{\ast
r},c^{\ast r}\right) \xi ^{q}.  \label{local_anti}
\end{equation}%
Then the effective action $\Gamma _{\mathrm{eff}}\left( B,\Sigma \right) $
with composite and background fields defined as%
\begin{equation}
\Gamma _{\mathrm{eff}}\left( B,\Sigma \right) =\left. \Gamma \left( B,\phi
,\Sigma ,\phi ^{\ast }\right) \right\vert _{\phi =\phi ^{\ast }=0}
\label{Gamma_eff}
\end{equation}%
satisfies an identity (see Appendix \ref{A.III}) related to (\ref{local_Z}),
\begin{equation}
\int d^{D}x\ \left\{ \left[ D_{\mu }^{pq}\left( B\right) \xi ^{q}\right]
\frac{\overrightarrow{\delta }}{\delta B_{\mu }^{p}}+gf^{\left\{ p\right\}
\hat{r}q}\Sigma _{\mu _{1}\cdots \mu _{l}}^{p_{1}\cdots \hat{r}\cdots
p_{k}}\xi ^{q}\frac{\overrightarrow{\delta }}{\delta \Sigma _{\mu _{1}\cdots
\mu _{l}}^{p_{1}\cdots p_{k}}}\right\} \Gamma _{\mathrm{eff}}\left( B,\Sigma
\right) =0,  \label{ident}
\end{equation}%
and is thereby invariant under the local transformations
\begin{equation}
\delta _{\xi }B_{\mu }^{p}=D_{\mu }^{pq}\left( B\right) \xi ^{q},\ \ \
\delta _{\xi }\Sigma _{\mu _{1}\cdots \mu _{l}}^{p_{1}\cdots
p_{k}}=gf^{\left\{ p\right\} \hat{r}q}\Sigma _{\mu _{1}\cdots \mu
_{l}}^{p_{1}\cdots \hat{r}\cdots p_{k}}\xi ^{q},  \label{back_gauge}
\end{equation}%
which consist of the initial gauge transformations for the background field $%
B_{\mu }^{p}$ and of the local $SU(N)$ transformations\ for the fields $%
\Sigma _{\mu _{1}\cdots \mu _{l}}^{p_{1}\cdots p_{k}}$.

Returning once again to the modified Ward identities, we point out that,
once the composite fields $\sigma ^{m}(\phi ,B)$ are absent, the formulas (%
\ref{mWIclalg}), (\ref{modWard_W}), (\ref{modWard_Gamma_1}) are reduced to
those involving the respective functionals $Z|_{L=0}$, $W|_{L=0}$, $\Gamma
|_{\Sigma =\sigma =0}$, which presents a new form of $\lambda $-dependent\
Ward identities, additional to the usual Ward identities ($\lambda =\mathrm{%
const}$) for these functionals. The deduction of the modified Ward
identities (\ref{mWIclalg}), (\ref{modWard_W}), (\ref{modWard_Gamma_1}) for
the generating functionals $Z$, $W$, $\Gamma $ of Green's functions with
composite and background fields, implying the respective usual Ward
identities (\ref{Ward_Z}), (\ref{Ward_W}), (\ref{Ward_Gamma_1}) by means of
finite FD BRST transformations, comprises the results of this section that
have a generic character. Finally, it should be noted that we have assumed
the existence of a \textquotedblleft deep\textquotedblright\ gauge-invariant
regularization preserving the Ward identities (see, e.g., \cite{Stepanyantz}%
), as we expect the corresponding renormalized generating functionals to
obey the same properties as the unrenormalized ones.

\section{Gauge Dependence Problem\label{sec4}}

\setcounter{equation}{0}\renewcommand{\theequation}{4.\arabic{equation}}

Let us study the gauge dependence of the generating functionals $Z\left(
B,J,L,\phi ^{\ast }\right) $, $W\left( B,J,L,\phi ^{\ast }\right) $, $\Gamma
\left( B,\phi ,\Sigma ,\phi ^{\ast }\right) $ of Green's functions with
composite and background fields. In this regard, the representations (\ref%
{lamDPsi}) and (\ref{mWIclalg}) also provide a relation which describes the
gauge dependence of $Z(B,J,L,\phi ^{\ast })=Z_{\Psi }$ for a finite change $%
\Psi \rightarrow \Psi +\Delta \Psi $:%
\begin{equation}
\Delta Z_{\Psi }=Z_{\Psi +\Delta \Psi }-Z_{\Psi }=\hat{\omega}{\lambda }%
\left( \frac{\hbar \overrightarrow{\delta }}{i\delta J},B\left\vert -\Delta
\Psi \right. \right) Z_{\Psi }=\frac{i}{\hbar }\hat{\omega}{\Delta }\Psi
\left( \frac{\hbar \overrightarrow{\delta }}{i\delta J},B\right) Z_{\Psi
}+o(\Delta \Psi ).  \label{GDInew1}
\end{equation}%
The corresponding finite change $\Delta W_{\Psi }=\Delta W\left( B,J,L,\phi
^{\ast }\right) $ can be presented as follows,\footnote{%
From $\Delta Z=\hat{A}Z$, with a certain operator $\hat{A}$, it follows that
$\Delta W=\frac{\hbar }{i}\left\langle \hat{A}_{W}\right\rangle =\frac{\hbar
}{i}\hat{A}_{W}1$, with $\hat{A}_{W}\,\ $given by $\hat{A}_{W}=e^{-i/\hbar W}%
\hat{A}e^{i/\hbar W}$.} with account taken\ of (\ref{unittransf}) and of the
usual Ward identity (\ref{Ward_W}) for $W_{\Psi }$, namely,
\begin{equation}
\Delta W_{\Psi }=\frac{\hbar }{i}\hat{\Omega}{\lambda }\left( \frac{\hbar
\overrightarrow{\delta }}{i\delta J}+\frac{\overrightarrow{\delta }W}{\delta
J},B\left\vert -\Delta \Psi \right. \right) =\hat{\Omega}\Delta \Psi \left(
\frac{\hbar \overrightarrow{\delta }}{i\delta J}+\frac{\overrightarrow{%
\delta }W}{\delta J},B\right) +o(\Delta \Psi ),  \label{Delta_W}
\end{equation}%
where $\hat{\Omega}$, given by (\ref{Omega}), is nilpotent, $\hat{\Omega}{}%
^{2}=0$, as a consequence of $\hat{\omega}^{2}=0$.

To obtain a finite change $\Delta \Gamma \left( B,\phi ,\Sigma ,\phi ^{\ast
}\right) $, we note that%
\begin{equation*}
\Delta \Gamma \left( B,\phi ,\Sigma ,\phi ^{\ast }\right) =\Delta W\left(
B,J,L,\phi ^{\ast }\right) ,
\end{equation*}%
as a general property of the Legendre transformation in the case of its
dependence on an external parameter $\eta $, namely, $\Delta \Gamma \left(
\eta \right) =\Delta W\left( \eta \right) $. Then $\Delta \Gamma \left(
B,\phi ,\Sigma ,\phi ^{\ast }\right) $ admits the representation%
\begin{eqnarray}
&&\ \Delta \Gamma =\frac{\hbar }{i}\hat{\omega}_{\Gamma }\langle \langle {%
\lambda }\left( B|-\Delta \Psi \right) \rangle \rangle =\delta \Gamma
+o\left( \langle \langle \Delta \Psi \rangle \rangle \right) ,
\label{Delta_Gamma} \\
&&\ \langle \langle {\lambda }\left( B|-\Delta \Psi \right) \rangle \rangle
\equiv {\lambda }(\hat{\phi},B|-\Delta \Psi ),\ \ \ \delta \Gamma \equiv
\hat{\omega}_{\Gamma }\langle \langle \Delta \Psi \rangle \rangle ,  \notag
\end{eqnarray}%
where $\hat{\phi}{}^{A}$ is given by (\ref{phi_hat}), while the operator $%
\hat{\omega}_{\Gamma }$ is a Legendre transform of $\hat{\Omega}$ in (\ref%
{Omega}), and thereby inherits the property of nilpotency: $\hat{\omega}%
_{\Gamma }^{2}=0$.

From (\ref{Delta_Gamma}), it follows, according to \cite{6, LOR}, that the
generating functional $\Gamma \left( B,\phi ,\Sigma ,\phi ^{\ast }\right) $\
of vertex Green's functions is gauge-independent, $\delta \Gamma =0$, on the
extremals%
\begin{equation}
\frac{\delta \Gamma }{\delta \phi ^{A}}=\frac{\delta \Gamma }{\delta \Sigma
^{m}}=0,  \label{ext}
\end{equation}%
so that the effective action $\Gamma _{\mathrm{eff}}=\Gamma _{\mathrm{eff}%
}\left( B,\Sigma \right) $\ with composite and background fields (\ref%
{Gamma_eff}) is gauge-independent, $\delta \Gamma _{\mathrm{eff}}=0$, on the
extremals (\ref{ext}), which is the principal result of this section.

\section{Loop Expansion\label{sec45}}

\setcounter{equation}{0}\renewcommand{\theequation}{5.\arabic{equation}}

Now we examine the procedure of a loop expansion for the effective action
(EA) with composite and background fields. The initial relation
\begin{eqnarray}
\exp \left\{ \frac{i}{\hbar }\Gamma \left( B,\Sigma ,\phi ,\phi ^{\ast
}\right) \right\} &=&\exp \left\{ \frac{i}{\hbar }\Gamma _{,m}\Sigma
^{m}\right\} \int d\widetilde{\phi }\exp \left\{ \frac{i}{\hbar }\left[ S_{%
\mathrm{ext}}\left( \widetilde{\phi },\phi ^{\ast },B\right) -F_{A}\left(
B,\Sigma ,\phi ,\phi ^{\ast }\right) \right. \right.  \notag \\
&&\left. \left. \times \left( \widetilde{\phi }^{A}-\phi ^{A}\right) -\Gamma
_{,m}\left( \sigma ^{m}\left( \widetilde{\phi },B\right) -\sigma ^{m}\left( {%
\phi },B\right) \right) \right] \right\} ,  \label{expext}
\end{eqnarray}%
according to (\ref{Z_ext}), (\ref{Gamma}), with allowance for the notation (%
\ref{phi_hat}) and a shift of variables\footnote{%
The change of variables (\ref{shiftvar}) is identified with the
background-quantum splitting used in \cite{lavrovmerzlikin}, where the
background component $\phi ^{A}$ is not to be confused with $B^{\mu }$.}
involving $\widetilde{\phi }{}^{A}$,%
\begin{equation}
\widetilde{\phi }{}^{A}\rightarrow \widetilde{\phi }{}^{A}+\phi ^{A},
\label{shiftvar}
\end{equation}%
acquires the form (see (\ref{phi_hat}) for the notation $F_{A}$)
\begin{eqnarray}
\exp \left\{ \frac{i}{\hbar }\Gamma \left( B,\Sigma ,\phi ,\phi ^{\ast
}\right) \right\} &=&\exp \left\{ \frac{i}{\hbar }\Gamma _{,m}\Sigma
^{m}\right\} \int d\widetilde{\phi }\exp \left\{ \frac{i}{\hbar }\left[ S_{%
\mathrm{ext}}\left( \widetilde{\phi }+\phi ,\phi ^{\ast },B\right) \right.
\right.  \notag \\
&&\left. \left. -F_{A}\widetilde{\phi }{}^{A}-\Gamma _{,m}\left( \sigma
^{m}\left( \widetilde{\phi }+\phi ^{A},B\right) -\sigma ^{m}\left( {\phi }%
,B\right) \right) \right] \right\} .  \label{nexpext}
\end{eqnarray}%
The representation (\ref{nexpext}) examined at a vanishing background $%
B^{\mu }$ reduces to the EA with composite fields \cite{lavrovmerzlikin},
albeit in the case of arbitrary (not limited to scalars) fields $\phi ^{A}$,
with a dependence of the composite fields $\sigma ^{m}$ on $\phi ^{A}$ being
generally more then quadratic.

We further assume the representation%
\begin{equation}
\Gamma \left( B,\Sigma ,\phi ,\phi ^{\ast }\right) =S_{\mathrm{ext}}\left(
\phi ,\phi ^{\ast },B\right) +\hbar \Gamma ^{(1)}\left( B,\Sigma ,\phi ,\phi
^{\ast }\right) +\Gamma _{2}\left( B,\Sigma ,\phi ,\phi ^{\ast }\right) ,
\label{LP}
\end{equation}%
with a one-loop effective action $\Gamma ^{(1)}$ and a functional $\Gamma
_{2}$ of order $\hbar ^{2}$, which includes all the two-particle-irreducible
vacuum graphs depending on the antifields $\phi ^{\ast }$, with the vertices
determined by a functional $S_{\mathrm{int}}(\widetilde{\phi },\phi ,\ldots
) $ $=$ $S_{\mathrm{int}}(\widetilde{\phi },\phi ,\phi ^{\ast },B)$ given by
the interaction part of the quantum action and the non-quadratic part due to
the composite fields:
\begin{eqnarray}
S_{\mathrm{int}}(\widetilde{\phi },\phi ,\ldots ) &\equiv &S_{\mathrm{ext}%
}\left( \widetilde{\phi }+\phi ,...\right) -S_{\mathrm{ext}}\left( \phi
,...\right) -S_{\mathrm{ext}}\left( \phi ,...\right) \left( \frac{%
\overleftarrow{\delta }}{\delta \phi ^{A}}+\frac{1}{2}\frac{\overleftarrow{%
\delta }}{\delta \phi ^{A}}\frac{\overleftarrow{\delta }}{\delta \phi ^{B}}%
\widetilde{\phi }{}^{B}\right) \widetilde{\phi }{}^{A}  \notag \\
&&-F_{A}\widetilde{\phi }{}^{A}-\Gamma _{,m}\left[ \sigma ^{m}\left(
\widetilde{\phi }+\phi ,B\right) -\sigma ^{m}\left( {\phi },B\right) \right]
.  \label{Sint}
\end{eqnarray}%
From the representation (\ref{compfield}), (\ref{Z_ext}), we obtain the
relation
\begin{equation}
\sum_{n\geq 2}\frac{1}{n!}\Lambda _{A_{n}...A_{1}}^{m}\left( \frac{\hbar }{i}%
\right) ^{n-1}\prod_{k=1}^{n}\frac{\overrightarrow{\delta }}{\delta J_{A_{k}}%
}Z\left( B,J,L,\phi ^{\ast }\right) =\frac{\overrightarrow{\delta }}{\delta
L_{m}}Z\left( B,J,L,\phi ^{\ast }\right) ,  \label{zwinter}
\end{equation}%
which can be recast\footnote{%
The dots \textquotedblleft $\ldots $\textquotedblright\ in (\ref{zwinter})
stand for a number of terms containing more than two derivatives $\frac{%
\overrightarrow{\delta }}{\delta J_{A_{k}}}$ entering as multipliers. These
contributions are related to the terms in (\ref{zwinter1}), (\ref{zwinter2})
which are also indicated by dots.} in terms of the generating functional $%
W\left( B,J,L,\phi ^{\ast }\right) $, $W=\left( \hbar /i\right) \ln Z$,%
\begin{eqnarray}
&&\ \sum_{n\geq 2}\frac{1}{n!}\Lambda _{A_{n}...A_{1}}^{m}\left( \frac{\hbar
}{i}\right) ^{n-1}\left[ \prod_{k=1}^{n}\frac{\overrightarrow{\delta }}{%
\delta J_{A_{k}}}W+\theta _{n,2}n\frac{i}{\hbar }\left( \prod_{k=1}^{n-1}%
\frac{\overrightarrow{\delta }}{\delta J_{A_{k}}}\right) W\frac{%
\overrightarrow{\delta }}{\delta J_{A_{n}}}W\right.  \notag \\
&&\ +\left. \ldots +\left( \frac{i}{\hbar }\right)
^{n-1}\prod_{k=1}^{n}\left( \frac{\overrightarrow{\delta }}{\delta J_{A_{k}}}%
W\right) \right] =\frac{\overrightarrow{\delta }}{\delta L_{m}}W,
\label{zwinter1}
\end{eqnarray}%
with the Heaviside symbol $\theta _{n,k}=\{0\ (n\leq k),1\ (n>k)\}$, as well
as in terms of the EA:
\begin{eqnarray}
&&\sum_{n\geq 2}\frac{1}{n!}\Lambda _{A_{n}...A_{1}}^{m}\left( \frac{\hbar }{%
i}\right) ^{n-1}\left[ \prod_{k=1}^{n-2}\left( G^{\prime \prime -1}\right)
^{A_{k}\mathsf{a}_{k}}\frac{\overrightarrow{\delta }}{\delta \Phi ^{\mathsf{a%
}_{k}}}\left( G^{\prime \prime -1}\right) ^{A_{n-1}A_{n}}\right.  \notag \\
&&+\left. \theta _{n,2}n\frac{i}{\hbar }\phi ^{A_{1}}\left(
\prod_{k=2}^{n-2}\left( G^{\prime \prime -1}\right) ^{A_{k}\mathsf{a}_{k}}%
\frac{\overrightarrow{\delta }}{\delta \Phi ^{\mathsf{a}_{k}}}\left(
G^{\prime \prime -1}\right) ^{A_{n-1}A_{n}}\right) \right.  \notag \\
&&+\left. \ldots +n\left( G^{\prime \prime -1}\right) ^{A_{1}A_{2}}\left(
\frac{i}{\hbar }\right) ^{n-2}\left( \prod_{k=3}^{n-3}\left( G^{\prime
\prime -1}\right) ^{A_{k}\mathsf{a}_{k}}\frac{\overrightarrow{\delta }}{%
\delta \Phi ^{\mathsf{a}_{k}}}\right) \left( G^{\prime \prime -1}\right)
^{A_{n-1}A_{n}}\right] =\Sigma ^{m}.  \label{zwinter2}
\end{eqnarray}%
From (\ref{nexpext}), we then find a representation for the one-loop
approximation,%
\begin{equation}
\Gamma ^{(1)}\left( B,\Sigma ,\phi ,\phi ^{\ast }\right) -\Gamma _{,m}\Sigma
^{m}=\frac{i}{2}\text{\textrm{sTr}\ }\ln \left[ S_{\mathrm{ext}}^{\prime
\prime }\left( \phi ,\phi ^{\ast },B\right) -\Gamma _{,m}(\sigma
^{m})^{\prime \prime }\left( \phi ^{A},B\right) \right] |_{\widetilde{\phi }%
=0}\ ,  \label{1lp}
\end{equation}%
being a functional Clairaut type equation (see \cite{lavrovmerzl2} for
details), with the variables $\phi $, $\phi ^{\ast }$, $B$ treated as
parameters. The equation (\ref{1lp}) can be solved as follows:%
\begin{equation}
2\Gamma ^{(1)}\left( B,\Sigma ,\phi ,\phi ^{\ast }\right) =\text{\textrm{%
sTr\ }}\ln \left[ (\Sigma ^{m}\Upsilon _{m})^{ab}\bar{S}_{\mathrm{ext}%
|bc}^{\prime \prime }\right] -i{}\text{\textrm{sTr}\ }\ln \left[ (\Sigma
^{m}\Upsilon _{m})^{ab}\right] +i{}\text{\textrm{sTr}\ }\ln \left( \bar{S}_{%
\mathrm{ext}|\beta \gamma }^{\prime \prime }\right) -i{}\delta
(0)(n_{+}-n_{-}),  \label{1lEA}
\end{equation}%
using a division of the discrete part of indices $\left( A;B\right) $ = $%
\left( x,a,\alpha ;y,b,\beta \right) $, $m=(x,\tilde{m})$, in the form $%
a,b=1,\ldots ,n=(n_{+},n_{-})$ and $\alpha ,\beta =n+1,\ldots
,N=(N_{+},N_{-})$, with the property $\tilde{m}=1,\ldots ,\frac{1}{2}%
n(n+1)\leq \frac{1}{2}N(N+1)$ implied by the supermatrices $(\sigma
^{m})^{\prime \prime }{}_{cd}$, as we impose on the supermatrices $(\Upsilon
_{m})^{ab}$ in (\ref{1lEA}) the following condition:%
\begin{equation}
(\Upsilon _{m})^{ab}(\sigma ^{m})^{\prime \prime }{}_{cd}\left( \phi
,B\right) |_{\widetilde{\phi }=0}=\frac{1}{2}\left( \delta _{c}^{a}\delta
_{d}^{b}+\delta _{c}^{b}\delta _{d}^{a}\right) .  \label{defS}
\end{equation}%
The supermatrices $\bar{S}_{\mathrm{ext}|bc}^{\prime \prime }$ and $\bar{S}_{%
\mathrm{ext}|\beta \gamma }^{\prime \prime }$ are given by%
\begin{eqnarray}
&&\bar{S}_{\mathrm{ext}|bc}^{\prime \prime }(x,y)={S}_{\mathrm{ext}%
|bc}^{\prime \prime }(x,y)-\int dz\ dz^{\prime }\,{S}_{\mathrm{ext}|b\gamma
}^{\prime \prime }(x,z)\left( {S}_{\mathrm{ext}}^{\prime \prime -1}\right)
^{\gamma \delta }(z,z^{\prime }){S}_{\mathrm{ext}|\delta c}^{\prime \prime
}(z^{\prime },y),  \notag \\
&&{S}_{\mathrm{ext}|BC}^{\prime \prime }=\frac{\overrightarrow{\delta }}{%
\delta \phi ^{C}}{S}_{\mathrm{ext}}\frac{\overleftarrow{\delta }}{\delta
\phi ^{B}}=\left(
\begin{array}{cc}
{S}_{\mathrm{ext}|bc}^{\prime \prime } & {S}_{\mathrm{ext}|b\gamma }^{\prime
\prime } \\
{S}_{\mathrm{ext}|\beta c}^{\prime \prime } & {S}_{\mathrm{ext}|\beta \gamma
}^{\prime \prime }%
\end{array}%
\right) (x,y),\ \ \left( B;C\right) =\left( b,\beta ,x;c,\gamma ,y\right) .
\label{tilS}
\end{eqnarray}%
For $A=(x,a)$, the third term in (\ref{1lEA}) is vanishing, $n=N$, with ${S}%
_{\mathrm{ext}|BC}^{\prime \prime }\equiv {S}_{\mathrm{ext}|bc}^{\prime
\prime }=\bar{S}_{\mathrm{ext}|bc}^{\prime \prime }$, as in \cite%
{lavrovmerzl2}, albeit for a model featuring gauge invariance.

\section{Gribov--Zwanziger Theory\label{sec5}}

\setcounter{equation}{0}\renewcommand{\theequation}{6.\arabic{equation}}

Let us extend the\ case of background fields to the concept of Gribov
horizon \cite{Gribov}, implemented in the Gribov--Zwanziger model \cite%
{Zwanziger} by using a composite field. We propose three descriptions for
the Gribov horizon introduction. To do so, we consider a Euclidean form%
\footnote{%
We use the metric signature $\eta _{\mu \nu }=\left( -,+,\ldots ,+\right) $
and carry out a Wick rotation, $x^{0}\rightarrow ix^{0}$, $%
A^{p|0}\rightarrow iA^{p|0}$, $S_{\mathrm{FP}}\rightarrow iS_{\mathrm{FP}}$.
In Euclidean metric, $A_{\mu }=A^{\mu }$, we maintain the summation
convention $A_{\mu }B_{\mu }=A_{\mu }B^{\mu }$.} of the Faddeev--Popov action%
 $S_{\mathrm{FP}}\left( \phi \right) $ for a Yang--Mills
theory (\ref{S_FP}), (\ref{S_0}), (\ref{Psi}) in Landau gauge $%
\chi _{\mathrm{L}}^{p}\left( \phi \right) =\partial ^{\mu }A_{\mu }^{p}$ and
examine a non-local horizon functional $H\left( A\right) $,%
\begin{equation}
H\left( A\right) =\gamma ^{2}\int d^{D}x\left[ \int d^{D}y\ f^{prt}gA_{\mu
}^{r}\left( x\right) \left( K^{-1}\right) ^{pq}\left( x;y\right)
f^{qst}gA^{s|\mu }\left( y\right) +D\left( N^{2}-1\right) \right] ,
\label{gribovH}
\end{equation}%
where $K^{-1}$ is the inverse,%
\begin{equation}
\int d^{D}z\left( K^{-1}\right) ^{pr}\left( x;z\right) \left( K\right)
^{rq}\left( z;y\right) =\int d^{D}z\left( K\right) ^{pr}\left( x;z\right)
\left( K^{-1}\right) ^{rq}\left( z;y\right) =\delta ^{pq}\delta \left(
x-y\right) ,  \label{K^-1}
\end{equation}%
of the Faddeev--Popov operator $K$ in terms of the gauge condition $\partial
^{\mu }A_{\mu }^{p}=0$,%
\begin{equation}
K^{pq}\left( x;y\right) =\left( \delta ^{pq}\partial ^{2}+gf^{prq}A_{\mu
}^{r}\partial ^{\mu }\right) \delta \left( x-y\right) ,\ \ \ K^{pq}\left(
x;y\right) =K^{qp}\left( y;x\right) ,  \label{K}
\end{equation}%
and $\gamma $ is a Gribov thermodynamic parameter \cite{Zwanziger}. The
latter is introduced in a self-consistent way by solving a gap equation
(horizon condition) for a Gribov--Zwanziger action $S_{\mathrm{GZ}}=S_{%
\mathrm{GZ}}\left( \phi \right) $,%
\begin{equation*}
\frac{\partial E_{\mathrm{vac}}}{\partial \gamma }=0,\ \ \ \exp \left(
-\hbar ^{-1}E_{\mathrm{vac}}\right) \equiv \int d\phi \ \exp \left( -\hbar
^{-1}S_{\mathrm{GZ}}\right) ,
\end{equation*}%
where $E_{\mathrm{vac}}$ is the vacuum energy, and the action $S_{\mathrm{GZ}%
}$ is given by%
\begin{equation}
S_{\mathrm{GZ}}\left( \phi \right) =S_{\mathrm{FP}}\left( \phi \right)
-H\left( A\right) \,.  \label{GZbab}
\end{equation}%
A generating functional of Green's functions $Z_{H}\left( J,L\right) $ with
composite fields for\ the quantum theory in question can be presented in
terms of a Faddeev--Popov action shifted by a constant value, $S_{\mathrm{FP}%
}\left( \phi \right) -H\left( 0\right) $,%
\begin{eqnarray}
&&Z_{H}\left( J,L\right) =\left. Z_{H}\left( J,\mathcal{L}\right)
\right\vert _{L_{0}=1},\ \ \ \mathcal{L}_{\text{\textsf{M}}}=\left(
L_{0},L_{m}\right) ,  \notag \\
&&Z_{H}\left( J,\mathcal{L}\right) =\int d\phi \ \exp \left\{ -\hbar ^{-1}%
\left[ S_{\mathrm{FP}}\left( \phi \right) -H\left( 0\right) +J_{A}\phi ^{A}+%
\mathcal{L}_{\text{\textsf{M}}}\sigma ^{\text{\textsf{M}}}\left( A\right) %
\right] \right\} ,  \label{represented}
\end{eqnarray}%
where $\mathcal{L}_{\text{\textsf{M}}}=(L_{0},L_{m})\left( x\right) $, $%
\epsilon \left( L_{0}\right) =\mathrm{gh}\left( L_{0}\right) =0$, are
sources to composite fields $\sigma ^{\text{\textsf{M}}}\left( A\right)
=(\sigma ^{0},\sigma ^{m})\left( A\right) $, and $\sigma ^{0}\left( A\right)
\equiv \sigma \left( A\right) $ is a non-local field,%
\begin{equation}
\sigma \left( A\right) \left( x\right) =\gamma ^{2}\int d^{D}y\
f^{trp}gA_{\mu }^{r}\left( x\right) (\tilde{K}^{-1})^{pq}\left( x;y\right)
f^{qst}gA^{s|\mu }\left( y\right) ,  \label{sigma_Gibov}
\end{equation}%
with $\tilde{K}^{-1}$ being the inverse,%
\begin{equation}
\int d^{D}z\ (\tilde{K}^{-1})^{pr}\left( x;z\right) (\tilde{K})^{rq}\left(
z;y\right) =\int d^{D}z\ (\tilde{K})^{pr}\left( x;z\right) (\tilde{K}%
^{-1})^{rq}\left( z;y\right) =\delta ^{pq}\delta \left( x-y\right) ,
\label{calligK^-1}
\end{equation}%
of an operator $\tilde{K}$ defined for a quantity $F^{p}=F^{p}\left(
x\right) $,%
\begin{equation}
\int d^{D}y\ (\tilde{K})^{pq}\left( x;y\right) F^{q}\left( y\right) \equiv %
\left[ \partial _{\mu },\left[ D^{\mu }\left( A\right) ,F\right] \right]
^{p}\left( x\right) ,\ \ A_{\mu }\left( x\right) =T^{p}A_{\mu }^{p}\left(
x\right) ,\ \ F\left( x\right) =T^{p}F^{p}\left( x\right) ,  \label{calligK}
\end{equation}%
which results in%
\begin{equation}
\tilde{K}^{pq}\left( x;y\right) =\partial ^{\mu }D_{\mu }^{pq}\left(
A\right) \delta \left( x-y\right) ,  \label{K_tilde}
\end{equation}%
and therefore reduces to the operator $K$ of (\ref{K}) as one takes into
account the Landau gauge condition, due to $S_{\mathrm{FP}}\left( \phi
\right) $, in the path integral (\ref{represented}). Note in conclusion that
one cannot absorb the constant term $H(0)$ into $\sigma \left( A\right)
\left( x\right) $ while preserving the basic definition (\ref{compfield})
for composite fields. Besides, it should be noted that the horizon and
therefore also the field $\sigma \left( A\right) \left( x\right) $ in (\ref%
{sigma_Gibov}) are not BRST-invariant.

\subsection{Background Horizon Term}

Let us now extend the generating functional $Z_{H}\left( J,\mathcal{L}%
\right) $ with a non-local composite field (\ref{represented}), (\ref%
{sigma_Gibov}), (\ref{calligK^-1}), (\ref{calligK}) to the case of a
background field $B_{\mu }$ equipped with a covariant derivative $D_{\mu
}\left( B\right) $ having the gauge properties (\ref{gauge_B}), (\ref%
{property}), by using the approach (\ref{intro1}), (\ref{rule}) as adapted
to Euclidean QFT, which implies a modification of derivatives $\partial
_{\mu }\rightarrow D_{\mu }\left( B\right) $ in (\ref{represented}),
according to%
\begin{equation}
Z_{H}\left( B,J,\mathcal{L}\right) =\left. Z_{H}\left( J,\mathcal{L}\right)
\right\vert _{\partial _{\mu }\rightarrow D_{\mu }\left( B\right) }=\int
d\phi \exp \left\{ -\hbar ^{-1}\left[ S_{\mathrm{FP}}\left( \phi ,B\right)
-H\left( 0\right) +J_{A}\phi ^{A}+\mathcal{L}_{\text{\textsf{M}}}\sigma ^{%
\text{\textsf{M}}}\left( A,B\right) \right] \right\} ,  \label{represented2}
\end{equation}%
where $S_{\mathrm{FP}}\left( \phi ,B\right) \,$is the Faddeev--Popov action
in the Landau background gauge $\chi _{\mathrm{L}}^{p}\left( \phi ,B\right)
=0$, see\ (\ref{background_gauges}), and $\sigma ^{0}\left( A,B\right)
\equiv \sigma \left( A,B\right) $ is a non-local composite field on a
background:
\begin{equation}
\sigma \left( A,B\right) \left( x\right) =\gamma ^{2}\int d^{D}y\
f^{trp}gA_{\mu }^{r}\left( x\right) (\tilde{K}_{B}^{-1})^{pq}\left(
x;y\right) f^{qst}gA^{s|\mu }\left( y\right) .  \label{sigma_Gibov2}
\end{equation}%
Here, $\tilde{K}_{B}^{-1}$ is a modified operator $\tilde{K}^{-1}$ as in (%
\ref{K^-1}), with the corresponding inverse $\tilde{K}_{B}$ determined by
the replacement $\tilde{K}\rightarrow \tilde{K}_{B}$,%
\begin{equation}
\int d^{D}y\ (\tilde{K})_{B}^{pq}\left( x;y\right) F^{q}\left( y\right)
\equiv \left[ D_{\mu }\left( B\right) ,\left[ D^{\mu }\left( A+B\right) ,F%
\right] \right] ^{p}\left( x\right) ,\ \ \ F\left( x\right) =F^{p}\left(
x\right) T^{p},  \label{calligK_B}
\end{equation}%
and having the manifest form%
\begin{equation}
(\tilde{K})_{B}^{pq}\left( x;y\right) =D_{\mu }^{pr}\left( B\right)
D^{rq|\mu }\left( A+B\right) \delta \left( x-y\right) .  \label{KBDcallig}
\end{equation}%
In the particular case, cf. (\ref{represented}),%
\begin{equation*}
Z_{H}\left( B,J,L\right) =\left. Z_{H}\left( B,J,\mathcal{L}\right)
\right\vert _{L_{0}=1}\ ,
\end{equation*}%
we arrive at the generating functional%
\begin{eqnarray}
&&Z_{H}\left( B,J,L\right) =\int d\phi \exp \left\{ -\hbar ^{-1}\left[ S_{%
\mathrm{GZ}}\left( \phi ,B\right) +J_{A}\phi ^{A}+L_{m}\sigma ^{m}\left(
A,B\right) \right] \right\} ,  \label{ZBJ_horizon} \\
&&S_{\mathrm{GZ}}\left( \phi ,B\right) \equiv S_{\mathrm{FP}}\left( \phi
,B\right) -H\left( A,B\right) ,  \notag
\end{eqnarray}%
with a non-local functional $H\left( A,B\right) $ given by%
\begin{equation}
H\left( A,B\right) =\gamma ^{2}\int d^{D}x\left[ \int d^{D}y\ f^{prt}gA_{\mu
}^{r}\left( x\right) \left( K_{B}^{-1}\right) ^{pq}\left( x;y\right)
f^{qst}gA^{s|\mu }\left( y\right) +D\left( N^{2}-1\right) \right] ,
\label{gribovHB}
\end{equation}%
where $K_{B}^{-1}$ is the inverse of an operator $K_{B}$ as in (\ref%
{calligK^-1}) which is identical with the operator $\tilde{K}_{B}$ in (\ref%
{calligK_B}) being expressed, due to $S_{\mathrm{FP}}\left( \phi ,B\right) $
in (\ref{ZBJ_horizon}), by using the background gauge condition $D_{\mu
}^{pq}\left( B\right) A^{q|\mu }=0$ and the properties of $f^{pqr}$,
including the Jacobi identity,%
\begin{equation}
K_{B}\left( x;y\right) =\left[ \partial ^{2}+g\left( \partial _{\mu }B^{\mu
}\right) +g\left( A_{\mu }+2B_{\mu }\right) \partial ^{\mu }+g^{2}\left(
A_{\mu }+B_{\mu }\right) B^{\mu }\right] \delta \left( x-y\right) ,
\label{K_B1}
\end{equation}%
where $A_{\mu }$, $B_{\mu }$ are matrices with the elements $\left( A_{\mu
}^{pq},B_{\mu }^{pq}\right) =f^{prq}\left( A_{\mu }^{r},B_{\mu }^{r}\right) $%
, and $K_{B}\left( x;y\right) $ is related to $\tilde{K}_{B}\left(
x;y\right) $ in (\ref{KBDcallig}) by the equality (see Appendix \ref{B.I})%
\begin{equation}
K_{B}\left( x;y\right) =\tilde{K}_{B}\left( x;y\right) -g\left[ D_{\mu
}\left( B\right) ,A^{\mu }\right] \delta \left( x-y\right) =D^{\mu }\left(
A+B\right) D_{\mu }\left( B\right) \delta \left( x-y\right) .  \label{K_B}
\end{equation}%
The operator $K_{B}$ is an extension of the original operator $K$ in (\ref{K}%
) and exhibits the properties%
\begin{equation}
\left. K_{B}\right\vert _{B=0}=K,\ \ \ \left( K_{B}\right) ^{pq}\left(
x;y\right) =\left( K_{B}\right) ^{qp}\left( y;x\right) ,  \label{K_prop}
\end{equation}%
where the latter can be verified by a straightforward calculation:%
\begin{equation*}
\int d^{D}y\ \left[ \left( K_{B}\right) ^{pq}\left( x;y\right) -\left(
K_{B}\right) ^{qp}\left( y;x\right) \right] F^{q}\left( y\right) =gf^{prq}%
\left[ D_{\mu }^{rs}\left( B\right) A^{s|\mu }\right] F^{q}\left( x\right)
=0.
\end{equation*}%
In view of (\ref{K_prop}), we interpret $S_{\mathrm{GZ}}\left( \phi
,B\right) $ as a Gribov--Zwanziger action on a background $B_{\mu }$, with a
non-local background horizon term $H\left( A,B\right) $ given by (\ref%
{gribovHB}), (\ref{K_B}).

Since the consideration involves the action functionals $S_{\mathrm{FP}%
}\left( \phi \right) $ and $S_{\mathrm{FP}}\left( \phi ,B\right) $, which
are invariant under respective global $SU\left( N\right) $ transformations
and localized $SU\left( N\right) $ transformations combined with gauge
transformations for $B_{\mu }^{p}$, it is natural to analyze the behavior of
the generating functionals $Z_{H}\left( J\right) $ and $Z_{H}\left(
B,J\right) $ with respect to these transformations. For such a purpose, it
is convenient to recast $Z_{H}\left( B,J\right) $ in a local form by
extending the configuration space along the lines of \cite{Kondo}.%
{\ }Namely, we introduce a set of commuting $(\bar{\varphi}_{\mu
}^{pq},\varphi _{\mu }^{pq})$ and anticommuting $(\bar{\omega}_{\mu
}^{pq},\omega _{\mu }^{pq})$ auxiliary fields, where $\bar{\varphi}_{\mu
}^{pq}$ and $\varphi _{\mu }^{pq}$ are mutually complex-conjugate,%
\begin{equation*}
\begin{tabular}{|c|c|c|c|c|}
\hline
& $\bar{\varphi}_{\mu }^{pq}$ & $\varphi _{\mu }^{pq}$ & $\bar{\omega}_{\mu
}^{pq}$ & $\omega _{\mu }^{pq}$ \\ \hline
$\epsilon $ & $0$ & $0$ & $1$ & $1$ \\ \hline
$\mathrm{gh}$ & $0$ & $0$ & $-1$ & $1$ \\ \hline
\end{tabular}%
\end{equation*}%
This allows one to construct the parameterization
\begin{equation}
\exp \left\{ \hbar ^{-1}\left[ H\left( A,B\right) -H\left( 0,B\right) \right]
\right\} =\int d\bar{\varphi}\ d\varphi \ d\bar{\omega}\ d\omega \exp \left[
-\hbar ^{-1}S_{\gamma }\left( A,B;\bar{\varphi}\ \varphi ,\bar{\omega}%
,\omega \right) \right] ,  \label{HS_gamma}
\end{equation}%
where%
\begin{equation}
S_{\gamma }=\int d^{D}x\ \left[ -\bar{\varphi}_{\mu }^{rp}K_{B}^{pq}\varphi
^{rq|\mu }+\bar{\omega}_{\mu }^{rp}K_{B}^{pq}\omega ^{rq|\mu }+i\gamma g\
f^{prq}A^{r|\mu }\left( \bar{\varphi}_{\mu }^{pq}+\varphi _{\mu
}^{pq}\right) \right] ,  \label{S_gamma}
\end{equation}%
as we imply%
\begin{equation*}
K_{B}^{pq}\varphi _{\mu }^{rq}\left( x\right) =\int d^{D}y\ K_{B}^{pq}\left(
x,y\right) \varphi _{\mu }^{rq}\left( y\right) ,\ \ \ K_{B}^{pq}\omega _{\mu
}^{rq}\left( x\right) =\int d^{D}y\ K_{B}^{pq}\left( x,y\right) \omega _{\mu
}^{rq}\left( y\right) .
\end{equation*}%
The auxiliary fields are regarded as BRST doublets \cite{Kondo}, so that the
Slavnov variation, being in our case the operator $\overleftarrow{s}_{%
\mathrm{q}}$\ of (\ref{mod_Slavnov}), can be extended as follows:%
\begin{equation*}
\left( \bar{\varphi}_{\mu }^{pq},\varphi _{\mu }^{pq},\bar{\omega}_{\mu
}^{pq},\omega _{\mu }^{pq}\right) \overleftarrow{s}_{\mathrm{q}}=\left(
0,\omega _{\mu }^{pq},-\bar{\varphi}_{\mu }^{pq},0\right) .
\end{equation*}%
These transformations, however, do not provide invariance for the functional
$S_{\gamma }$:%
\begin{equation*}
S_{\gamma }\overleftarrow{s}_{\mathrm{q}}\not=0.
\end{equation*}%
In the extended configuration space $\Phi =\left( \phi ,\bar{\varphi},\
\varphi ,\bar{\omega},\omega \right) $, the generating functional $%
Z_{H}\left( B,J\right) $ given by the restriction $L_{m}=0$ in (\ref%
{ZBJ_horizon}) acquires the form%
\begin{equation}
Z_{H}\left( B,J\right) =\int d\Phi \exp \left\{ -\hbar ^{-1}\left[ S_{%
\mathrm{GZ}}\left( \Phi ,B\right) +J_{A}\phi ^{A}\right] \right\} ,
\label{ZBJ_horizon2}
\end{equation}%
where the local action $S_{\mathrm{GZ}}\left( \Phi ,B\right) $ of the
Gribov--Zwanziger theory on a background reads (notice the antisymmetry of $%
f^{pqr}$)%
\begin{eqnarray}
S_{\mathrm{GZ}}\left( \Phi ,B\right) &=&S_{\mathrm{FP}}\left( \phi ,B\right)
-H\left( 0,B\right) -i\gamma g\int d^{D}x\ \mathrm{Tr\ }A^{\mu }\left( \bar{%
\varphi}_{\mu }-\varphi _{\mu }^{\mathrm{T}}\right)  \notag \\
&&+\int d^{D}x\ d^{D}y\ \mathrm{Tr}\left[ -\bar{\varphi}^{\mu }\left(
x\right) K_{B}\left( x,y\right) \varphi _{\mu }^{\mathrm{T}}\left( y\right) +%
\bar{\omega}^{\mu }\left( x\right) K_{B}\left( x,y\right) \omega _{\mu }^{%
\mathrm{T}}\left( y\right) \right] .  \label{S_GZ}
\end{eqnarray}%
The action $S_{\mathrm{GZ}}\left( \Phi ,B\right) $ is invariant under the
global $SU\left( N\right) $ transformations
\begin{equation}
\left( A_{\mu },B_{\mu },b,\bar{c},c,\bar{\varphi}_{\mu },\bar{\omega}_{\mu
},\omega _{\mu }^{\mathrm{T}}\right) \overset{U}{\rightarrow }U\left( A_{\mu
},B_{\mu },b,\bar{c},c,\bar{\varphi}_{\mu },\bar{\omega}_{\mu },\omega _{\mu
}^{\mathrm{T}}\right) U^{-1}.  \label{finite}
\end{equation}%
Indeed, due to the unitarity of $U$, we also have%
\begin{equation*}
\bar{\varphi}_{\mu }\overset{U}{\rightarrow }U\bar{\varphi}_{\mu
}U^{-1}\Longrightarrow \varphi _{\mu }^{\mathrm{T}}\overset{U}{\rightarrow }%
U\varphi _{\mu }^{\mathrm{T}}U^{-1},
\end{equation*}%
and the manifest expression (\ref{K_B1}) for $K_{B}\left( x,y\right) $
consistently implies
\begin{equation}
K_{B}\left( x,y\right) \overset{U}{\rightarrow }UK_{B}\left( x;y\right)
U^{-1}=\left( K^{\prime }\right) _{B}\left( x;y\right) ,\ \ \ \left(
K^{\prime }\right) _{B}\left( x;y\right) =\left( K^{\prime }\right) _{B}^{%
\mathrm{T}}\left( y;x\right) .  \label{KBU}
\end{equation}%
The infinitesimal form of field transformations (\ref{finite}), given by%
\begin{equation}
\delta _{\varsigma }F^{pq}=gf^{prs}F^{rq}\varsigma
^{s}+gf^{qrs}F^{pr}\varsigma ^{s},\ \ \ F^{pq}=\left( A_{\mu },B_{\mu },b,%
\bar{c},c,\bar{\varphi}_{\mu },\varphi _{\mu }^{\mathrm{T}},\bar{\omega}%
_{\mu },\omega _{\mu }^{\mathrm{T}}\right) ^{pq},  \label{phitilde}
\end{equation}%
produces a unit Jacobian (notice the antisymmetry of $f^{pqr}$) in the
integration measure of (\ref{ZBJ_horizon2}) and leaves invariant the
functional $Z_{H}\left( B,J\right) $ under infinitesimal global $SU\left(
N\right) $ transformations of the background field $B_{\mu }$ and the
sources $J_{A}$, having the adjoint representation form%
\begin{equation}
\delta _{\varsigma }G^{pq}=gf^{prs}G^{rq}\varsigma
^{s}+gf^{qrs}G^{pr}\varsigma ^{s},\ \ \ G^{pq}=\left( B_{\mu },J_{\mu \left(
A\right) },J_{\left( b\right) },J_{\left( \bar{c}\right) },J_{\left(
c\right) }\right) ^{pq}.  \label{B_sources_adj}
\end{equation}%
This behavior of $Z_{H}\left( B,J\right) $ includes the invariance of the
restricted generating functional $Z_{H}\left( J\right) $ under the global $%
SU\left( N\right) $ transformations of the sources and can also be
established directly in the non-local form by using the properties of $%
(K_{B}^{-1})\left( x;y\right) $,%
\begin{equation*}
(K_{B}^{-1})\left( x,y\right) \overset{U}{\rightarrow }U(K_{B}^{-1})\left(
x;y\right) U^{-1},\ \ \ \left( K^{-1}\right) _{B}\left( x;y\right)
=(K^{-1})_{B}^{\mathrm{T}}\left( y;x\right) ,
\end{equation*}%
implied by (\ref{K^-1}), (\ref{KBU}) and providing a global $SU\left(
N\right) $ invariance of the original $H\left( A\right) =H\left( A,0\right) $
and the background-modified $H\left( A,B\right) $ horizon functionals in (%
\ref{gribovH}), (\ref{gribovHB}),%
\begin{equation*}
\left( A_{\mu },B_{\mu }\right) \overset{U}{\rightarrow }\left( A_{\mu
},B_{\mu }\right) ^{\prime }=U\left( A_{\mu },B_{\mu }\right) U^{-1},\ \ \
H\left( A^{\prime },B^{\prime }\right) =H\left( A,B\right) .
\end{equation*}%
Notably, it turns out that the global $SU\left( N\right) $ invariance of the
background functional $Z_{H}\left( J,B\right) $ does not translate itself
into a local symmetry. To prove this point, let us subject the integrand in (%
\ref{ZBJ_horizon2}) to an infinitesimal local change of variables with a
unitary matrix $V=V\left( \xi \right) $, $\xi =\xi \left( x\right) $,%
\begin{equation}
\Phi \overset{V}{\rightarrow }\Phi ^{\prime },\ \ \ B_{\mu }\overset{V}{%
\rightarrow }B_{\mu }^{\prime }=VB_{\mu }V^{-1}+g^{-1}V\left( \partial _{\mu
}V^{-1}\right) ,  \label{even_though}
\end{equation}%
which implies a unit Jacobian and induces a variation $\delta _{\xi }S_{%
\mathrm{GZ}}\left( \Phi ,B\right) =\delta _{\xi }S_{K}\left( \Phi ,B\right) $
in (\ref{S_GZ}),%
\begin{equation*}
S_{K}\left( \Phi ,B\right) \equiv \int d^{D}x\ d^{D}y\ \mathrm{Tr}\left[ -%
\bar{\varphi}^{\mu }\left( x\right) K_{B}\left( x,y\right) \varphi _{\mu }^{%
\mathrm{T}}\left( y\right) +\bar{\omega}^{\mu }\left( x\right) K_{B}\left(
x,y\right) \omega _{\mu }^{\mathrm{T}}\left( y\right) \right] ,
\end{equation*}%
due to the parameterization term $S_{\gamma }\left( \Phi ,B\right) $ in (\ref%
{HS_gamma}), (\ref{S_gamma}),%
\begin{equation*}
S_{\mathrm{GZ}}\left( \Phi ^{\prime },B^{\prime }\right) =S_{\mathrm{FP}%
}\left( \phi ,B\right) -H\left( 0,B\right) +S_{\gamma }\left( \Phi ^{\prime
},B^{\prime }\right) .
\end{equation*}%
Using the explicit form of $K_{B}\left( x,y\right) $ given by (\ref{K_B}),
we find\footnote{%
One uses a repeated integration by parts and the antisymmetry of $f^{pqr}$
to remove the delta-function $\delta \left( x-y\right) $ absorbed in $%
K_{B}\left( x,y\right) $ and to recast $S_{K}\left( \Phi ,B\right) $ in the
form (\ref{S_gamma_local}).}%
\begin{equation}
S_{K}\left( \Phi ,B\right) =\int d^{D}x\ \mathrm{Tr}\left[ -\bar{\varphi}%
^{\mu }D_{\nu }\left( A+B\right) D^{\nu }\left( B\right) \varphi _{\mu }^{%
\mathrm{T}}+\bar{\omega}^{\mu }D_{\nu }\left( A+B\right) D^{\nu }\left(
B\right) \omega _{\mu }^{\mathrm{T}}\right] ,  \label{S_gamma_local}
\end{equation}%
so that the presence of extra derivatives $\partial _{\mu }V^{-1}$ and $%
\partial ^{2}V^{-1}$ in the transformed expression%
\begin{eqnarray*}
S_{K}\left( \Phi ^{\prime },B^{\prime }\right) &=&\int d^{D}x\ \mathrm{Tr}%
\left[ -V\bar{\varphi}^{\mu }D_{\nu }\left( A+B\right) D^{\nu }\left(
B\right) \varphi _{\mu }^{\mathrm{T}}V^{-1}+V\bar{\omega}^{\mu }D_{\nu
}\left( A+B\right) D^{\nu }\left( B\right) \omega _{\mu }^{\mathrm{T}}V^{-1}%
\right] \\
&\not=&\int d^{D}x\ \mathrm{Tr}\left[ -V^{-1}V\bar{\varphi}^{\mu }D_{\nu
}\left( A+B\right) D^{\nu }\left( B\right) \varphi _{\mu }^{\mathrm{T}%
}+V^{-1}V\bar{\omega}^{\mu }D_{\nu }\left( A+B\right) D^{\nu }\left(
B\right) \omega _{\mu }^{\mathrm{T}}\right] ,
\end{eqnarray*}%
leads to%
\begin{equation*}
S_{K}\left( \Phi ^{\prime },B^{\prime }\right) \not=S_{K}\left( \Phi
,B\right) ,
\end{equation*}%
which also implies a local non-invariance of the background horizon
functional,%
\begin{equation*}
H\left( A,B\right) \overset{V}{\rightarrow }H\left( A^{\prime },B^{\prime
}\right) \not=H\left( A,B\right) .
\end{equation*}%
As a consequence, we conclude that the background functional $Z_{H}\left(
B,J\right) $ is not left invariant by the gauge transformations of the
background field $B_{\mu }$ combined with the local $SU\left( N\right) $
transformations of the sources $J_{A}$, since the latter do not compensate
the variation $\delta _{\xi }S_{\mathrm{GZ}}\left( \Phi ,B\right) \not=0$,
due to $\delta _{\xi }\left( J_{A}\phi ^{A}\right) =0$. In other words, the
global invariance of $Z_{H}\left( J\right) $ is not inherited by a related
local symmetry of $Z_{H}\left( B,J\right) $ in the theory (\ref{ZBJ_horizon}%
), (\ref{gribovHB}), (\ref{K_B}).

\subsection{Locally Invariant Horizon Term}

The local non-invariance of the functional $Z_{H}\left( B,J\right) $ can be
traced back to the fact that the background $B_{\mu \text{ }}$ has been
incorporated directly into the non-local horizon term via (\ref{K_B}),
whereas the emergence of the auxiliary fields $(\bar{\varphi}_{\mu },\varphi
_{\mu })$ and $(\bar{\omega}_{\mu },\omega _{\mu })$ as a means of
parameterizing the term $H\left( A,B\right) $ does not provide them with a
covariant derivative in the form (\ref{rule}), as one can observe from (\ref%
{S_gamma_local}). To resolve this issue, it is natural to examine an
alternative way of introducing a background, namely, by using a local
parameterization of the original term $H\left( A\right) $ prior to the point
the background has been incorporated. To do so, we consider the expressions (%
\ref{HS_gamma}), (\ref{S_gamma}), (\ref{ZBJ_horizon2}), (\ref{S_GZ})
restricted to $B_{\mu }=0$ and present the functional $Z_{H}\left( J\right) $
in (\ref{represented}) as follows:%
\begin{equation*}
Z_{H}\left( J\right) =\int d\Phi \ \exp \left\{ -\hbar ^{-1}\left[ S_{%
\mathrm{GZ}}\left( \Phi \right) +J_{A}\phi ^{A}\right] \right\} ,\ S_{%
\mathrm{GZ}}\left( \Phi \right) =S_{\mathrm{GZ}}\left( \Phi ,0\right) ,\
\Phi =\left( \phi ,\bar{\varphi},\ \varphi ,\bar{\omega},\omega \right) .
\end{equation*}%
For a treatment of the auxiliary fields $\left( \bar{\varphi},\bar{\omega}%
\right) \ $and $\left( \varphi ^{\mathrm{T}},\omega ^{\mathrm{T}}\right) $
on equal footing, notice that the action $S_{\mathrm{GZ}}\left( \Phi \right)
$ in the above integrand, with the Landau gauge condition $\partial _{\mu
}A^{\mu }=0$ absorbed in the factor $\exp \left[ -\hbar S_{\mathrm{FP}%
}\left( \phi \right) \right] $, is equivalent to an action $\mathcal{S}_{%
\mathrm{GZ}}\left( \Phi \right) $ arising from the replacement of $K\left(
x,y\right) $ by $\mathcal{K}\left( x,y\right) $, defined as a symmetrization:%
\begin{equation}
\mathcal{K}\left( x,y\right) \equiv \frac{1}{2}\left[ K\left( x,y\right) +%
\tilde{K}\left( x,y\right) \right] ,\ \ \ \tilde{K}\left( x,y\right)
=K\left( x,y\right) +g\left[ \partial _{\mu },A^{\mu }\right] \delta \left(
x-y\right) .  \label{sym}
\end{equation}%
The action $\mathcal{S}_{\mathrm{GZ}}\left( \Phi \right) $ reads\ (see
Appendix \ref{B.II})%
\begin{eqnarray}
\mathcal{S}_{\mathrm{GZ}}\left( \Phi \right) &=&S_{\mathrm{FP}}\left( \phi
\right) -H\left( 0\right) +S_{\mathcal{K}}\left( \Phi \right) -i\gamma g\int
d^{D}x\ \mathrm{Tr\ }A^{\mu }\left( \bar{\varphi}_{\mu }-\varphi _{\mu }^{%
\mathrm{T}}\right) ,  \label{S_1} \\
S_{\mathcal{K}}\left( \Phi \right) &\equiv &\frac{1}{2}\int d^{D}x\ \mathrm{%
Tr}\left\{ \left[ D^{\nu }\left( A\right) ,\bar{\varphi}^{\mu }\right]
\partial _{\nu }\varphi _{\mu }^{\mathrm{T}}+\left( \partial ^{\nu }\bar{%
\varphi}^{\mu }\right) \left[ D_{\nu }\left( A\right) ,\varphi _{\mu }^{%
\mathrm{T}}\right] \right\} -\left( \bar{\varphi},\varphi \rightarrow \bar{%
\omega},\omega \right) ,  \notag
\end{eqnarray}%
and implies a natural introduction of a background, $\mathcal{S}_{\mathrm{GZ}%
}\left( \Phi \right) \rightarrow \mathcal{S}_{\mathrm{GZ}}\left( \Phi
,B\right) $, in the form (\ref{rule}),%
\begin{eqnarray}
\mathcal{S}_{\mathrm{GZ}}\left( \Phi ,B\right) &=&S_{\mathrm{FP}}\left( \phi
,B\right) -H\left( 0\right) +S_{\mathcal{K}}\left( \Phi ,B\right) -i\gamma
g\int d^{D}x\ \mathrm{Tr\ }A^{\mu }\left( \bar{\varphi}_{\mu }-\varphi _{\mu
}^{\mathrm{T}}\right) ,  \label{S_GZ2} \\
S_{\mathcal{K}}\left( \Phi ,B\right) &=&\frac{1}{2}\int d^{D}x\ \mathrm{Tr}%
\left\{ \left[ D^{\nu }\left( A+B\right) ,\bar{\varphi}^{\mu }\right] \left[
D_{\nu }\left( B\right) ,\varphi _{\mu }^{\mathrm{T}}\right] +\left[ D^{\nu
}\left( B\right) ,\bar{\varphi}^{\mu }\right] \left[ D_{\nu }\left(
A+B\right) ,\varphi _{\mu }^{\mathrm{T}}\right] \right\}  \notag \\
&&-\left( \bar{\varphi},\varphi \rightarrow \bar{\omega},\omega \right)
\equiv \int d^{D}x\ (-\bar{\varphi}_{\mu }^{pq}\mathcal{K}%
_{B}^{pq|rs}\varphi ^{rs|\mu }+\bar{\omega}_{\mu }^{pq}\mathcal{K}%
_{B}^{pq|rs}\omega ^{rs|\mu }).  \notag
\end{eqnarray}%
Using the notation%
\begin{equation*}
\int d^{D}x\ F^{pq}\mathcal{K}_{B}^{pq|rs}G^{rs}\equiv \int d^{D}x\ d^{D}y\
F^{pq}\left( x\right) \mathcal{K}_{B}^{pq|rs}\left( x;y\right) G^{rs}\left(
y\right) ,\ \ \epsilon \left( F\right) =\epsilon \left( G\right) ,
\end{equation*}%
for the expression%
\begin{equation*}
-\frac{1}{2}\int d^{D}x\ \mathrm{Tr}\left\{ \left[ D_{\mu }\left( A+B\right)
,F\right] \left[ D^{\mu }\left( B\right) ,G\right] +\left[ D_{\mu }\left(
B\right) ,F\right] \left[ D^{\mu }\left( A+B\right) ,G\right] \right\} ,
\end{equation*}%
we find, due to the (anti)symmetry of the latter under $F\leftrightarrow G$,
the following property:%
\begin{equation*}
\mathcal{K}_{B}^{pq|rs}\left( x;y\right) =\mathcal{K}_{B}^{rs|pq}\left(
y;x\right) .
\end{equation*}%
Thereby, we interpret $\mathcal{S}_{\mathrm{GZ}}\left( \Phi ,B\right) $ in (%
\ref{S_GZ2}) as an alternative local Gribov--Zwanziger action on the
background $B_{\mu }$, with the corresponding background horizon functional $%
\mathcal{H}\left( A,B\right) $ given by
\begin{equation}
\exp \left\{ \hbar ^{-1}\left[ \mathcal{H}\left( A,B\right) -\mathcal{H}%
\left( 0,B\right) \right] \right\} =\int d\bar{\varphi}\ d\varphi \ d\bar{%
\omega}\ d\omega \exp \left[ -\hbar ^{-1}\mathcal{S}_{\gamma }\left( \Phi
,B\right) \right] ,\ \ \mathcal{H}\left( 0,B\right) \equiv H\left( 0\right) ,
\label{new_horizon}
\end{equation}%
where%
\begin{eqnarray*}
\mathcal{S}_{\gamma }\left( \Phi ,B\right) &=&S_{\mathcal{K}}\left( \Phi
,B\right) -i\gamma {}g\int d^{D}x\ \mathrm{Tr\ }A^{\mu }\left( \bar{\varphi}%
_{\mu }-\varphi _{\mu }^{\mathrm{T}}\right) \\
&=&\int d^{D}x\ \left[ -\bar{\varphi}_{\mu }^{pq}\mathcal{K}%
_{B}^{pq|rs}\varphi ^{rs|\mu }+\bar{\omega}_{\mu }^{pq}\mathcal{K}%
_{B}^{pq|rs}\omega ^{rs|\mu }+i\gamma {}g\ f^{prq}A^{r|\mu }\left( \bar{%
\varphi}_{\mu }^{pq}+\varphi _{\mu }^{pq}\right) \right] .
\end{eqnarray*}%
The action $\mathcal{S}_{\mathrm{GZ}}\left( \Phi ,B\right) $ in (\ref{S_GZ2}%
) is manifestly invariant with respect to the local transformations (\ref%
{even_though}), which produces a unit Jacobian in the infinitesimal case and
implies an invariance of the background generating functional%
\begin{equation}
Z_{\mathcal{H}}\left( B,J\right) =\int d\Phi \ \exp \left\{ -\hbar ^{-1}%
\left[ \mathcal{S}_{\mathrm{GZ}}\left( \Phi ,B\right) +J_{A}\Phi ^{A}\right]
\right\} ,\ \ \ Z_{\mathcal{H}}\left( 0,J\right) =Z_{H}\left( J\right) ,
\label{ZHBJnew}
\end{equation}%
under the following local transformations of the sources and the background
field:%
\begin{equation}
\delta _{\xi }B_{\mu }^{p}=D_{\mu }^{pq}\left( B\right) \xi ^{q},\ \ \
\delta _{\xi }(J_{\mu \left( A\right) },J_{\left( b\right) },J_{\left( \bar{c%
}\right) },J_{\left( c\right) })^{p}=gf^{prq}(J_{\mu \left( A\right)
},J_{\left( b\right) },J_{\left( \bar{c}\right) },J_{\left( c\right)
})^{r}\xi ^{q}.  \label{ZHBinv}
\end{equation}%
This also means a local invariance of the alternative horizon functional $%
\mathcal{H}\left( A,B\right) $ in (\ref{new_horizon}),%
\begin{equation*}
\delta _{\xi }\mathcal{H}\left( A,B\right) =0,\ \ \delta _{\xi }B_{\mu
}^{p}=D_{\mu }^{pq}\left( B\right) \xi ^{q},\ \ \delta _{\xi }A_{\mu
}^{p}=gf^{prq}A_{\mu }^{r}\xi ^{q}.
\end{equation*}%
As we introduce the generating functionals of connected $W_{\mathcal{H}%
}\left( B,J\right) $ and vertex $\Gamma _{\mathcal{H}}\left( B,\phi \right) $
Green's functions on a background,%
\begin{equation}
Z_{\mathcal{H}}=\exp \left( -\hbar ^{-1}W_{\mathcal{H}}\right) ,\ \ \Gamma _{%
\mathcal{H}}\left( B,\phi \right) =W_{\mathcal{H}}\left( B,J\right)
-J_{A}\phi ^{A},\ \ \phi ^{A}=\frac{\overrightarrow{\delta }}{\delta J_{A}}%
W_{\mathcal{H}},\ \ J_{A}=-\Gamma _{\mathcal{H}}\frac{\overleftarrow{\delta }%
}{\delta \phi ^{A}},  \label{defin}
\end{equation}%
the invariance of $Z_{\mathcal{H}}\left( B,J\right) $ with respect to (\ref%
{ZHBinv}) can be recast as the invariance of $\Gamma _{\mathcal{H}}\left(
B,\phi \right) $ under the following local transformations, $\delta _{\xi
}\Gamma _{\mathcal{H}}=0$ (see Appendix \ref{B.III}),%
\begin{equation}
\delta _{\xi }B_{\mu }^{p}=D_{\mu }^{pq}\left( B\right) \xi ^{q},\ \ \
\delta _{\xi }(A_{\mu },b,\bar{c},c)^{p}=gf^{prq}(A_{\mu },b,\bar{c}%
,c)^{r}\xi ^{q},  \label{local_tr}
\end{equation}%
which consist of the gauge transformations for the background field $B_{\mu
} $ and of the local $SU(N)$ transformations\ for the quantum fields $\phi
^{A} $. These symmetry properties are readily generalized to the case of
extended functionals $Z_{\mathcal{H}}\left( B,\mathfrak{J}\right) $,\ $W_{%
\mathcal{H}}\left( B,\mathfrak{J}\right) $, $\Gamma _{\mathcal{H}}\left(
B,\Phi \right) $, where $\mathfrak{J}$ are sources to the fields $\Phi $,
with the invariance $\delta _{\xi }Z_{\mathcal{H}}=\delta _{\xi }W_{\mathcal{%
H}}=\delta _{\xi }\Gamma _{\mathcal{H}}=0$ under the gauge transformations
of $B_{\mu }$ combined with the local $SU(N)$ transformations\ of $\mathfrak{%
J}$ or $\Phi $, so that the background effective action $\Gamma _{\mathrm{eff%
}}\left( B\right) $\ for the Gribov--Zwanziger model defined as%
\begin{equation}
\Gamma _{\mathrm{eff}}\left( B\right) =\left. \Gamma _{\mathcal{H}}\left(
B,\Phi \right) \right\vert _{\Phi =0}  \label{GZ_eff}
\end{equation}%
is invariant, $\delta _{\xi }\Gamma _{\mathrm{eff}}=0$, under the gauge
transformations of the background field $B_{\mu }$.

\subsection{Local BRST Invariant Horizon Term}

By considering a gauge-invariant horizon $H\left( A^{\mathrm{h}}\right)
=\left. H\left( A\right) \right\vert _{A=A^{\mathrm{h}}}$ of \cite{Capri-1},
involving non-local transverse fields $A_{\mu }^{\mathrm{h}}$=$(A^{\mathrm{h}%
})_{\mu }^{p}T^{p}$, the case of the background term becomes simplified
using gauge- and BRST-invariant fields $A_{\mu }^{\mathrm{h}}$, defined%
\footnote{%
In (\ref{gitrans}), we maintain the notation for the interaction constant $g$
consistent with (\ref{consistent}). The same is implied in (\ref{Stuek})
below.} according to \cite{SemenovTyanshan}, $A_{\mu }=A_{\mu }^{\mathrm{h}%
}+A_{\mu }^{\mathrm{L}}$,
\begin{eqnarray}
&&A_{\mu }^{\mathrm{h}}=\left( \eta _{\mu \nu }-\frac{\partial _{\mu
}\partial _{\nu }}{\partial ^{2}}\right) \left( A^{\nu }-g\left[ \frac{%
\partial A}{\partial ^{2}},A^{\nu }-\frac{1}{2}\partial ^{\nu }\frac{%
\partial A}{\partial ^{2}}\right] \right) +O(A^{3}),\ \ A_{\mu }^{\mathrm{h}}%
\overleftarrow{s}=0,  \label{gitrans} \\
&&H(A)=H(A^{\mathrm{h}})+\gamma ^{2}\int d^{D}x\,d^{D}y\ R^{p}(x,y)\partial
^{\mu }A_{\mu }^{p}(y),\ \ H(A^{\mathrm{h}})\overleftarrow{s}=0,
\label{gihor}
\end{eqnarray}%
with a non-local function $R^{p}(x,y)$ of \cite{Capri-1}. Due to this
structure, the second term in $H(A)$ can be added to the gauge-fixing term $%
b^{p}\partial ^{\mu }A_{\mu }^{p}$ of the Faddeev--Popov action $S_{0}+\Psi
\overleftarrow{s}$ in a way reduced to a change of variables in $Z_{H,\Psi }$
given by the shift $b^{p}\rightarrow b^{p}+\gamma ^{2}R^{p}$ with a unit
Jacobian, which entirely removes the dependence on the BRST symmetry
breaking term entering $Z_{H,\Psi }$. Thereby, the action
\begin{equation}
\tilde{S}_{\mathrm{GZ}}(\phi )=S_{0}+\int d^{D}x\ (\bar{c}^{p}\partial ^{\mu
}A_{\mu }^{p})\overleftarrow{s}+H(A^{\mathrm{h}}),\ \ \phi ^{A}%
\overleftarrow{s}=(D_{\mu }^{pq}c^{q},0,b^{p},g/2f^{pqr}c^{q}c^{r}),
\label{brstinvgz}
\end{equation}%
provides independence under $R_{\xi }$-gauges in the YM theory and the
Standard Model \cite{SMMR,MoshReshIzv}, with the Faddeev--Popov operator $%
(K)^{pq}(x,y)$ being unaltered and the BRST symmetry unaffected, in which
case one may expect the theory to be unitary in the framework of the
Faddeev--Popov quantization rules \cite{FP}. The same results concerning the
issues of unitarity and gauge-independence can be presented using the above
fields $A_{\mu }^{\mathrm{h}}$ in a description of the Gribov--Zwanziger
theory when the horizon functional is localized \cite{Zwanziger,Kondo} using
a quartet of auxiliary fields $\phi _{\mathrm{aux}}$=$(\bar{\varphi}_{\mu
}^{pq},\varphi _{\mu }^{pq},\bar{\omega}_{\mu }^{pq},\omega _{\mu }^{pq})$
having opposite Grassmann parities, $\epsilon (\varphi ,\bar{\varphi})$=$%
\epsilon (\omega ,\bar{\omega})+1$=$0$. Using the previously employed
parameterization \cite{Kondo} of the gauge-invariant horizon $H\left( A^{%
\mathrm{h}}\right) $ in terms of $\phi _{\mathrm{aux}}$, namely, by setting $%
B_{\mu }=0$ and replacing $K_{B}^{pq}(A)\rightarrow K^{pq}(A^{\mathrm{h}})$
in (\ref{HS_gamma}), (\ref{S_gamma}), we have
\begin{eqnarray}
S_{\mathrm{GZ}}(\phi ,\phi _{\mathrm{aux}}) &=&S_{0}(A)+\int d^{D}x(\bar{c}%
^{p}\partial ^{\mu }A_{\mu }^{p})\overleftarrow{s}+\tilde{S}_{\gamma }(A^{%
\mathrm{h}},\phi _{\mathrm{aux}}),  \label{Sgamma} \\
\tilde{S}_{\gamma }(A^{\mathrm{h}},\phi _{\mathrm{aux}}) &=&\int d^{D}x\left[
-\bar{\varphi}_{\mu }^{rp}K^{pq}(A^{\mathrm{h}})\varphi ^{rq|\mu }+\bar{%
\omega}_{\mu }^{rp}K^{pq}(A^{\mathrm{h}})\omega ^{rq|\mu }\right.  \notag \\
&&+\left. i\gamma g\ f^{prq}(A^{\mathrm{h}})^{r|\mu }\left( \bar{\varphi}%
_{\mu }^{pq}+\varphi _{\mu }^{pq}\right) +\gamma ^{2}D(N{}^{2}-1)\right] .
\label{Sgamma1}
\end{eqnarray}%
The part $\tilde{S}_{\gamma }$ additional to the Faddeev--Popov action is
manifestly invariant under the BRST transformations (\ref{BRST}) combined
with a trivial form of BRST transformations for the auxiliary fields, $\phi
_{\mathrm{aux}}\overleftarrow{s}=0$, suggested for the first time in \cite%
{MoshReshIzv}.

Despite a formally localized description, the Gribov--Zwanziger (GZ) action $%
S_{\mathrm{GZ}}(\phi ,\phi _{\mathrm{aux}})$ in (\ref{Sgamma}) remains in
fact non-local due to the presence of the non-local field $A_{\mu }^{\mathrm{%
h}}$. To render the action local, we use a parameterization in terms of a
Stueckelberg-like field $\zeta ^{p}$ introduced in \cite{localtransGZ} with
the help of a matrix-valued field $h^{pq}$ defined by%
\begin{equation}
A_{\mu }^{\mathrm{h}}=g^{-1}hD_{\mu }(A)h^{-1}=hA_{\mu
}h^{-1}+g^{-1}h\partial _{\mu }h^{-1},\ \ \mathtt{\ }h=\exp \left( -g\zeta
^{p}T^{p}\right) ,  \label{Stuek}
\end{equation}%
and subject to the transversality condition, implying (\ref{gitrans}),
\begin{equation}
\partial ^{\mu }A_{\mu }^{\mathrm{h}}=0.  \label{transAh}
\end{equation}%
Given this, a completely local and BRST-invariant GZ action can be
determined in an extended configuration space parameterized by the fields\
\begin{equation}
\mathbf{\Phi }^{\mathcal{A}}=\left( A^{p|\mu },b^{p},\bar{c}^{p},c^{p};\bar{%
\varphi}^{pq|\mu },\varphi ^{pq|\mu },\bar{\omega}^{pq|\mu },\omega ^{pq|\mu
};h^{pq}(\zeta ),\tau ^{p},\bar{\eta}^{p},\eta ^{p}\right) ,
\label{localconfspace}
\end{equation}%
where $\epsilon \left( \bar{\eta}\right) =\epsilon \left( \eta \right)
=\epsilon \left( \tau \right) +1=1$, and has the form
\begin{eqnarray}
S_{\mathrm{loc,GZ}}(\phi ,\phi _{\mathrm{aux}}) &=&S_{\mathrm{FP}}(\phi
)-H(0)+\bar{S}_{\gamma }(A^{\mathrm{h}},\phi _{\mathrm{aux}},h(\zeta ),\tau ,%
\bar{\eta},\eta ),  \label{hSgamma} \\
\bar{S}_{\gamma } &=&{\tilde{S}}_{\gamma }+\int d^{D}x\left[ \tau
^{p}\partial ^{\mu }(A^{\mathrm{h}})_{\mu }^{p}-{\bar{\eta}}^{p}K^{pq}(A^{%
\mathrm{h}})\eta ^{q}\right] .  \label{hSgamma1}
\end{eqnarray}%
A generating functional for the local BRST-invariant horizon is then given by%
\begin{equation}
Z_{\mathrm{loc,}H}\left( J\right) =\int d\Phi \ d\Phi _{\mathrm{loc}}\ \exp
\left\{ -\hbar ^{-1}\left[ S_{\mathrm{loc,GZ}}\left( \Phi \right) +J_{A}\phi
^{A}\right] \right\} ,\ \ d\Phi _{\mathrm{loc}}=\left( d\zeta ,d\tau ,d\bar{%
\eta},d\eta \right)  \label{Zlocal}
\end{equation}%
(which is readily extended, along the lines of (\ref{ZBJ_horizon}), to a
generating functional $Z_{\mathrm{loc,}H}\left( J,L\right) $ with local
composite fields), with the integrand being invariant under the following
BRST transformations:
\begin{equation}
\mathbf{\Phi }^{\mathcal{A}}\overleftarrow{s}=\left( D^{pq|\mu }\left(
A\right)
c^{q},0,b^{p},g/2f^{pqr}c^{q}c^{r};0,0,0,0;gc^{s}(T^{s})^{pr}h^{rq},0,0,0%
\right) .  \label{totlocBRST}
\end{equation}%
In a matrix form, the transformation $\delta h=h(\zeta )\overleftarrow{s}%
\lambda $ can be presented in terms of the field $\zeta $ as follows:%
\begin{equation}
g\delta \zeta =-gc\lambda +\left. \mathcal{Z}\right\vert _{\delta \zeta
=-c\lambda }\equiv gj(\zeta )c\lambda ,\ \ \delta \zeta =\zeta
\overleftarrow{s}\lambda ,  \label{delta_zeta}
\end{equation}%
where $\mathcal{Z}$ is given by the Baker--Campbell--Hausdorff formula%
\begin{equation*}
\mathcal{Z}=\frac{1}{2}\left[ \mathcal{X},\mathcal{Y}\right] +\frac{1}{12}%
\left\{ \left[ \mathcal{X},\left[ \mathcal{X},\mathcal{Y}\right] \right] +%
\left[ \mathcal{Y},\left[ \mathcal{Y},\mathcal{X}\right] \right] \right\}
+\cdots \ ,\ \ \exp \mathcal{X}\exp \mathcal{Y}=\exp \left( \mathcal{X}+%
\mathcal{Y}+\mathcal{Z}\right) ,
\end{equation*}%
corresponding to the explicit values%
\begin{equation*}
\mathcal{X}=-g\left( \zeta +\delta \zeta \right) ,\ \ \mathcal{Y}=g\zeta ,\
\ \left[ \mathcal{X},\mathcal{Y}\right] =g^{2}\left[ \zeta ,\delta \zeta %
\right] .
\end{equation*}%
The expression (\ref{delta_zeta}) for $\delta \zeta $ presents $\zeta ^{p}%
\overleftarrow{s}$ as an expansion in powers of $g$, which is also an
explicit power series in $\zeta ^{p}$. For instance, the linear
approximation has the form%
\begin{equation*}
\zeta ^{p}\overleftarrow{s}=j^{pq}(\zeta )c^{q},\ \ j^{pq}=-\delta ^{pq}+%
\frac{g}{2}f^{prq}\zeta ^{r}+O(g^{2}).
\end{equation*}%
The BRST-invariant GZ theory with the action (\ref{hSgamma}) can be
naturally extended to a background-dependent GZ action $\mathcal{S}_{\mathrm{%
loc,GZ}}\left( \mathbf{\Phi },B\right) $, along the lines of the
representation (\ref{S_GZ2}),%
\begin{eqnarray}
\mathcal{S}_{\mathrm{loc,GZ}}\left( \mathbf{\Phi },B\right) &=&S_{\mathrm{FP}%
}\left( \phi ,B\right) -H\left( 0\right) +S_{\mathrm{loc,}\mathcal{K}}\left(
\mathbf{\Phi },B\right) -i\gamma g\int d^{D}x\ \mathrm{Tr\ }A_{\mu }^{%
\mathrm{h}}\left( \bar{\varphi}^{\mu }-\varphi ^{\mu \mathrm{T}}\right) ,
\label{S_GZ2L} \\
S_{\mathrm{loc,}\mathcal{K}}\left( \mathbf{\Phi },B\right) &=&\frac{1}{2}%
\int d^{D}x\ \mathrm{Tr}\left\{ \left[ D^{\nu }\left( A^{\mathrm{h}%
}+B\right) ,\bar{\varphi}^{\mu }\right] \left[ D_{\nu }\left( B\right)
,\varphi _{\mu }^{\mathrm{T}}\right] +\left[ D^{\nu }\left( B\right) ,\bar{%
\varphi}^{\mu }\right] \left[ D_{\nu }\left( A^{\mathrm{h}}+B\right)
,\varphi _{\mu }^{\mathrm{T}}\right] \right\}  \notag \\
&&-\left( \bar{\varphi},\varphi ^{\mathrm{T}}\rightarrow \bar{\omega},\omega
^{\mathrm{T}}\right) +2\left( \bar{\varphi},\varphi ^{\mathrm{T}}\rightarrow
\bar{\eta},\eta \right) +2\int d^{D}x\ \mathrm{Tr\ }\tau \left[ D^{\mu
}(B),A_{\mu }^{\mathrm{h}}\right]  \notag \\
&\equiv &\int d^{D}x\left[ -\bar{\varphi}_{\mu }^{pq}\mathcal{K}_{\mathrm{%
loc,}B}^{pq|rs}\varphi ^{rs|\mu }+\bar{\omega}_{\mu }^{pq}\mathcal{K}_{%
\mathrm{loc,}B}^{pq|rs}\omega ^{rs|\mu }-\bar{\eta}^{p}\mathcal{K}_{\mathrm{%
loc,}B}^{pq}\eta ^{q}+\tau ^{p}D_{\mu }^{pq}(B)(A^{\mathrm{h}})^{q|\mu }%
\right] .  \notag
\end{eqnarray}%
This action is background-invariant, including the corresponding generating
functional of Green's functions,%
\begin{eqnarray}
Z_{\mathrm{loc,}H}\left( B,\mathbf{J},L,\mathbf{\Phi }^{\ast }\right)
&=&\int d\Phi \ d\Phi _{\mathrm{loc}}\exp \left\{ -\hbar ^{-1}\left[ S_{%
\mathrm{loc,GZ}}\left( \mathbf{\Phi },B\right) +\zeta ^{\ast p}j^{pq}(\zeta
)c^{q}\right. \right.  \notag \\
&&\left. \left. +\mathbf{J}_{\mathcal{A}}\mathbf{\Phi }^{\mathcal{A}%
}+L_{m}\sigma ^{m}\left( \Phi ,B\right) \right] \right\} ,\ \ Z_{\mathrm{loc,%
}H}\left( 0,\mathbf{J},0,0\right) |_{\mathbf{J}=J}=Z_{\mathrm{loc,}H}\left(
J\right) ,  \label{Zlocalbrst}
\end{eqnarray}%
so that an effective action $\Gamma _{\mathrm{loc,eff}}\left( B,\Sigma
\right) $ for the GZ theory featuring a local BRST-invariant horizon with
background and composite fields,%
\begin{equation}
\Gamma _{\mathrm{loc,eff}}\left( B,\Sigma \right) =\left. \Gamma _{\mathrm{%
loc}}\left( B,\mathbf{\Phi },\Sigma ,\mathbf{\Phi }^{\ast }\right)
\right\vert _{\mathbf{\Phi }=\mathbf{\Phi }^{\ast }=0},  \label{Gamma_effL}
\end{equation}%
proves to be invariant under the local transformations (\ref{back_gauge}).
This is a first main result of the present subsection.

For the generating functionals of Green's functions $Z_{\mathrm{loc,}H}$ and
$W_{\mathrm{loc,}H}$, related by $Z_{\mathrm{loc,}H}$\ $=\ e^{-\hbar ^{-1}W_{%
\mathrm{loc,}H}}$ and depending on $\left( B,\mathbf{J},L,\mathbf{\Phi }%
^{\ast }\right) $ in (\ref{Zlocalbrst}), as well as for the effective action
$\Gamma _{\mathrm{loc}}\left( B,\mathbf{\Phi },\Sigma ,\mathbf{\Phi }^{\ast
}\right) $ in (\ref{Gamma_effL}) obtained by a Legendre transform of $W_{%
\mathrm{loc,}H}$ along the lines of \textbf{\ }(\ref{Gamma}), we can derive
modified Ward identities in the respective forms (\ref{mWIclalg}), (\ref%
{modWard_W}), (\ref{modWard_Gamma_1}), as well as the usual Ward identities (%
\ref{Ward_Z}), (\ref{Ward_W}), (\ref{Ward_Gamma_1}), with appropriate
Grassmann-odd operators $\hat{\omega}_{H}$, $\hat{\Omega}_{H}$, $\hat{\omega}%
_{\Gamma ,H}$. These identities are deduced starting from the FD BRST
transformations (\ref{totlocBRST}), $\Delta \Phi ^{\mathcal{A}}=\Phi ^{%
\mathcal{A}}\overleftarrow{s}\lambda (\Phi )$, with a Grassmann-odd FD
functional $\lambda (\Phi )$, further background-extended as $\overleftarrow{%
s}\rightarrow \overleftarrow{s}_{\mathrm{q}}$, $\lambda (\Phi )\rightarrow
\lambda (\Phi ,B)$. The operators $\hat{\omega}_{H}$, $\hat{\Omega}_{H}$, $%
\hat{\omega}_{\Gamma ,H}$ are constructed as their counterparts $\hat{\omega}
$, $\hat{\Omega}$, $\hat{\omega}_{\Gamma }$ of (\ref{omega}), (\ref{Omega}),
(\ref{omegaG}), albeit with a GZ action $S_{\mathrm{loc,GZ}}$ defined in a
space of variables which is larger than that for the Faddeev--Popov action $%
S_{\mathrm{FP}}$. For instance, the operator $\hat{\omega}_{H}$ is given by%
\begin{equation}
\hat{\omega}_{H}=\left[ \mathbf{J}_{\mathcal{A}}+\delta _{A\mathcal{A}%
}L_{m}\sigma _{,A}^{m}\left( \frac{\hbar }{i}\frac{\overrightarrow{\delta }}{%
\delta J},B\right) \right] \frac{\overrightarrow{\delta }}{\delta \mathbf{%
\Phi }_{\mathcal{A}}^{\ast }},\ \ \hat{\omega}_{H}^{2}=0.  \label{omegaH}
\end{equation}%
A study of the gauge-dependence problem following the receipt of Section~\ref%
{sec4} leads to the representations (\ref{GDInew1}), (\ref{Delta_W}), (\ref%
{Delta_Gamma}) for the respective finite variations $\Delta Z_{\Psi }^{%
\mathrm{loc}}\ \equiv \ \Delta Z_{\mathrm{loc,}H}\left( B,\mathbf{J},L,%
\mathbf{\Phi }^{\ast }\right) $, $\Delta W_{\Psi }^{\mathrm{loc}}\ \equiv \
\Delta W_{\mathrm{loc,}H}\left( B,\mathbf{J},L,\mathbf{\Phi }^{\ast }\right)
$, $\Delta \Gamma _{\Psi }^{\mathrm{loc}}\ \equiv \ \Delta \Gamma _{\mathrm{%
loc,}H}\left( B,\mathbf{\Phi },L,\mathbf{\Phi }^{\ast }\right) $ generated
by finite variations of the gauge Fermion $\Delta \Psi $, so that $\Delta
Z_{\Psi }^{\mathrm{loc}}=Z_{\Psi +\Delta \Psi }^{\mathrm{loc}}-Z_{\Psi }^{%
\mathrm{loc}}$,%
\begin{eqnarray}
\ \Delta Z_{\Psi }^{\mathrm{loc}} &=&\hat{\omega}_{H}{\lambda }\left( \frac{%
\hbar \overrightarrow{\delta }}{i\delta \mathbf{J}},B\left\vert -\Delta \Psi
\right. \right) Z_{\Psi }^{\mathrm{loc}}=\frac{i}{\hbar }\hat{\omega}_{H}{%
\Delta }\Psi \left( \frac{\hbar \overrightarrow{\delta }}{i\delta \mathbf{J}}%
,B\right) Z_{\Psi }^{\mathrm{loc}}+o(\Delta \Psi ),  \label{GDInew11} \\
\ \Delta W_{\Psi }^{\mathrm{loc}} &=&\frac{\hbar }{i}\hat{\Omega}_{H}{%
\lambda }\left( \frac{\hbar \overrightarrow{\delta }}{i\delta \mathbf{J}}+%
\frac{\overrightarrow{\delta }W^{\mathrm{loc}}}{\delta \mathbf{J}}%
,B\left\vert -\Delta \Psi \right. \right) =\hat{\Omega}_{H}\Delta \Psi
\left( \frac{\hbar \overrightarrow{\delta }}{i\delta \mathbf{J}}+\frac{%
\overrightarrow{\delta }W^{\mathrm{loc}}}{\delta \mathbf{J}},B\right)
+o(\Delta \Psi ),  \label{Delta_W1} \\
\ \Delta \Gamma _{\Psi }^{\mathrm{loc}} &=&\frac{\hbar }{i}\hat{\omega}%
_{\Gamma }\langle \langle {\lambda }\left( B|-\Delta \Psi \right) \rangle
\rangle =\delta \Gamma _{\mathrm{loc,}H}+o\left( \langle \langle \Delta \Psi
\rangle \rangle \right) .  \label{Delta_Gamma1}
\end{eqnarray}%
As a result, the EA with composite and background fields for the GZ action $%
S_{\mathrm{loc,GZ}}$ determined by the local BRST-invariant horizon does not
depend on a variation of the gauge condition on the extremals $\delta \Gamma
_{\mathrm{loc,}H}/{\delta \mathbf{\Phi }^{\mathcal{A}}}=\delta \Gamma _{%
\mathrm{loc,}H}/{\delta \Sigma ^{m}}=0$. Thereby, we can state that the
Gribov horizon defined using a composite field (being added to the
Faddeev--Popov quantum action) and the horizon defined without such a field
lead to different forms of the mass shell for the respective EA. This is a
second main result of the present subsection.

By choosing local BRST-invariant composite fields $\sigma ^{m}=\sigma
^{m}\left( A,A^{\mathrm{h}},B\right) $, related in the case of $D=4$ to an
emergence of dimension-two condensates, with $Z_{\mathrm{loc,}H}\left(
B,J,L\right) $ defined along the lines of (\ref{ZBJ_horizon}) according to%
\begin{equation}
\left( \sigma ^{1},\sigma ^{2}\right) =\frac{1}{2}\left( \mathrm{Tr\ }A_{\mu
}^{\mathrm{h}}A^{\mathrm{h}\mu },2\mathrm{Tr}\left[ \bar{\varphi}^{\mu }{%
\varphi }_{\mu }^{\mathrm{T}}-\bar{\omega}^{\mu }{\omega }_{\mu }^{\mathrm{T}%
}\right] \right) ,\ \ \left( \sigma ^{1},\sigma ^{2}\right) =\left( \sigma
^{1},\sigma ^{2}\right) \left( x\right) ,  \label{compbackbrst}
\end{equation}%
we arrive (for $L_{1}\left( x\right) =m^{2}$ and $L_{2}\left( x\right)
=-M^{2}$) at a refined GZ action $S_{\mathrm{RGZ}}$ in Landau gauge. Using
FD BRST transformations relating the integrands of generating functionals of
Green's functions in Landau gauge and arbitrary $R_{\xi }$-gauges, we obtain
from (\ref{S_GZ2L}) a refined GZ action $S_{\mathrm{RGZ}}^{\mathrm{LCG}}$ in
covariant gauges; see \cite{1708.01543}, Eq. (34). Thereby, one can extend
the related study of renormalizability \cite{1708.01543} for the resulting
quantum action and generating functional $Z_{\mathrm{loc,}H}\left(
B,J,L\right) $ in all orders of perturbation theory to the case of arbitrary
local composite fields. This is a third main result of the present
subsection.

\section{Two Dimensional Gravity with Dynamical Torsion\label{sec6}}

\setcounter{equation}{0}\renewcommand{\theequation}{7.\arabic{equation}}

Consider a theory of two-dimensional gravity with dynamical torsion
described in terms of a zweibein $e_{\mu }^{i}$ and a Lorentz connection $%
\omega _{\mu }$ by the action \cite{VK}%
\begin{equation}
S_{0}(e,\omega )=\int {d^{2}}x\;e\left( \frac{1}{16\alpha }R_{\mu \nu
}{}^{ij}R^{\mu \nu }{}_{ij}-\frac{1}{8\beta }T_{\mu \nu }{}^{i}T^{\mu \nu
}{}_{i}-\gamma \right) ,  \label{2dim_grav}
\end{equation}%
where $\alpha $, $\beta $, $\gamma $ are constant parameters. For indices of
quantities transforming by the local Lorentz group, we use Latin characters:
$i$, $j$, $k\ldots $ $(i=0,1)$; $\varepsilon ^{ij}$ is a constant
antisymmetric second-rank pseudo-tensor subject to the normalization
condition $\varepsilon ^{01}=1$. Greek characters stand for indices of
quantities transforming as (pseudo-)tensors under the general coordinate
transformations: $\lambda $, $\mu $, $\nu \ldots $ $(\lambda =0,1)$. The
Latin indices are raised and lowered by the Minkowski metric $\eta _{ij}$ $%
(+,-)$ and the Greek indices, by the metric tensor $g_{\mu \nu }=\eta
_{ij}e_{\mu }^{i}e_{\nu }^{j}$. Besides, the following notation is used:%
\begin{eqnarray}
e &=&\mathrm{det\ }e_{\mu }^{i},  \notag \\
R_{\mu \nu }{}^{ij} &=&\varepsilon ^{ij}R_{\mu \nu },\;\;R_{\mu \nu
}=\partial _{\mu }\omega _{\nu }-(\mu \leftrightarrow \nu ),  \label{notate}
\\
T_{\mu \nu }{}^{i} &=&\partial _{\mu }e_{\nu }^{i}+\varepsilon ^{ij}\omega
_{\mu }e_{\nu j}-(\mu \leftrightarrow \nu ).  \notag
\end{eqnarray}%
The action (\ref{2dim_grav}) is invariant under the local Lorentz
transformations $e_{\mu }^{i}\rightarrow e_{\mu }^{^{\prime }i}$, $\omega
_{\mu }\rightarrow \omega ^{^{\prime }}{}_{\mu }$%
\begin{eqnarray}
e_{\mu }^{^{\prime }i} &=&(\Lambda e_{\mu })^{i},  \notag \\
(\Omega ^{^{\prime }}{}_{\mu })_{j}^{i} &=&(\Lambda \Omega _{\mu }\Lambda
^{-1})_{j}^{i}+(\Lambda \partial _{\mu }\Lambda ^{-1})_{j}^{i},\;\;\;(\Omega
_{\mu })_{j}^{i}\equiv \varepsilon ^{ik}\eta _{kj}\omega _{\mu },
\label{Lorentz}
\end{eqnarray}%
or, infinitesimally, with a parameter $\zeta $,%
\begin{equation}
\delta e_{\mu }^{i}=\varepsilon ^{ij}e_{\mu j}\zeta ,\;\;\delta \omega _{\mu
}=-\partial _{\mu }\zeta ,  \label{Lor_infinite}
\end{equation}%
as well as under the general coordinate transformations, $x\rightarrow
x^{^{\prime }}=x^{^{\prime }}(x)$,%
\begin{eqnarray}
e_{\mu }^{i} &\rightarrow &e_{\mu }^{^{\prime }i}(x^{^{\prime }})=\frac{%
\partial x^{\lambda }}{\partial x^{^{\prime }\mu }}e_{\lambda }^{i}(x),
\notag \\
\omega _{\mu } &\rightarrow &\omega _{\mu }^{^{\prime }}(x^{^{\prime }})=%
\frac{\partial x^{\lambda }}{\partial x^{^{\prime }\mu }}\omega _{\lambda
}(x),  \label{gen_coord}
\end{eqnarray}%
implying the infinitesimal field variations, with some parameters $\xi ^{\mu
}$,%
\begin{equation}
\delta e_{\mu }^{i}=e_{\nu }^{i}\partial _{\mu }\xi ^{\nu }+(\partial _{\nu
}e_{\mu }^{i})\xi ^{\nu },\ \ \ \delta \omega _{\mu }=\omega _{\nu }\partial
_{\mu }\xi ^{\nu }+(\partial _{\nu }\omega _{\mu })\xi ^{\nu }.
\label{form_trans}
\end{equation}%
The gauge transformations (\ref{Lor_infinite}), (\ref{form_trans}) form a
closed algebra:
\begin{eqnarray}
{[}\delta _{\zeta (1)},\;\delta _{\zeta (2)}{]} &=&0,  \notag \\
{[}\delta _{\xi (1)},\;\delta _{\xi (2)}{]} &=&\delta _{\xi (1,2)},
\label{close-alg} \\
{[}\delta _{\zeta },\;\delta _{\xi }{]} &=&\delta _{\zeta ^{^{\prime }}},
\notag
\end{eqnarray}%
where
\begin{equation*}
\xi ^{\mu }{}_{(1,2)}=\xi ^{\nu }{}_{(1)}\partial _{\nu }\xi ^{\mu
}{}_{(2)}-(\partial _{\nu }\xi ^{\mu }{}_{(1)})\xi ^{\nu
}{}_{(2)},\;\;\;\zeta ^{^{\prime }}=(\partial _{\mu }\zeta )\xi ^{\mu }.
\end{equation*}%
so that the Faddeev--Popov method applies to the given theory, with the
total configuration space $\phi ^{A}$\ given by the classical fields $%
(e_{\mu }^{i},\omega _{\mu })$, as well as by the Faddeev--Popov ghosts ($%
\overline{c}$, $c$, $\overline{c}^{\mu }$, $c^{\mu }$) and the
Nakanishi--Lautrup fields ($b$, $b^{\mu }$), according to the respective
number of gauge parameters $\zeta $, $\xi ^{\mu }$\ in (\ref{Lor_infinite}),
(\ref{form_trans}). The fields $\phi ^{A}=(e_{\mu }^{i},\omega _{\mu
};b,b^{\mu };\overline{c},\overline{c}^{\mu },c,c^{\mu })$\ possess the
following Grassmann parity and ghost number:%
\begin{equation*}
\begin{tabular}{|c|c|c|c|c|}
\hline
& $\left( e_{\mu }^{i},\omega _{\mu }\right) $ & $\left( b,b^{\mu }\right) $
& $\left( \overline{c},\overline{c}^{\mu }\right) $ & $\left( c,c^{\mu
}\right) $ \\ \hline
$\epsilon $ & $0$ & $0$ & $1$ & $1$ \\ \hline
$\mathrm{gh}$ & $0$ & $0$ & $-1$ & $1$ \\ \hline
\end{tabular}%
\end{equation*}%
Let us present a quantum theory for (\ref{2dim_grav}), (\ref{Lor_infinite}),
(\ref{form_trans}), (\ref{close-alg}) in the background field method by
following the treatment \cite{LM}, based on an ansatz for the vacuum
functional (see also \cite{Abbott}) which corresponds to $Z\left( B\right) $%
\ of (\ref{rule2}) in the case of Yang--Mills theories with the
Nakanishi--Lautrup fields\ eliminated using some background gauges. Namely,
we assign to the initial classical fields the sets of quantum $Q$ and
background $B$ fields, which, in view of further convenience, we denote by $%
Q=(q_{\mu }^{i},q_{\mu })$ and $B=(e_{\mu }^{i},\omega _{\mu })$, with the
associated metric tensor $g_{\mu \nu }$ and the notation $e$, $(\Omega _{\mu
})_{j}^{i}\ $in (\ref{notate}), (\ref{Lorentz}) being related to the
background fields alone. Let us also associate the gauge transformations (%
\ref{Lor_infinite}), (\ref{form_trans}) with two kinds of infinitesimal
transformations, namely, background $\delta _{\mathrm{b}}$ and quantum $%
\delta _{\mathrm{q}}$, introduced by analogy with \ref{bq}, so that the
action $S_{0}(Q+B)$ in (\ref{2dim_grav}) should be left invariant under both
kinds of these transformations:%
\begin{eqnarray}
&&%
\begin{array}{ll}
\delta _{\mathrm{b}}e_{\mu }^{i}=\varepsilon ^{ij}e_{\mu j}\zeta +e_{\nu
}^{i}\partial _{\mu }\xi ^{\nu }+(\partial _{\nu }e_{\mu }^{i})\xi ^{\nu },
& \delta _{\mathrm{b}}q_{\mu }^{i}=\varepsilon ^{ij}q_{\mu j}\zeta +q_{\nu
}^{i}\partial _{\mu }\xi ^{\nu }+(\partial _{\nu }q_{\mu }^{a})\xi ^{\nu },
\\
\delta _{\mathrm{b}}\omega _{\mu }=-\partial _{\mu }\zeta +\omega _{\nu
}\partial _{\mu }\xi ^{\nu }+(\partial _{\nu }\omega _{\mu })\xi ^{\nu }, &
\delta _{\mathrm{b}}q_{\mu }=q_{\nu }\partial _{\mu }\xi ^{\nu }+(\partial
_{\nu }q_{\mu })\xi ^{\nu },%
\end{array}
\label{b2} \\
&&  \notag \\
&&%
\begin{array}{ll}
\delta _{\mathrm{q}}e_{\mu }^{i}=0, & \delta _{\mathrm{q}}q_{\mu
}^{i}=\varepsilon ^{ij}(e_{\mu j}+q_{\mu j})\zeta +(e_{\nu }^{i}+q_{\nu
}^{i})\partial _{\mu }\xi ^{\nu }+(\partial _{\nu }e_{\mu }^{i}+\partial
_{\nu }q_{\mu }^{i})\xi ^{\nu }, \\
\delta _{\mathrm{q}}\omega _{\mu }=0, & \delta _{\mathrm{q}}q_{\mu
}=-\partial _{\mu }\zeta +(\omega _{\nu }+q_{\nu })\partial _{\mu }\xi ^{\nu
}+(\partial _{\nu }\omega _{\mu }+\partial _{\nu }q_{\mu })\xi ^{\nu }.%
\end{array}
\label{q2}
\end{eqnarray}%
Following \cite{LM}, we introduce an analogue \cite{Abbott} of the
generating functional of Green's functions, as we denote $(\bar{c},\bar{c}%
^{\mu },c,c^{\mu })=(\overline{C},C)$,%
\begin{equation}
Z(B,J)=\int dQ\;d\overline{C}\;dC\;\exp \left\{ \frac{i}{\hbar }\left[
S_{0}(Q+B)+S_{\mathrm{gf}}(Q,B)+S_{\mathrm{gh}}(Q,B;\overline{C},C)+JQ\right]
\right\} ,  \label{ZVK}
\end{equation}%
where $J=(J_{i}^{\mu },J^{\mu })$ are sources to the quantum fields $%
Q=(q_{\mu }^{i},q_{\mu })$, and the functional $S_{\mathrm{gf}}=S_{\mathrm{gf%
}}(Q,B)$ is determined by some background gauge functions $\chi $, $\chi
_{\mu }$ for the respective gauge parameters $\zeta $, $\xi ^{\mu }$,
according to the condition of invariance under the background
transformations, $\delta _{\mathrm{b}}S_{\mathrm{gf}}=0$, with the ghost
term $S_{\mathrm{gh}}=S_{\mathrm{gh}}(Q,B;\overline{C},C)$ given by the rule
\begin{equation}
S_{\mathrm{gh}}=\int d^{2}x\;\left. (\overline{c}\delta _{\mathrm{q}}\chi +%
\overline{c}^{\mu }\delta _{\mathrm{q}}\chi _{\mu })\right\vert _{(\zeta
,\xi ^{\mu })\rightarrow (c,c^{\mu })}.  \label{26}
\end{equation}%
The background gauge functions $\chi =\chi (Q,B)$ and $\chi _{\mu }=\chi
_{\mu }(Q,B)$ will be chosen, according to \cite{LM}, as linear in the
quantum fields $Q=(q_{\mu }^{i},q_{\mu })$,%
\begin{equation}
\chi =eg^{\mu \nu }\nabla _{\mu }q_{\nu }\;,\ \ \ \chi _{\mu }=eg^{\lambda
\nu }e_{\mu i}\nabla _{\lambda }q_{\nu }^{i}\;,  \label{chi}
\end{equation}%
where $e$, $g^{\mu \nu }$\ are determined by the background fields $e_{\mu
}^{i}$\ ($g^{\mu \lambda }g_{\lambda \nu }=\delta _{\nu }^{\mu }$, $g_{\mu
\nu }=\eta _{ij}e_{\mu }^{i}e_{\nu }^{j}$, $e=\mathrm{det}\ e_{\mu }^{i}$),
and $\nabla _{\mu }$ is a covariant derivative, whose action on an arbitrary
(psedo-)tensor field $T_{\mu _{1}\ldots \mu _{k}}^{\nu _{1}\ldots \nu
_{l}}{}_{i_{1}\ldots i_{m}}^{j_{1}\ldots j_{n}}$\ is given in terms of $%
(\Omega _{\mu })_{j}^{i}=\varepsilon ^{ik}\eta _{kj}\omega _{\mu }$ and the
Christoffel symbols $\Gamma _{\mu \nu }^{\lambda }$%
\begin{equation}
\Gamma _{\mu \nu }^{\lambda }=\frac{1}{2}g^{\lambda \sigma }(\partial _{\nu
}g_{\mu \sigma }+\partial _{\mu }g_{\nu \sigma }-\partial _{\sigma }g_{\mu
\nu })  \label{Christoffel}
\end{equation}%
by the rule%
\begin{eqnarray}
\nabla _{\mu }T_{\mu _{1}\ldots \mu _{k}}^{\nu _{1}\ldots \nu
_{l}}{}_{i_{1}\ldots i_{m}}^{j_{1}\ldots j_{n}} &=&\partial _{\mu }T_{\mu
_{1}\ldots \mu _{k}}^{\nu _{1}\ldots \nu _{l}}{}_{i_{1}\ldots
i_{m}}^{j_{1}\ldots j_{n}}-\Gamma _{\mu \left\{ \mu \right\} }^{\hat{\lambda}%
}T_{\mu _{1}\ldots \hat{\lambda}\ldots \mu _{k}}^{\nu _{1}\ldots \nu
_{l}}{}_{i_{1}\ldots i_{m}}^{j_{1}\ldots j_{n}}+\Gamma _{\mu \hat{\lambda}%
}^{\left\{ \nu \right\} }T_{\mu _{1}\ldots \mu _{k}}^{\nu _{1}\ldots \hat{%
\lambda}\ldots \nu _{l}}{}_{i_{1}\ldots i_{m}}^{j_{1}\ldots j_{n}}  \notag \\
&&+\ (\Omega _{\mu })_{\hat{p}}^{\left\{ j\right\} }T_{\mu _{1}\ldots \mu
_{k}}^{\nu _{1}\ldots \nu _{l}}{}_{i_{1}\ldots i_{m}}^{j_{1}\ldots \hat{p}%
\ldots j_{n}}-(\Omega _{\mu })_{\left\{ i\right\} }^{\hat{p}}T_{\mu
_{1}\ldots \mu _{k}}^{\nu _{1}\ldots \nu _{l}}{}_{i_{1}\ldots \hat{p}\ldots
i_{m}}^{j_{1}\ldots j_{n}}\;,  \label{tensor}
\end{eqnarray}%
with the notation (\ref{as_we_denote}), so that the covariant derivative $%
\nabla _{\mu }$ has the usual properties ($F$, $G$ are arbitrary
(psedo-)tensor fields)\
\begin{equation}
\nabla _{\sigma }g_{\mu \nu }=\nabla _{\sigma }g^{\mu \nu }=0,\ \ \ \nabla
_{\mu }(FG)=F\nabla _{\mu }G+(\nabla _{\mu }F)G.  \label{prop}
\end{equation}%
The above ingredients allow one to construct the gauge-fixing term $S_{%
\mathrm{gf}}$ as a functional being quadratic in $\chi $, $\chi ^{\mu }$
(with certain numeric parameters $a$, $b$)
\begin{equation}
S_{\mathrm{gf}}=\frac{1}{2}\int d^{2}x\;e^{-1}\left( a\chi ^{2}+b\chi _{\mu
}\chi ^{\mu }\right)  \label{29}
\end{equation}%
and invariant under the local Lorentz transformations%
\begin{eqnarray}
e_{\mu }^{^{\prime }i} &=&(\Lambda e_{\mu })^{i},\;\;q_{\mu }^{^{\prime
}i}=(\Lambda q_{\mu })^{i},  \notag \\
(\Omega ^{^{\prime }}{}_{\mu })_{b}^{i} &=&(\Lambda \Omega _{\mu }\Lambda
^{-1})_{j}^{i}+(\Lambda \partial _{\mu }\Lambda ^{-1})_{j}^{i},\;\;q_{\mu
}^{^{\prime }}=q_{\mu },  \label{30}
\end{eqnarray}%
as well as under the general coordinate transformations, $x\rightarrow
x^{^{\prime }}=x^{^{\prime }}(x)$,%
\begin{eqnarray}
&&e_{\mu }^{^{\prime }i}(x^{^{\prime }})=\frac{\partial x^{\lambda }}{%
\partial x^{^{\prime }\mu }}e_{\lambda }^{i}(x),\;\;\;\omega _{\mu
}^{^{\prime }}(x^{^{\prime }})=\frac{\partial x^{\lambda }}{\partial
x^{^{\prime }\mu }}\omega _{\lambda }(x),  \notag \\
&&q_{\mu }^{^{\prime }i}(x^{^{\prime }})=\frac{\partial x^{\lambda }}{%
\partial x^{^{\prime }\mu }}q_{\lambda }^{i}(x),\;\;\;q_{\mu }^{^{\prime
}}(x^{^{\prime }})=\frac{\partial x^{\lambda }}{\partial x^{^{\prime }\mu }}%
q_{\lambda }(x).  \label{31}
\end{eqnarray}%
Indeed, the infinitesimal form of field transformations implied by (\ref{30}%
) and (\ref{31}) is identical with the background transformations (\ref{b2}%
), which satisfies the requirement $\delta _{\mathrm{b}}S_{\mathrm{gf}}=0$.
Given this and the fact that the non-vanishing quantum transformations (\ref%
{q2}), with allowance for (\ref{tensor}), can be represented as ($\zeta
\rightarrow c$, $\xi ^{\mu }\rightarrow c^{\mu }$)
\begin{eqnarray*}
\delta _{\mathrm{q}}q_{\mu }^{i} &=&\varepsilon ^{ij}(e_{\mu j}+q_{\mu
j})c+(e_{\nu }^{i}+q_{\nu }^{i})\nabla _{\mu }c^{\nu }+(\nabla _{\nu }e_{\mu
}^{i}+\nabla _{\nu }q_{\mu }^{i})c^{\nu }-\varepsilon ^{ij}\omega _{\nu
}(e_{\mu j}+q_{\mu b})c^{\nu }, \\
\delta _{\mathrm{q}}q_{\mu } &=&-\nabla _{\mu }c+(\omega _{\nu }+q_{\nu
})\nabla _{\mu }c^{\nu }+(\nabla _{\nu }\omega _{\mu }+\nabla _{\nu }q_{\mu
})c^{\nu },
\end{eqnarray*}%
the ghost contribution $S_{\mathrm{gh}}$ in (\ref{26}) acquires the form
\begin{eqnarray}
S_{\mathrm{gh}} &=&\int d^{2}x\;e\left\{ -\overline{c}\nabla _{\mu }\nabla
^{\mu }c+\overline{c}\nabla ^{\mu }[(\nabla _{\nu }\omega _{\mu }+\nabla
_{\nu }q_{\mu })c^{\nu }+(\omega _{\nu }+q_{\nu })\nabla _{\mu }c^{\nu
}]\right.  \notag \\
&&+\varepsilon ^{ij}\overline{c}^{\mu }e_{\mu i}\nabla ^{\nu }[(e_{\nu
j}+q_{\nu j})(c-\omega _{\lambda }c^{\lambda })]  \notag \\
&&+\left. \overline{c}^{\mu }e_{\mu i}\nabla ^{\nu }[(\nabla _{\lambda
}e_{\nu }^{i}+\nabla _{\lambda }q_{\nu }^{i})c^{\lambda }+(e_{\lambda
}^{i}+q_{\lambda }^{i})\nabla _{\nu }c^{\lambda }]\right\} .  \label{32}
\end{eqnarray}%
The quantum action in (\ref{ZVK}), determined by (\ref{2dim_grav}), (\ref%
{chi}), (\ref{29}), (\ref{32}), proves to be invariant (as well as the
integrand in (\ref{ZVK}) at the vanishing sources $J=0$, within the usual
assumption $\left. \delta \left( x\right) =\partial _{\mu }\delta \left(
x\right) \right\vert _{x=0}=0$) under the background transformations (\ref%
{b2}), combined with a set of compensating local transformations for the
ghost fields \cite{LM},%
\begin{equation}
\begin{array}{ll}
\delta \overline{c}=(\partial _{\mu }\overline{c})\xi ^{\mu }, & \delta
\overline{c}^{\mu }=-\overline{c}^{\nu }\partial _{\nu }\xi ^{\mu
}+(\partial _{\nu }\overline{c}^{\mu })\xi ^{\nu }, \\
\delta c=-c^{\mu }\partial _{\mu }\zeta +(\partial _{\mu }c)\xi ^{\mu }, &
\delta {c}^{\mu }=-{c}^{\nu }\partial _{\nu }\xi ^{\mu }+(\partial _{\nu }{c}%
^{\mu })\xi ^{\nu }.%
\end{array}
\label{33}
\end{equation}%
As a consequence of (\ref{b2}), (\ref{33}), the generating functional $%
Z(B,J) $ in (\ref{ZVK}) is invariant \cite{LM} under the initial gauge
transformations (\ref{Lor_infinite}), (\ref{form_trans}) of the background
fields $B=(e_{\mu }^{i},\omega _{\mu })$, combined with the following local
transformations of the sources $J=(J_{i}^{\mu },J^{\mu })$:
\begin{equation}
\delta J_{i}^{\mu }=-\varepsilon _{i}^{k}J_{k}^{\mu }\zeta -J_{i}^{\nu
}\partial _{\nu }\xi ^{\mu }+\partial _{\nu }(J_{i}^{\mu }\xi ^{\nu }),\ \ \
\delta J^{\mu }=-J^{\nu }\partial _{\nu }\xi ^{\mu }+\partial _{\nu }(J^{\mu
}\xi ^{\nu }),\ \ \ \varepsilon _{j}^{i}\equiv \varepsilon ^{ik}\eta _{kj}\ .
\label{34}
\end{equation}%
On the one hand, this ensures the property
\begin{equation*}
\delta \left( JQ\right) =\int d^{2}x\ \partial _{\mu }F^{\mu }\ ,\ \ \
F^{\mu }\left( x\right) \equiv \left( J_{i}^{\nu }q_{\nu }^{i}+J^{\nu
}q_{\nu }\right) \xi ^{\mu }
\end{equation*}%
for the source term $JQ$ in (\ref{ZVK}), and, on the other hand, this
extends a tensor transformation law for the sources $J_{i}^{\mu }\ $and$\
J^{\mu }$, at the infinitesimal level, by including the respective
contributions $J_{i}^{\mu }\partial _{\nu }\xi ^{\nu }$ and $J^{\mu
}\partial _{\nu }\xi ^{\nu }$,%
\begin{equation*}
\delta J_{i}^{\mu }=\left[ -\varepsilon _{i}^{k}J_{k}^{\mu }\zeta
-J_{i}^{\nu }\partial _{\nu }\xi ^{\mu }+(\partial _{\nu }J_{i}^{\mu })\xi
^{\nu }\right] +J_{i}^{\mu }\partial _{\nu }\xi ^{\nu },\ \ \ \delta J^{\mu
}=\left[ -J^{\nu }\partial _{\nu }\xi ^{\mu }+(\partial _{\nu }J^{\mu })\xi
^{\nu }\right] +J^{\mu }\partial _{\nu }\xi ^{\nu }.
\end{equation*}%
Due to the invariance of $Z(B,J)=\exp \{\left( i/h\right) W(B,J)\}$ under (%
\ref{Lor_infinite}), (\ref{form_trans}), (\ref{34}), one achieves an
invariance \cite{LM} of the functional $\Gamma =\Gamma (B,Q)$ given by%
\begin{equation}
\Gamma (B,Q)=W(B,J)-JQ,\;\;\;Q=\frac{\delta W}{\delta J},\;\;\;J=-\frac{%
\delta \Gamma }{\delta Q},\ \ \ Q=(q_{\mu }^{i},q_{\mu })  \label{35}
\end{equation}%
under the background transformations (\ref{b2}) of the fields $B$ and $Q$
(see Appendix \ref{C.I})%
\begin{equation}
\delta _{\mathrm{b}}\Gamma =\int d^{2}x\left[ \frac{\delta \Gamma }{\delta
B\left( x\right) }\delta _{\mathrm{b}}B\left( x\right) +\frac{\delta \Gamma
}{\delta Q\left( x\right) }\delta _{\mathrm{b}}Q\left( x\right) \right] =0,
\label{delta_b_Gamma}
\end{equation}%
which implies that the effective action $\Gamma _{\mathrm{eff}}(B)$ of the
background field method defined by%
\begin{equation*}
\Gamma _{\mathrm{eff}}(B)=\Gamma (B,Q)|_{Q=0}\ ,
\end{equation*}%
is invariant under the gauge transformations (\ref{Lor_infinite}), (\ref%
{form_trans}) of the background fields $B=(e_{\mu }^{i},\omega _{\mu })$.

Let us proceed to extend the generating functional (\ref{ZVK}), suggested in
\cite{LM}, with the entire quantum action now denoted by $S(Q,B;\overline{C}%
,C)$, to a functional $Z(B,J,L)$, as we introduce some background-dependent
composite fields $\sigma ^{m}\left( Q,B\right) $ with sources$\ L_{m}$,%
\begin{equation*}
\sigma ^{m}\left( Q,B\right) =\sigma _{\mu _{1}\cdots \mu _{l}}^{i_{1}\cdots
i_{k}}\left( Q\left( x\right) ,B\left( x\right) \right) ,\ \
L_{m}=L_{i_{1}\cdots i_{k}}^{\mu _{1}\cdots \mu _{l}}\,\left( x\right) ,\ \
m=\left( x,i_{1},\ldots ,i_{k},\mu _{1},\ldots ,\mu _{l}\right) ,
\end{equation*}%
namely,%
\begin{equation}
Z(B,J,L)=\int dQ\ d\overline{C}\ dC\ \exp \left\{ \frac{i}{\hbar }\left[
S(Q,B;\overline{C},C)+JQ+L_{m}\sigma ^{m}\left( Q,B\right) \right] \right\} .
\label{ZBJL}
\end{equation}%
In doing so, we require that the extended functional $Z(B,J,L)$ should\
inherit the local symmetry of $Z(B,J)$ under the transformations (\ref%
{Lor_infinite}), (\ref{form_trans}), (\ref{34}) of the background fields $%
B=(e_{\mu }^{i},\omega _{\mu })$ and the sources $J=(J_{i}^{\mu },J^{\mu })$%
. To this end, we demand that the composite fields $\sigma _{\mu _{1}\cdots
\mu _{n}}^{i_{1}\cdots i_{m}}(x)=\sigma _{\mu _{1}\cdots \mu
_{n}}^{i_{1}\cdots i_{m}}\left( Q(x),B(x)\right) $ transform as tensors with
respect to the Lorentz (\ref{30}) and general coordinate (\ref{31})
transformations of the quantum $Q$ and background $B$ fields,%
\begin{eqnarray}
\sigma _{\mu _{1}\cdots \mu _{n}}^{^{\prime }i_{1}\cdots i_{m}}(x)
&=&\Lambda _{j_{1}}^{i_{1}}\cdots \Lambda _{j_{m}}^{i_{m}}\sigma _{\mu
_{1}\cdots \mu _{n}}^{j_{1}\cdots j_{m}}(x)\ ,  \notag \\
\sigma _{\mu _{1}\cdots \mu _{n}}^{^{\prime }i_{1}\cdots i_{m}}(x^{^{\prime
}}) &=&\frac{\partial x^{\nu _{1}}}{\partial x^{^{\prime }\mu _{1}}}\cdots
\frac{\partial x^{\nu _{n}}}{\partial x^{^{\prime }\mu _{n}}}\sigma _{\nu
_{1}\cdots \nu _{n}}^{i_{1}\cdots i_{m}}(x)\ ,\ \ \ x^{^{\prime
}}=x^{^{\prime }}(x)\ .  \label{tensors}
\end{eqnarray}%
In general, a composite field $\sigma _{\mu _{1}\cdots \mu
_{n}}^{i_{1}\cdots i_{m}}\left( Q,B\right) $ subject to (\ref{tensors})\ is
multiplicative with respect to the quantum fields $Q=(q_{\mu }^{i},q_{\mu })$
and the background field ingredients $e_{\mu }^{i}$, $g_{\mu \nu }$, $R_{\mu
\nu }{}^{ij}$, $T_{\mu \nu }{}^{i}$, see (\ref{notate}); besides, it may
contain a number of background covariant derivatives $\nabla _{\mu }$
acting\ according to (\ref{tensor}), (\ref{Christoffel}), with the
properties (\ref{prop}). It is obvious, however, that the composite fields
subject to the restriction $\sigma _{\mu _{1}\cdots \mu _{n}}^{i_{1}\cdots
i_{m}}\left( Q,0\right) \not=0$ are allowed to contain the background fields
$B$ only via covariant derivatives $\nabla _{\mu }$, given in terms of $%
\Gamma _{\mu \nu }^{\lambda }$, $(\Omega _{\mu })_{j}^{i}$ and acting on $%
q_{\mu }^{i}$, $q_{\mu }$, namely,%
\begin{equation*}
\nabla _{\mu }q_{\nu }^{i}=\partial _{\mu }q_{\nu }^{i}-\Gamma _{\mu \nu
}^{\lambda }q_{\lambda }^{i}+(\Omega _{\mu })_{j}^{i}q_{\nu }^{j}\ ,\ \ \
\nabla _{\mu }q_{\nu }=\partial _{\mu }q_{\nu }-\Gamma _{\mu \nu }^{\lambda
}q_{\lambda }\ .
\end{equation*}%
Infinitesimally, the transformations (\ref{tensors}) correspond to local
tensor variations $\delta \sigma _{\mu _{1}\cdots \mu _{n}}^{i_{1}\cdots
i_{m}}$ with parameters $\zeta $ and $\xi ^{\mu }$,%
\begin{equation}
\delta \sigma _{\mu _{1}\cdots \mu _{n}}^{i_{1}\cdots i_{m}}=\varepsilon _{%
\hat{p}}^{\left\{ i\right\} }\sigma _{\mu _{1}\cdots \mu _{n}}^{i_{1}\cdots
\hat{p}\cdots i_{m}}\zeta +\sigma _{\mu _{1}\cdots \hat{\nu}\cdots \mu
_{n}}^{i_{1}\cdots i_{m}}\partial _{\left\{ \mu \right\} }\xi ^{\hat{\nu}%
}+(\partial _{\nu }\sigma _{\mu _{1}\cdots \mu _{n}}^{i_{1}\cdots i_{m}})\xi
^{\nu },  \label{tens_var}
\end{equation}%
Given this assumption and the invariance of the vacuum functional in (\ref%
{ZBJL}) under the background transformations (\ref{b2}) combined with the
compensating local transformations (\ref{33}) of the ghost fields, the
extended generating functional $Z(B,J,L)$ in (\ref{ZBJL})\ proves to be
invariant under the initial gauge transformations (\ref{Lor_infinite}), (\ref%
{form_trans}) of the background fields $B=(e_{\mu }^{i},\omega _{\mu })$
combined with the local transformations (\ref{34}) of the sources $J=\left(
J_{i}^{\mu },J^{\mu }\right) $ and some local transformations of the sources$%
\ L_{i_{1}\cdots i_{m}}^{\mu _{1}\cdots \mu _{n}}$, namely,%
\begin{equation}
\delta L_{i_{1}\cdots i_{m}}^{\mu _{1}\cdots \mu _{n}}=-\varepsilon
_{\left\{ i\right\} }^{\hat{p}}L_{i_{1}\cdots \hat{p}\cdots i_{m}}^{\mu
_{1}\cdots \mu _{n}}\zeta -L_{i_{1}\cdots i_{m}}^{\mu _{1}\cdots \hat{\nu}%
\cdots \mu _{n}}\partial _{\hat{\nu}}\xi ^{\left\{ \mu \right\} }+\partial
_{\nu }(L_{i_{1}\cdots i_{m}}^{\mu _{1}\cdots \mu _{n}}\xi ^{\nu })\ ,
\label{delta_L}
\end{equation}%
which differs from the (infinitesimal) tensor transformation law by the
contribution $L_{i_{1}\cdots i_{m}}^{\mu _{1}\cdots \mu _{n}}\partial _{\nu
}\xi ^{\nu }$,%
\begin{equation*}
\delta L_{i_{1}\cdots i_{m}}^{\mu _{1}\cdots \mu _{n}}=\left[ -\varepsilon
_{\left\{ i\right\} }^{\hat{p}}L_{i_{1}\cdots \hat{p}\cdots i_{m}}^{\mu
_{1}\cdots \mu _{n}}\zeta -L_{i_{1}\cdots i_{m}}^{\mu _{1}\cdots \hat{\nu}%
\cdots \mu _{n}}\partial _{\hat{\nu}}\xi ^{\left\{ \mu \right\} }+(\partial
_{\nu }L_{i_{1}\cdots i_{m}}^{\mu _{1}\cdots \mu _{n}})\xi ^{\nu }\right]
+L_{i_{1}\cdots i_{m}}^{\mu _{1}\cdots \mu _{n}}\partial _{\nu }\xi ^{\nu }\
,
\end{equation*}%
and provides for the source term $L_{m}\sigma ^{m}$ in (\ref{ZBJL}) the
corresponding property%
\begin{equation*}
\delta \left( L_{m}\sigma ^{m}\right) =\int d^{2}x\ \partial _{\mu }G^{\mu
}\ ,\ \ \ G^{\mu }\equiv L_{i_{1}\cdots i_{m}}^{\nu _{1}\cdots \nu
_{n}}\sigma _{\nu _{1}\cdots \nu _{n}}^{i_{1}\cdots i_{m}}\xi ^{\mu }\ .
\end{equation*}%
The invariance of $Z(B,J,L)$ and the subsequent invariance of $%
W(B,J,L)=\left( h/i\right) \ln Z(B,J,L)$ can be recast, with the
corresponding variations $\delta B$, $\delta J$, $\delta L$ given by (\ref%
{Lor_infinite}), (\ref{form_trans}), (\ref{34}), (\ref{delta_L}), in the
form, $Y=\left\{ Z,W\right\} $,%
\begin{equation}
\int d^{2}x\ \left[ \delta B\left( x\right) \frac{\delta }{\delta B\left(
x\right) }+\delta J\left( x\right) \frac{\delta }{\delta J\left( x\right) }%
+\delta L_{i_{1}\cdots i_{m}}^{\mu _{1}\cdots \mu _{n}}\left( x\right) \frac{%
\delta }{\delta L_{i_{1}\cdots i_{m}}^{\mu _{1}\cdots \mu _{n}}\left(
x\right) }\right] Y\left( B,J,L\right) =0.  \label{asacons}
\end{equation}%
Let us consider a functional $\Gamma =\Gamma (B,Q,\Sigma )$\ given by the
double Legendre transformation%
\begin{equation}
\Gamma \left( B,Q,\Sigma \right) =W\left( B,J,L\right) -JQ-L_{m}\left[
\sigma ^{m}\left( Q,B\right) +\Sigma ^{m}\right]  \label{Gamma1}
\end{equation}%
in terms of additional fields $\Sigma ^{m}=\Sigma _{\mu _{1}\cdots \mu
_{n}}^{i_{1}\cdots i_{m}}\left( x\right) $,
\begin{equation*}
Q=\frac{\delta W}{\delta J},\ \ \Sigma ^{m}=\frac{\delta W}{\delta L_{m}}%
-\sigma ^{m}\left( \frac{\delta W}{\delta J},B\right) ,\ \ -J=\frac{\delta
\Gamma }{\delta Q}+L_{m}\frac{\delta \sigma ^{m}}{\delta Q},\ \ -L_{m}=\frac{%
\delta \Gamma }{\delta \Sigma ^{m}}\ .
\end{equation*}%
Then, the effective action $\Gamma _{\mathrm{eff}}\left( B,\Sigma \right) $
with composite and background fields,%
\begin{equation}
\Gamma _{\mathrm{eff}}\left( B,\Sigma \right) =\left. \Gamma \left(
B,Q,\Sigma \right) \right\vert _{Q=0}\ ,  \label{eff_comp_back}
\end{equation}%
is invariant, as a consequence of (\ref{asacons}), under\ a set of local
transformations given by the gauge transformations (\ref{Lor_infinite}), (%
\ref{form_trans}) of the background fields $B=(e_{\mu }^{i},\omega _{\mu })$%
\ combined with the infinitesimal local tensor transformations (see Appendix %
\ref{C.II})%
\begin{equation}
\delta \Sigma _{\mu _{1}\cdots \mu _{n}}^{i_{1}\cdots i_{m}}=\varepsilon _{%
\hat{p}}^{\left\{ i\right\} }\Sigma _{\mu _{1}\cdots \mu _{n}}^{i_{1}\cdots
\hat{p}\cdots i_{m}}\zeta +\Sigma _{\mu _{1}\cdots \hat{\nu}\cdots \mu
_{n}}^{i_{1}\cdots i_{m}}\partial _{\left\{ \mu \right\} }\xi ^{\hat{\nu}%
}+(\partial _{\nu }\Sigma _{\mu _{1}\cdots \mu _{n}}^{i_{1}\cdots i_{m}})\xi
^{\nu }  \label{infinit_loc_tens}
\end{equation}%
of the additional fields $\Sigma _{\mu _{1}\cdots \mu _{n}}^{i_{1}\cdots
i_{m}}$, cf. (\ref{tensors}), (\ref{tens_var}).

\section{Summary\label{sec7}}

\setcounter{equation}{0}

The present work has been devoted to quantum non-Abelian gauge models with
composite and background fields. According to the principal research issues
listed in Introduction, the following tasks have been completed:

1. A generating functional (\ref{Z_mod}), (\ref{Z_ext}) of Green's functions
has been introduced for composite and background fields in Yang--Mills
theories. The corresponding symmetry properties have been investigated, as
well as the properties of a generating functional (\ref{Gamma}), (\ref%
{Gamma_eff}) of vertex Green's functions (effective action). These
properties can be expressed in a differential form as the relations (\ref%
{local_Z}), (\ref{Ward_Z}), (\ref{Ward_Gamma_1}), (\ref{ident}), where (\ref%
{local_Z}), (\ref{ident}) reflect the gauge transformations (\ref{set_local}%
), (\ref{back_gauge}), which consist of local $SU(N)$ transformations
accompanied by gauge transformations of a background field $B_{\mu }$,
whereas (\ref{Ward_Z}), (\ref{Ward_Gamma_1}) are related to the BRST
symmetry transformations (\ref{mod_Slavnov}), (\ref{BRST_mod}) with a
modified Slavnov variation $\overleftarrow{s}_{\mathrm{q}}$ depending on $%
B_{\mu }$.

2. On the basis of the above BRST transformations, we have proposed, for the
first time, a set of finite field-dependent BRST (FD BRST) transformations (%
\ref{FFDBRST}), including a background field dependence, and have studied
their properties; see Appendix~\ref{A.I}.

3. Using the finite FD BRST transformations, we have investigated the
related (modified) Ward identities (\ref{smWIclalg}), (\ref{mWIclalg1}), (%
\ref{modWard_W}), (\ref{modWard_Gamma_1}), depending on an FD parameter, as
well as the gauge dependence (\ref{GDInew1}), (\ref{Delta_W}), (\ref%
{Delta_Gamma}) of the above generating functionals with composite and
background fields. It should be noticed that the modified Ward identities
for a constant anticommuting parameter are reduced to the familiar
identities (\ref{Ward_Z}), (\ref{Ward_W}), (\ref{Ward_Gamma_1}) of \cite{6,
LOR}. A gauge variation of the effective action has been found in terms of a
nilpotent operator (\ref{omegaG}) depending on the composite and background
fields, and the conditions (\ref{ext}) of on-shell gauge-independence have
been established. A procedure of loop expansion for the effective action
with composite and background fields has been examined to determine the
representation (\ref{1lp}) for a one-loop effective action.

4. The Gribov--Zwanziger theory \cite{Zwanziger} has been examined, being a
quantum Yang--Mills theory incorporating the presence of a Gribov horizon
\cite{Gribov} in terms of a non-local composite field. The theory \cite%
{Zwanziger} has been extended (\ref{ZHBJnew}) by introducing a background
field $B_{\mu }$ and shown to provide an effective action (\ref{defin}), (%
\ref{GZ_eff}) invariant under the gauge transformations of $B_{\mu }$. The
same result is shown to hold for the effective action (\ref{Gamma_effL}) of
a GZ theory having a local BRST-invariant horizon with background and
additional local composite fields. A quantum action has been suggested,
having a local BRST-invariant horizon (\ref{S_GZ2L}) with background and
composite fields. The corresponding generating functional of Green's
functions extends the scope of the study \cite{1708.01543}, devoted to
renormalizability in the presence of local BRST-invariant quadratic
composite fields (\ref{compbackbrst}), to the case of arbitrary local
composite fields in the background formalism. The problem of gauge
independence has been studied for the effective action (EA) with composite
and background fields, starting from the Gribov--Zwanziger action (\ref%
{S_GZ2L}). It has been shown (\ref{Delta_Gamma1}) that the EA does not
depend on a variation of the gauge condition on the extremals. This makes it
possible to conclude that the Gribov horizon, when defined with a composite
field (added to the Faddeev--Popov quantum action) and without such a field,
leads to different mass-shell conditions. The only representation using the
Gribov--Zwanziger quantum action that is physically relevant is the one with
an on-shell non-vanishing Gribov mass parameter $\gamma $.

5. The model of two-dimensional gravity with dynamical torsion by Volovich
and Katanaev \cite{VK} has been considered, being quantized according to the
background field method in \cite{LM} and featuring a gauge-invariant
effective action, due to (\ref{delta_b_Gamma}). The quantized
two-dimensional gravity \cite{LM} has been generalized to the presence of
composite fields (\ref{ZBJL}), and the corresponding effective action (\ref%
{Gamma1}), (\ref{eff_comp_back}) with composite and background fields has
been found to be gauge-invariant under (\ref{Lor_infinite}), (\ref%
{form_trans}), (\ref{infinit_loc_tens}), in a way similar to the Yang--Mills
case, cf. (\ref{back_gauge}).

Possible applications of the approach developed in the present work can be
the following. The present study of Yang--Mills theories can be employed to
include the QCD gauge theory of strong interactions with the $SU(3)$ gauge
group for the purpose of describing hadron particles (mesons and baryons) as
composite fields. The part related to the two-dimensional gravity with
dynamical torsion can be turned to the advantage of dealing with the
so-called Generalized Lagrange space (for metric fields), so as to exploit
its properties of curvature, torsion and deflection in order to take into
account the asymmetries and anisotropies emerging in physical phenomena
mostly at the cosmological level. The suggested background gauge-invariant
effective action for the Gribov--Zwanziger theory appears to be promising as
a next point in a renormalization analysis of the Gribov--Zwanziger model,
as one accounts for both the non-local and localized BRST-invariant Gribov
horizon in the background formalism, while extending the scope of \cite%
{1708.01543}. Finally, the general approach to Yang--Mills theories with
composite and background fields can be extended to the case of
field-dependent BRST--anti-BRST symmetry along the lines of \cite%
{MR1,MR2,SMMR}.

\section*{Acknowledgments}

{The authors are indebted to the referee for helpful criticism.}
A.A.R. appreciates the partial support granted by the Ministry of Education
of the Russian Federation, project No. FEWF-2020-0003. R.A.C. is grateful
for the hospitality extended to him by the Department of Nuclear Physics,
Institute of Physics, University of S\~{a}o Paulo.

\appendix

\section{Yang--Mills Theory\label{A}}

\setcounter{equation}{0} \renewcommand{\theequation}{A.\arabic{equation}}%
\renewcommand{\theparagraph}{A.\Roman{paragraph}}

\subsection{Background FD BRST transformations}

\label{A.I}

The transformations of the fields $\left( A_{\mu },b,\bar{c},c\right) $
induced by a Slavnov generator $\overleftarrow{s}_{\mathrm{q}}$\ , which
depends on a background field $B^{\mu }$ in (\ref{mod_Slavnov}), and
parameterized by a finite FD Grassmann-odd functional $\lambda (\phi ,B)$ in
(\ref{FFDBRST}) represent finite FD BRST transformations with a background
field in Yang--Mills theories. The Jacobian $\mathrm{Sdet}\left\Vert {\delta
\phi ^{\prime }}/{\delta \phi }\right\Vert $ for a change of variables
induced by the transformations (\ref{FFDBRST}) in the path integral (\ref%
{Z_ext}) can be expressed according to the recipe \cite{Reshetnyak,LL1},%
\begin{equation*}
\mathrm{Sdet}\left\Vert {\delta \phi ^{\prime }}/{\delta \phi }\right\Vert
=\exp \left[ \mathrm{Str}\,\mathrm{ln}\left( \delta _{B}^{A}+\frac{%
\overleftarrow{\delta }(\phi ^{A}\overleftarrow{s}_{\mathrm{q}}\lambda )}{%
\delta \phi {}^{B}}\right) \right] =\exp \left[ -\mathrm{Str}\sum_{n=1}\frac{%
(-1)^{n}}{n}\left( \frac{\delta (\phi ^{A}\overleftarrow{s}_{\mathrm{q}%
}\lambda )}{\delta \phi {}^{B}}\right) ^{n}\right]
\end{equation*}%
\begin{equation}
=\exp \left[ -\sum_{n=1}^{\infty }\frac{(-1)^{n}}{n}\mathrm{Str}\left( \phi
^{A}\overleftarrow{s}_{\mathrm{q}}\lambda \frac{\overleftarrow{\delta }}{%
\delta \phi {}^{B}}\right) ^{n}\right] =\left[ 1+\lambda (\phi ,B)%
\overleftarrow{s}_{\mathrm{q}}\right] ^{-1}.  \label{JacobianB}
\end{equation}%
In calculating the Jacobian, we have used the properties $\frac{%
\overleftarrow{\delta }(\phi ^{A}\overleftarrow{s}_{\mathrm{q}})}{\delta
\phi {}^{B}}(\phi ^{B}\overleftarrow{s}_{\mathrm{q}})=0$ and $\frac{%
\overleftarrow{\delta }(\phi ^{A}\overleftarrow{s}_{\mathrm{q}})}{\delta
\phi {}^{A}}\lambda =0$, due to the antisymmetry of the structure constants $%
f^{pqr}$. For a vanishing background field $B^{\mu }$, the Jacobian (\ref%
{JacobianB}) assumes the usual form \cite{LL1}.

The invariance of the quantum action $S_{\mathrm{ext}}\left( \phi ,\phi
^{\ast },B\right) $ in (\ref{Z_ext}) with respect to the finite FD BRST
transformations (\ref{FFDBRST}) implies that a change $\phi ^{A}\rightarrow
\phi ^{\prime A}=\phi ^{A}[1+\overleftarrow{s}_{\mathrm{q}}\lambda (\phi
,B)] $ induces in (\ref{Z_ext}) a transformation of the integrand $\mathcal{I%
}_{\phi ,\phi ^{\ast },B}^{\Psi }\ $, namely,%
\begin{eqnarray}
\mathcal{I}_{\phi +\phi \overleftarrow{s}_{\mathrm{q}}\lambda ,\phi ^{\ast
},B}^{\Psi } &=&d\phi \ \exp \left( \ln \mathrm{Sdet}\left\Vert {\delta \phi
^{\prime }}/{\delta \phi }\right\Vert \right) \exp \left[ \left( i/\hbar
\right) S_{\mathrm{ext}}\left( \phi +\phi \overleftarrow{s}_{\mathrm{q}%
}\lambda ,\phi ^{\ast },B\right) \right]  \notag \\
&=&d\phi \ \exp \left\{ \left( i/\hbar \right) \left[ S_{\mathrm{ext}}\left(
\phi +\phi \overleftarrow{s}_{\mathrm{q}}\lambda ,\phi ^{\ast },B\right)
-i\hbar \ln \mathrm{Sdet}\left\Vert {\delta \phi ^{\prime }}/{\delta \phi }%
\right\Vert \right] \right\} ,  \label{superJ2}
\end{eqnarray}%
and therefore%
\begin{equation}
\mathcal{I}_{\phi +\phi \overleftarrow{s}_{\mathrm{q}}\lambda ,\phi ^{\ast
},B}^{\Psi }=d\phi \ \exp \left\{ \left( i/\hbar \right) \left[ S_{\mathrm{%
ext}}\left( \phi ,\phi ^{\ast },B\right) +i\hbar \ \mathrm{\ln }\left(
1+\lambda \overleftarrow{s}_{\mathrm{q}}\right) \right] \right\} \ .
\label{superJ3}
\end{equation}%
Due to the explicit form of the initial quantum action $S_{\mathrm{ext}%
}^{\Psi }=S_{0}(A)+\Psi \left( \phi ,B\right) \overleftarrow{s}_{\mathrm{q}%
}+\phi _{A}^{\ast }(\phi ^{A}\overleftarrow{s}_{\mathrm{q}})$ in (\ref{S_FP}%
), (\ref{Z_ext}), the BRST-exact contribution $i\hbar \,\mathrm{\ln }\left(
1+\lambda (\phi ,B)\overleftarrow{s}_{\mathrm{q}}\right) $ to $S_{\mathrm{ext%
}}^{\Psi }$ can then be interpreted as a change of the gauge-fixing Fermion
made in the original integrand $\mathcal{I}_{\phi ,\phi ^{\ast },B}^{\Psi }\
$:%
\begin{align}
& i\hbar \ \mathrm{\ln }\left( 1+\lambda (\phi ,B)\overleftarrow{s}_{\mathrm{%
q}}\right) =\left( \Delta \Psi \right) \overleftarrow{s}_{\mathrm{q}}
\label{superJ3m} \\
& \Longrightarrow \mathcal{I}_{\phi +\phi \overleftarrow{s}_{\mathrm{q}%
}\lambda (\phi ,B),\phi ^{\ast },B}^{\Psi }=d\phi \ \exp \left\{ \left(
i/\hbar \right) \left[ S_{0}+\left( \Psi +\Delta \Psi \right) \overleftarrow{%
s}_{\mathrm{q}}+\phi _{A}^{\ast }(\phi ^{A}\overleftarrow{s}_{\mathrm{q}})%
\right] \right\} =\mathcal{I}_{\phi ,\phi ^{\ast },B}^{\Psi +\Delta \Psi },
\label{superJ41}
\end{align}%
with a certain $\Delta \Psi \left( \phi ,B|\lambda \right) $, whose
correspondence to $\lambda \left( \phi ,B\right) $ is established by the
relation (\ref{superJ3m}), which is a familiar compensation equation for an
unknown parameter $\lambda (\phi ,B)$ now including a certain background $%
B^{\mu }$, and which implies the gauge-independence of the vacuum
functional, $Z_{\Psi }(B,\phi ^{\ast })=Z_{\Psi +\Delta \Psi }(B,\phi ^{\ast
})$, in (\ref{gindvacbcomp}). An explicit solution of (\ref{superJ3m})
satisfying the solvability condition due to the BRST exactness of both sides
(up to BRST exact terms) is given by (\ref{lamDPsi}).

\subsection{Legendre transformation. Differential consequences}

\label{A.II}

The operator $\hat{\omega}_{\Gamma }$ in (\ref{omegaG}) is obtained from $%
\hat{\Omega}$ in (\ref{modWard_W}) with the help of a Legendre
transformation and differential consequences of the usual Ward identities (%
\ref{Ward_Z}), (\ref{Ward_W}) for $Z$, $W$, by using differentiation with
respect to $J_{A}$, $L_{m}$, namely,%
\begin{eqnarray}
&&\hat{\Omega}\left. \phi ^{A}\right\vert _{_{J,L}}=\frac{\overrightarrow{%
\delta }\Gamma }{\delta \phi _{A}^{\ast }}(-1)^{\epsilon _{A}}-\frac{i}{%
\hbar }\left[ \Gamma \frac{\overleftarrow{\delta }}{\delta \Sigma ^{m}}%
\left( \sigma _{,C}^{m}(\hat{\phi},B)\frac{\overrightarrow{\delta }\Gamma }{%
\delta \phi _{C}^{\ast }}\right) ,\phi ^{A}\right\} ,  \label{Omegaphi} \\
&&\hat{\Omega}\left. \Sigma ^{m}\right\vert _{J,L}=\left[ \sigma _{,A}^{m}(%
\hat{\phi},B)-\sigma _{,A}^{m}(\phi ,B)\right] \frac{\overrightarrow{\delta }%
\Gamma }{\delta \phi _{A}^{\ast }}-\frac{i}{\hbar }\left[ \Gamma \frac{%
\overleftarrow{\delta }}{\delta \Sigma ^{m}}\left( \sigma _{,C}^{m}(\hat{\phi%
},B)\frac{\overrightarrow{\delta }\Gamma }{\delta \phi _{C}^{\ast }}\right)
,\Sigma ^{m}\right\}  \notag \\
&&\ \ -\frac{i}{\hbar }\left[ \Gamma \frac{\overleftarrow{\delta }}{\delta
\Sigma ^{n}}\left( \sigma _{,C}^{n}(\hat{\phi},B)\frac{\overrightarrow{%
\delta }\Gamma }{\delta \phi _{C}^{\ast }}\right) ,\sigma ^{m}({\phi }%
,B)\right\}  \notag \\
&&\ \ +\frac{i}{\hbar }(-1)^{\epsilon (\sigma ^{m})+\epsilon (\phi
^{D})}\sigma _{,D}^{m}({\phi },B)\left[ \Gamma \frac{\overleftarrow{\delta }%
}{\delta \Sigma ^{n}}\left( \sigma _{,C}^{n}(\hat{\phi},B)\frac{%
\overrightarrow{\delta }\Gamma }{\delta \phi _{C}^{\ast }}\right) ,\phi
^{D}\right\}  \notag \\
&&\ \ +(-1)^{\epsilon (\sigma ^{m})+\epsilon (\phi ^{D})\epsilon (\phi
^{A})} \left[ \sigma _{,D}^{m}({\phi },B),\Gamma \frac{\overleftarrow{\delta
}}{\delta \Sigma ^{n}}\sigma _{,A}^{n}(\hat{\phi},B)\right\} \left(
G^{\prime \prime -1}\right) ^{A\mathsf{a}}\left( \frac{\overrightarrow{%
\delta }}{\delta \Phi ^{\mathsf{a}}}\frac{\overrightarrow{\delta }\Gamma }{%
\delta \phi _{D}^{\ast }}\right) ,  \label{OmegaSigma}
\end{eqnarray}%
as we calculate the variational derivatives according to the definitions (%
\ref{Gamma}),%
\begin{equation}
\left. \frac{\overrightarrow{\delta }}{\delta \phi _{A}^{\ast }}\right\vert
_{J,L}=\left. \frac{\overrightarrow{\delta }}{\delta \phi _{A}^{\ast }}%
\right\vert _{\phi ,\Sigma }+\left. \frac{\overrightarrow{\delta }\phi ^{B}}{%
\delta \phi _{A}^{\ast }}\frac{\overrightarrow{\delta }}{\delta \phi ^{B}}%
\right\vert _{\phi ^{\ast },\Sigma }+\left. \frac{\overrightarrow{\delta }%
\Sigma ^{m}}{\delta \phi _{A}^{\ast }}\frac{\overrightarrow{\delta }}{\delta
\Sigma ^{m}}\right\vert _{\phi ^{\ast },\phi },  \label{definit}
\end{equation}%
with allowance for (\ref{phi_hat}).

\subsection{Background effective action. Gauge invariance}

\label{A.III}

The invariance of $Z\left( B,J,L,\phi ^{\ast }\right) $ in (\ref{Z_ext}) and
that of the related functional $W\left( B,J,L,\phi ^{\ast }\right) $ under
the set of local transformations (\ref{set_local}), (\ref{local_anti})
translates itself into an extension of (\ref{local_Z}), as we denote $%
Y=\left\{ Z,W\right\} $,%
\begin{eqnarray}  \label{Ybackinv}
&&\ \int d^{D}x\ \left\{ \left[ D_{\mu }^{pq}\left( B\right) \xi ^{q}\right]
\frac{\overrightarrow{\delta }}{\delta B_{\mu }^{p}}+g\xi ^{q}f^{\left\{
p\right\} \hat{r}q}L_{\mu _{1}\cdots \mu _{j}}^{p_{1}\cdots \hat{r}\cdots
p_{i}}\frac{\overrightarrow{\delta }}{\delta L_{\mu _{1}\cdots \mu
_{j}}^{p_{1}\cdots p_{i}}}\right. \\
&&\ +g\xi ^{q}f^{prq}\left( J_{\left( A\right) }^{r|\mu }\frac{%
\overrightarrow{\delta }}{\delta J_{\left( A\right) }^{p|\mu }}+J_{\left(
b\right) }^{r}\frac{\overrightarrow{\delta }}{\delta J_{\left( b\right) }^{p}%
}+J_{\left( \bar{c}\right) }^{r}\frac{\overrightarrow{\delta }}{\delta
J_{\left( \bar{c}\right) }^{p}}+J_{\left( c\right) }^{r}\frac{%
\overrightarrow{\delta }}{\delta J_{\left( c\right) }^{p}}\right)  \notag \\
&&\ +\left. g\xi ^{q}f^{prq}\left( A_{\mu }^{\ast r}\frac{\overrightarrow{%
\delta }}{\delta A_{\mu }^{\ast p}}+b^{\ast r}\frac{\overrightarrow{\delta }%
}{\delta b^{\ast p}}+\bar{c}^{\ast r}\frac{\overrightarrow{\delta }}{\delta
\bar{c}^{\ast p}}+c^{\ast r}\frac{\overrightarrow{\delta }}{\delta c^{\ast p}%
}\right) \right\} Y\left( B,J,L,\phi ^{\ast }\right) =0.  \notag
\end{eqnarray}%
The functional $\Gamma \left( B,\phi ,\Sigma ,\phi ^{\ast }\right) $ in (\ref%
{Gamma}) satisfies the subsequent identity%
\begin{eqnarray}
&&\ \int d^{D}x\ \left\{ \left[ D_{\mu }^{pq}\left( B\right) \xi ^{q}\right]
\frac{\overrightarrow{\delta }}{\delta B_{\mu }^{p}}-g\xi ^{q}f^{\left\{
p\right\} \hat{r}q}\left[ \Sigma _{\mu _{1}\cdots \mu _{l}}^{p_{1}\cdots
p_{k}}+\sigma _{\mu _{1}\cdots \mu _{l}}^{p_{1}\cdots p_{k}}\left( \phi
,B\right) \right] \frac{\overrightarrow{\delta }}{\delta \Sigma _{\mu
_{1}\cdots \mu _{l}}^{p_{1}\cdots \hat{r}\cdots p_{k}}}\right.
\label{Gbackinv} \\
&&\ -g\xi ^{q}f^{prq}\left( A^{p|\mu }\frac{\overrightarrow{\delta }}{\delta
A^{r|\mu }}+b^{p}\frac{\overrightarrow{\delta }}{\delta b^{r}}+\bar{c}^{p}%
\frac{\overrightarrow{\delta }}{\delta \bar{c}^{r}}+c^{p}\frac{%
\overrightarrow{\delta }}{\delta c^{r}}\right)  \notag \\
&&\ +g\xi ^{q}f^{prq}\left( A_{\mu }^{\ast r}\frac{\overrightarrow{\delta }}{%
\delta A_{\mu }^{\ast p}}+b^{\ast r}\frac{\overrightarrow{\delta }}{\delta
b^{\ast p}}+\bar{c}^{\ast r}\frac{\overrightarrow{\delta }}{\delta \bar{c}%
^{\ast p}}+c^{\ast r}\frac{\overrightarrow{\delta }}{\delta c^{\ast p}}%
\right)  \notag \\
&&\ +\left. g\xi ^{q}f^{prq}\left( A^{p|\mu }\frac{\overrightarrow{\delta }}{%
\delta A^{r|\mu }}+b^{p}\frac{\overrightarrow{\delta }}{\delta b^{r}}+\bar{c}%
^{p}\frac{\overrightarrow{\delta }}{\delta \bar{c}^{r}}+c^{p}\frac{%
\overrightarrow{\delta }}{\delta c^{r}}\right) \sigma ^{m}\left( \phi
,B\right) \frac{\overrightarrow{\delta }}{\delta \Sigma ^{m}}\right\} \Gamma
\left( B,\phi ,\Sigma ,\phi ^{\ast }\right) =0,  \notag
\end{eqnarray}%
where the terms containing the derivatives of $\sigma ^{m}=\sigma ^{m}\left(
\phi ,B\right) $ over $\phi ^{A}=\left( A^{r|\mu },b^{r},\bar{c}%
^{r},c^{r}\right) \left( x\right) $ are understood in the form%
\begin{equation*}
\frac{\overrightarrow{\delta }\sigma ^{m}}{\delta \phi ^{A}}\frac{%
\overrightarrow{\delta }}{\delta \Sigma ^{m}}=\int d^{D}y\ \frac{%
\overrightarrow{\delta }}{\delta \phi ^{A}}\sigma _{\mu _{1}\cdots \mu
_{l}}^{p_{1}\cdots p_{k}}\left( y\right) \frac{\overrightarrow{\delta }}{%
\delta \Sigma _{\mu _{1}\cdots \mu _{l}}^{p_{1}\cdots p_{k}}\left( y\right) }%
.
\end{equation*}%
Then the functional $\Gamma _{\mathrm{eff}}\left( B,\Sigma \right) $ defined
by (\ref{Gamma_eff}) satisfies the relation (\ref{ident}) as a consequence
of the equality%
\begin{equation}
f^{\left\{ p\right\} \hat{r}q}\Sigma _{\mu _{1}\cdots \mu _{l}}^{p_{1}\cdots
p_{k}}\frac{\overrightarrow{\delta }}{\delta \Sigma _{\mu _{1}\cdots \mu
_{l}}^{p_{1}\cdots \hat{r}\cdots p_{k}}}=-f^{\left\{ p\right\} \hat{r}%
q}\Sigma _{\mu _{1}\cdots \mu _{l}}^{p_{1}\cdots \hat{r}\cdots p_{k}}\frac{%
\overrightarrow{\delta }}{\delta \Sigma _{\mu _{1}\cdots \mu
_{l}}^{p_{1}\cdots p_{k}}},  \label{cf1}
\end{equation}%
implied by the notation $f^{\left\{ p\right\} \hat{r}q}$ in (\ref{shorthand}%
) and by the antisymmetry of the structure constants:%
\begin{equation}
f^{p_{s}r_{s}q}\Sigma _{\mu _{1}\cdots \mu _{l}}^{p_{1}\cdots p_{s}\cdots
p_{k}}\frac{\overrightarrow{\delta }}{\delta \Sigma _{\mu _{1}\cdots \mu
_{l}}^{p_{1}\cdots r_{s}\cdots p_{k}}}=-f^{r_{s}p_{s}q}\Sigma _{\mu
_{1}\cdots \mu _{l}}^{p_{1}\cdots p_{s}\cdots p_{k}}\frac{\overrightarrow{%
\delta }}{\delta \Sigma _{\mu _{1}\cdots \mu _{l}}^{p_{1}\cdots r_{s}\cdots
p_{k}}}=-f^{p_{s}r_{s}q}\Sigma _{\mu _{1}\cdots \mu _{l}}^{p_{1}\cdots
r_{s}\cdots p_{k}}\frac{\overrightarrow{\delta }}{\delta \Sigma _{\mu
_{1}\cdots \mu _{l}}^{p_{1}\cdots p_{s}\cdots p_{k}}}.  \label{cf2}
\end{equation}

\section{Gribov--Zwanziger Theory\label{B}}

\setcounter{equation}{0} \renewcommand{\theequation}{B.\arabic{equation}}%
\renewcommand{\theparagraph}{B.\Roman{paragraph}}

\subsection{Background-dependent Faddeev--Popov operator}

\label{B.I}

From the expression (\ref{KBDcallig}) for $(\tilde{K})_{B}^{pq}\left(
x;y\right) $ written in the matrix form%
\begin{equation}
\tilde{K}_{B}\left( x;y\right) =D_{\mu }\left( B\right) D^{\mu }\left(
A+B\right) \delta \left( x-y\right)  \label{KBreprs}
\end{equation}%
it follows that%
\begin{equation*}
\tilde{K}_{B}\left( x;y\right) =\left[ \partial ^{2}+g\left( \partial _{\mu
}A^{\mu }\right) +g\left( \partial _{\mu }B^{\mu }\right) +g\left( A^{\mu
}+2B^{\mu }\right) \partial _{\mu }+g^{2}B_{\mu }\left( A_{\mu }+B_{\mu
}\right) \right] \delta \left( x-y\right) .
\end{equation*}%
Subtracting from this matrix the expression for $K_{B}\left( x;y\right) $
given by (\ref{K_B1}), we find%
\begin{equation}
\tilde{K}_{B}\left( x;y\right) -K_{B}\left( x;y\right) =g\left\{ \left(
\partial _{\mu }A^{\mu }\right) +g\left[ B_{\mu },A^{\mu }\right] \right\}
\delta \left( x-y\right) =g\left[ D_{\mu }\left( B\right) ,A^{\mu }\right]
\delta \left( x-y\right) .  \label{diffKreprs}
\end{equation}%
where, in virtue of the Jacobi identity for the structure constants $f^{pqr}$%
,%
\begin{equation*}
\left[ D_{\mu }\left( B\right) ,A^{\mu }\right] ^{pq}=\left( \partial _{\mu
}A^{\mu }+g\left[ B_{\mu },A^{\mu }\right] \right) ^{pq}=f^{prq}D_{\mu
}^{rs}\left( B\right) A^{s|\mu }.
\end{equation*}%
which proves that the matrix $\tilde{K}_{B}\left( x;y\right) $ defined by (%
\ref{KBDcallig}) does indeed reduce to $K_{B}\left( x;y\right) $ in (\ref%
{K_B1}) under the background gauge condition $D_{\mu }^{pq}\left( B\right)
A^{q|\mu }=0$, and also that the two matrices are related by (\ref{K_B}).

\subsection{Gribov--Zwanziger local action. Alternative representation}

\label{B.II}

Consider the integral expression
\begin{equation}
\left\langle F,G\right\rangle \equiv \int d^{D}x\ d^{D}y\ \mathrm{Tr\ }%
F\left( x\right) \mathcal{K}\left( x,y\right) G\left( y\right) ,
\label{int_exp}
\end{equation}%
constructed using the matrix $\mathcal{K}\left( x,y\right) $ in (\ref{sym}),%
\begin{equation*}
\mathcal{K}\left( x,y\right) =\frac{1}{2}\left[ K\left( x,y\right) +\tilde{K}%
\left( x,y\right) \right] \delta \left( x-y\right) ,
\end{equation*}%
where the matrix elements of $K\left( x,y\right) $ and $\tilde{K}\left(
x,y\right) $ are given by (\ref{K}), (\ref{K_tilde}), which implies%
\begin{equation}
\mathcal{K}\left( x,y\right) =\frac{1}{2}\left[ D^{\nu }\left( A\right)
\partial _{\nu }+\partial _{\nu }D^{\nu }\left( A\right) \right] \delta
\left( x-y\right) .  \label{Kreprs}
\end{equation}%
The expression (\ref{int_exp}) then transforms into
\begin{equation}
\left\langle F,G\right\rangle =\frac{1}{2}\int d^{D}x\ \mathrm{Tr\ }F\left[
D^{\nu }\left( A\right) \partial _{\nu }+\partial _{\nu }D^{\nu }\left(
A\right) \right] G,  \label{FGreprs}
\end{equation}%
and integration by parts results in%
\begin{equation}
\left\langle F,G\right\rangle =\frac{1}{2}\int d^{D}x\ \mathrm{Tr}\left[
-\left( \partial ^{\nu }F\right) \partial _{\nu }G-\left( \partial _{\nu
}F\right) \partial ^{\nu }G+gFA^{\nu }\partial _{\nu }G-g\left( \partial
_{\nu }F\right) A^{\nu }G\right] .  \label{exp}
\end{equation}%
As we rewrite $FA^{\nu }=-\left[ A^{\nu },F\right] +A^{\nu }F$,$\ A^{\nu }G=%
\left[ A^{\nu },G\right] -GA^{\nu }$ and use a permutation under the sign of
$\mathrm{Tr}$, the expression (\ref{exp}) transforms into%
\begin{equation*}
\left\langle F,G\right\rangle =-\frac{1}{2}\int d^{D}x\ \mathrm{Tr}\left\{ %
\left[ D^{\nu }\left( A\right) ,F\right] \partial _{\nu }G+\left( \partial
_{\nu }F\right) \left[ D^{\nu }\left( A\right) ,G\right] -gA^{\nu }\left[
F\partial _{\nu }G+\left( \partial _{\nu }F\right) G\right] \right\} .
\end{equation*}%
Applying this result to the settings%
\begin{equation*}
\left( F,G\right) =\left( \bar{\varphi}^{\mu },\varphi _{\mu }^{\mathrm{T}%
}\right) ,\ \ \ \left( F,G\right) =\left( \bar{\omega}^{\mu },\omega _{\mu
}^{\mathrm{T}}\right)
\end{equation*}%
made in the path integral (\ref{ZBJ_horizon2}), (\ref{S_GZ}) restricted to
the case $B_{\mu }=0$, where $K\left( x,y\right) \rightarrow \mathcal{K}%
\left( x,y\right) $ due to the Landau gauge condition $\partial _{\nu
}A^{\nu }=0$, which implies (after integrating by parts)%
\begin{equation*}
\int d^{D}x\ \mathrm{Tr\ }A^{\nu }\left[ F\partial _{\nu }G+\left( \partial
_{\nu }F\right) G\right] =0,
\end{equation*}%
we find that the action $S_{\mathrm{GZ}}\left( \Phi \right) $ in (\ref%
{represented}) is indeed equivalent to $\mathcal{S}_{\mathrm{GZ}}\left( \Phi
\right) $ given by (\ref{S_1}).

\subsection{Background gauge invariance of $W$, $\Gamma$}

\label{B.III}

The invariance of the functional $Z_{\mathcal{H}}=Z_{\mathcal{H}}\left(
B,J\right) $ in (\ref{ZHBJnew}) under the local transformations (\ref{ZHBinv}%
) in terms of the related functional $W_{\mathcal{H}}=W_{\mathcal{H}}\left(
B,J\right) $ reads%
\begin{equation}
\int d^{D}x\ \left\{ \left[ D_{\mu }^{pq}\left( B\right) \xi ^{q}\right]
\frac{\overrightarrow{\delta }}{\delta B_{\mu }^{p}}+g\xi ^{q}f^{prq}\left(
J_{\left( A\right) }^{r|\mu }\frac{\overrightarrow{\delta }}{\delta
J_{\left( A\right) }^{p|\mu }}+J_{\left( b\right) }^{r}\frac{\overrightarrow{%
\delta }}{\delta J_{\left( b\right) }^{p}}+J_{\left( \bar{c}\right) }^{r}%
\frac{\overrightarrow{\delta }}{\delta J_{\left( \bar{c}\right) }^{p}}%
+J_{\left( c\right) }^{r}\frac{\overrightarrow{\delta }}{\delta J_{\left(
c\right) }^{p}}\right) \right\} W_{\mathcal{H}}=0.  \label{WbackGZ}
\end{equation}%
and translates itself for the functional $\Gamma _{\mathcal{H}}=\Gamma _{%
\mathcal{H}}\left( B,\phi \right) $ in (\ref{defin}) as follows:%
\begin{equation}
\int d^{D}x\ \left\{ \left[ D_{\mu }^{pq}\left( B\right) \xi ^{q}\right]
\frac{\overrightarrow{\delta }}{\delta B_{\mu }^{p}}+g\xi ^{q}f^{prq}\left(
A_{\mu }^{r}\frac{\overrightarrow{\delta }}{\delta A_{\mu }^{p}}+b^{r}\frac{%
\overrightarrow{\delta }}{\delta b^{p}}+\bar{c}^{r}\frac{\overrightarrow{%
\delta }}{\delta \bar{c}^{p}}+c^{r}\frac{\overrightarrow{\delta }}{\delta
c^{p}}\right) \right\} \Gamma _{\mathcal{H}}=0,  \label{GbackGZ}
\end{equation}%
which implies the invariance of $\Gamma _{\mathcal{H}}\left( B,\phi \right) $
under the local transformations (\ref{local_tr}).

\section{Volovich--Katanaev Model\label{C}}

\setcounter{equation}{0} \renewcommand{\theequation}{C.\arabic{equation}}%
\renewcommand{\theparagraph}{C.\Roman{paragraph}}

\subsection{Background effective action. Gauge invariance}

\label{C.I}

The invariance of $W=W\left( B,J\right) $ under (\ref{Lor_infinite}), (\ref%
{form_trans}), (\ref{34}) implies%
\begin{eqnarray}
&&\int d^{2}x\left\{ \left[ \varepsilon ^{ij}e_{\mu j}\zeta +e_{\nu
}^{i}\partial _{\mu }\xi ^{\nu }+(\partial _{\nu }e_{\mu }^{i})\xi ^{\nu }%
\right] \frac{\delta W}{\delta e_{\mu }^{i}}+\left[ -\partial _{\mu }\zeta
+\omega _{\nu }\partial _{\mu }\xi ^{\nu }+(\partial _{\nu }\omega _{\mu
})\xi ^{\nu }\right] \frac{\delta W}{\delta \omega _{\mu }}\right.  \notag \\
&&-\left. \left[ \varepsilon _{i}^{k}J_{k}^{\mu }\zeta +J_{i}^{\nu }\partial
_{\nu }\xi ^{\mu }-\partial _{\nu }(J_{i}^{\mu }\xi ^{\nu })\right] \frac{%
\delta W}{\delta J_{i}^{\mu }}-\left[ J^{\nu }\partial _{\nu }\xi ^{\mu
}-\partial _{\nu }(J^{\mu }\xi ^{\nu })\right] \frac{\delta W}{\delta J^{\mu
}}\right\} =0  \label{WbackVK}
\end{eqnarray}%
and transforms into a relation for $\Gamma =\Gamma \left( B,Q\right) $
defined by (\ref{35}),%
\begin{eqnarray}
&&\int d^{2}x\left\{ \left[ \varepsilon ^{ij}e_{\mu j}\zeta +e_{\nu
}^{i}(\partial _{\mu }\xi ^{\nu })+(\partial _{\nu }e_{\mu }^{i})\xi ^{\nu }%
\right] \frac{\delta \Gamma }{\delta e_{\mu }^{i}}+\left[ -\partial _{\mu
}\zeta +\omega _{\nu }(\partial _{\mu }\xi ^{\nu })+(\partial _{\nu }\omega
_{\mu })\xi ^{\nu }\right] \frac{\delta \Gamma }{\delta \omega _{\mu }}%
\right.  \notag \\
&&+\left. q_{\mu }^{i}\left[ \varepsilon _{i}^{k}\frac{\delta \Gamma }{%
\delta q_{\mu }^{k}}\zeta +\frac{\delta \Gamma }{\delta q_{\nu }^{i}}%
\partial _{\nu }\xi ^{\mu }-\partial _{\nu }\left( \frac{\delta \Gamma }{%
\delta q_{\mu }^{i}}\xi ^{\nu }\right) \right] +q^{\mu }\left[ \frac{\delta
\Gamma }{\delta q_{\nu }}\partial _{\nu }\xi ^{\mu }-\partial _{\nu }\left(
\frac{\delta \Gamma }{\delta q_{\mu }}\xi ^{\nu }\right) \right] \right\} =0,
\label{GbackVK}
\end{eqnarray}%
which is integrated by parts to result in (\ref{delta_b_Gamma}).

\subsection{Background gauge invariance of $W,\Gamma $ with composite fields}

\label{C.II}

Let us introduce the notation, $p_{k}\in \left\{ p_{1},\ldots ,p_{m}\right\}
$, $\nu _{k}\in \left\{ \nu _{1},\ldots ,\nu _{n}\right\} $,

\begin{eqnarray}
F_{\left\{ i\right\} }^{\hat{p}}T_{i_{1}\cdots \hat{p}\cdots i_{m}}^{\mu
_{1}\cdots \mu _{n}} &=&\sum_{p_{k}}F_{i_{k}}^{p_{k}}T_{i_{1}\cdots
p_{k}\cdots i_{m}}^{\mu _{1}\cdots \mu _{n}}\ ,\ \ \ G_{\hat{\nu}}^{\left\{
\mu \right\} }T_{i_{1}\cdots i_{m}}^{\mu _{1}\cdots \hat{\nu}\cdots \mu
_{n}}=\sum_{\nu _{k}}G_{\nu _{k}}^{\mu _{k}}T_{i_{1}\cdots i_{m}}^{\mu
_{1}\cdots \nu _{k}\cdots \mu _{n}}\ ,  \notag \\
F_{\hat{p}}^{\left\{ i\right\} }T_{\mu _{1}\cdots \mu _{n}}^{i_{1}\cdots
\hat{p}\cdots i_{m}} &=&\sum_{p_{k}}F_{p_{k}}^{i_{k}}T_{\mu _{1}\cdots \mu
_{n}}^{i_{1}\cdots p_{k}\cdots i_{m}}\ ,\ \ \ G_{\left\{ \mu \right\} }^{%
\hat{\nu}}T_{\mu _{1}\cdots \hat{\nu}\cdots \mu _{n}}^{i_{1}\cdots
i_{m}}=\sum_{\nu _{k}}G_{\mu _{k}}^{\nu _{k}}T_{\mu _{1}\cdots \nu
_{k}\cdots \mu _{n}}^{i_{1}\cdots i_{m}}\ .  \label{as_we_denote}
\end{eqnarray}%
Then the invariance of $W=W\left( B,J,L\right) $ under (\ref{Lor_infinite}),
(\ref{form_trans}), (\ref{34}), (\ref{delta_L}) implies%
\begin{eqnarray}
&&\int d^{2}x\left\{ \left[ \varepsilon ^{ij}e_{\mu j}\zeta +e_{\nu
}^{i}(\partial _{\mu }\xi ^{\nu })+(\partial _{\nu }e_{\mu }^{i})\xi ^{\nu }%
\right] \frac{\delta W}{\delta e_{\mu }^{i}}+\left[ -\partial _{\mu }\zeta
+\omega _{\nu }(\partial _{\mu }\xi ^{\nu })+(\partial _{\nu }\omega _{\mu
})\xi ^{\nu }\right] \frac{\delta W}{\delta \omega _{\mu }}\right.  \notag \\
&&-\left[ \varepsilon _{i}^{k}J_{k}^{\mu }\zeta +J_{i}^{\nu }(\partial _{\nu
}\xi ^{\mu })-\partial _{\nu }(J_{i}^{\mu }\xi ^{\nu })\right] \frac{\delta W%
}{\delta J_{i}^{\mu }}-\left[ J^{\nu }(\partial _{\nu }\xi ^{\mu })-\partial
_{\nu }(J^{\mu }\xi ^{\nu })\right] \frac{\delta W}{\delta J^{\mu }}  \notag
\\
&&\ -\left. [\varepsilon _{\left\{ i\right\} }^{\hat{p}}L_{i_{1}\cdots \hat{p%
}\cdots i_{m}}^{\mu _{1}\cdots \mu _{n}}\zeta +L_{i_{1}\cdots i_{m}}^{\mu
_{1}\cdots \hat{\nu}\cdots \mu _{n}}(\partial _{\hat{\nu}}\xi ^{\left\{ \mu
\right\} })-\partial _{\nu }(L_{i_{1}\cdots i_{m}}^{\mu _{1}\cdots \mu
_{n}}\xi ^{\nu })]\frac{\delta W}{\delta L_{i_{1}\cdots i_{m}}^{\mu
_{1}\cdots \mu _{n}}}\right\} =0  \label{invar_prop}
\end{eqnarray}%
and transforms into a relation for $\Gamma =\Gamma (B,Q,\Sigma )$ defined by
(\ref{Gamma1}),
\begin{eqnarray}
&&\int d^{2}x\left\{ \left[ \varepsilon ^{ij}e_{\mu j}\zeta +e_{\nu
}^{i}(\partial _{\mu }\xi ^{\nu })+(\partial _{\nu }e_{\mu }^{i})\xi ^{\nu }%
\right] \frac{\delta \Gamma }{\delta e_{\mu }^{i}}+\left[ -\partial _{\mu
}\zeta +\omega _{\nu }(\partial _{\mu }\xi ^{\nu })+(\partial _{\nu }\omega
_{\mu })\xi ^{\nu }\right] \frac{\delta \Gamma }{\delta \omega _{\mu }}%
\right.  \notag \\
&&\ +\left( \Sigma _{\mu _{1}\cdots \mu _{n}}^{i_{1}\cdots i_{m}}+\sigma
_{\mu _{1}\cdots \mu _{n}}^{i_{1}\cdots i_{m}}\right) \left[ \varepsilon
_{\left\{ i\right\} }^{\hat{p}}\frac{\delta \Gamma }{\delta \Sigma _{\mu
_{1}\cdots \mu _{n}}^{i_{1}\cdots \hat{p}\cdots i_{m}}}\zeta +\frac{\delta
\Gamma }{\delta \Sigma _{\mu _{1}\cdots \hat{\nu}\cdots \mu
_{n}}^{i_{1}\cdots i_{m}}}\partial _{\hat{\nu}}\xi ^{\left\{ \mu \right\}
}-\partial _{\nu }\left( \frac{\delta \Gamma }{\delta \Sigma _{\mu
_{1}\cdots \mu _{n}}^{i_{1}\cdots i_{m}}}\xi ^{\nu }\right) \right]  \notag
\\
&&+\ q_{i}^{\mu }\left[ \varepsilon _{i}^{k}\left( \frac{\delta \Gamma }{%
\delta q_{\mu }^{k}}-\frac{\delta \Gamma }{\delta \Sigma ^{m}}\frac{\delta
\sigma ^{m}}{\delta q_{\mu }^{k}}\right) \zeta +\left( \frac{\delta \Gamma }{%
\delta q_{\nu }^{i}}-\frac{\delta \Gamma }{\delta \Sigma ^{m}}\frac{\delta
\sigma ^{m}}{\delta q_{\nu }^{i}}\right) \partial _{\nu }\xi ^{\mu
}-\partial _{\nu }\left( \frac{\delta \Gamma }{\delta q_{\mu }^{i}}\xi ^{\nu
}-\frac{\delta \Gamma }{\delta \Sigma ^{m}}\frac{\delta \sigma ^{m}}{\delta
q_{\mu }^{i}}\xi ^{\nu }\right) \right]  \notag \\
&&+\ \left. q^{\mu }\left[ \left( \frac{\delta \Gamma }{\delta q_{\nu }}-%
\frac{\delta \Gamma }{\delta \Sigma ^{m}}\frac{\delta \sigma ^{m}}{\delta
q_{\nu }}\right) \partial _{\nu }\xi ^{\mu }-\partial _{\nu }\left( \frac{%
\delta \Gamma }{\delta q_{\mu }}\xi ^{\nu }-\frac{\delta \Gamma }{\delta
\Sigma ^{m}}\frac{\delta \sigma ^{m}}{\delta q_{\mu }}\xi ^{\nu }\right) %
\right] \right\} =0,  \label{big_one}
\end{eqnarray}%
where the terms containing the derivatives $\frac{\delta \sigma ^{m}}{\delta
q_{\mu }^{i}}$, $\frac{\delta \sigma ^{m}}{\delta q_{\mu }}$ of the
composite fields $\sigma ^{m}\left( Q,B\right) $ are understood in the form%
\begin{equation*}
\frac{\delta \Gamma }{\delta \Sigma ^{m}}\frac{\delta \sigma ^{m}}{\delta Q}%
=\int d^{2}y\ \frac{\delta \Gamma }{\delta \Sigma _{\mu _{1}\cdots \mu
_{n}}^{i_{1}\cdots i_{m}}\left( y\right) }\frac{\delta \sigma _{\mu
_{1}\cdots \mu _{n}}^{i_{1}\cdots i_{m}}\left( y\right) }{\delta Q\left(
x\right) }\ ,\ \ \ Q\left( x\right) =(q_{\mu }^{i},q_{\mu })\left( x\right) .
\end{equation*}%
The functional $\Gamma _{\mathrm{eff}}=\Gamma _{\mathrm{eff}}\left( B,\Sigma
\right) $ defined by (\ref{eff_comp_back}) satisfies the identity%
\begin{eqnarray}
\hspace{-0.5em}&\hspace{-0.5em}&\hspace{-0.5em}\int d^{2}x\left\{ \left[
\varepsilon ^{ij}e_{\mu j}\zeta +e_{\nu }^{i}(\partial _{\mu }\xi ^{\nu
})+(\partial _{\nu }e_{\mu }^{i})\xi ^{\nu }\right] \frac{\delta }{\delta
e_{\mu }^{i}}+\left[ -\partial _{\mu }\zeta +\omega _{\nu }(\partial _{\mu
}\xi ^{\nu })+(\partial _{\nu }\omega _{\mu })\xi ^{\nu }\right] \frac{%
\delta }{\delta \omega _{\mu }}\right.  \notag \\
\hspace{-0.5em}&\hspace{-0.5em}&\hspace{-0.5em} +\left. \left[ \varepsilon
_{\left\{ i\right\} }^{\hat{p}}\Sigma _{\mu _{1}\cdots \mu
_{n}}^{i_{1}\cdots i_{m}}\zeta \frac{\delta }{\delta \Sigma _{\mu _{1}\cdots
\mu _{n}}^{i_{1}\cdots \hat{p}\cdots i_{m}}}+\Sigma _{\mu _{1}\cdots \mu
_{n}}^{i_{1}\cdots i_{m}}(\partial _{\hat{\nu}}\xi ^{\left\{ \mu \right\} })%
\frac{\delta }{\delta \Sigma _{\mu _{1}\cdots \hat{\nu}\cdots \mu
_{n}}^{i_{1}\cdots i_{m}}}+(\partial _{\nu }\Sigma _{\mu _{1}\cdots \mu
_{n}}^{i_{1}\cdots i_{m}})\xi ^{\nu }\frac{\delta }{\delta \Sigma _{\mu
_{1}\cdots \mu _{n}}^{i_{1}\cdots i_{m}}}\right] \right\} \Gamma _{\mathrm{%
eff}}=0,  \label{GcompbackVK}
\end{eqnarray}%
obtained by setting $\sigma ^{m}\left( 0,B\right) =0$ and integrating by
parts in (\ref{big_one}). Using the latter property and the following
consequences, cf. (\ref{cf1}), (\ref{cf2}), of the notation (\ref%
{as_we_denote}),%
\begin{eqnarray*}
\varepsilon _{\left\{ i\right\} }^{\hat{p}}\Sigma _{\mu _{1}\cdots \mu
_{n}}^{i_{1}\cdots i_{m}}\frac{\delta }{\delta \Sigma _{\mu _{1}\cdots \mu
_{n}}^{i_{1}\cdots \hat{p}\cdots i_{m}}} &=&\varepsilon _{\hat{p}}^{\left\{
i\right\} }\Sigma _{\mu _{1}\cdots \mu _{n}}^{i_{1}\cdots \hat{p}\cdots
i_{m}}\frac{\delta }{\delta \Sigma _{\mu _{1}\cdots \mu _{n}}^{i_{1}\cdots
i_{m}}}\ , \\
\ \Sigma _{\mu _{1}\cdots \mu _{n}}^{i_{1}\cdots i_{m}}(\partial _{\hat{\nu}%
}\xi ^{\left\{ \mu \right\} })\frac{\delta }{\delta \Sigma _{\mu _{1}\cdots
\hat{\nu}\cdots \mu _{n}}^{i_{1}\cdots i_{m}}} &=&\Sigma _{\mu _{1}\cdots
\hat{\nu}\cdots \mu _{n}}^{i_{1}\cdots i_{m}}(\partial _{\left\{ \mu
\right\} }\xi ^{\hat{\nu}})\frac{\delta }{\delta \Sigma _{\mu _{1}\cdots \mu
_{n}}^{i_{1}\cdots i_{m}}}\ ,
\end{eqnarray*}%
by virtue of $\varepsilon _{k}^{j}=\varepsilon _{j}^{k}$ ($\varepsilon
_{1}^{0}=\varepsilon _{0}^{1}=-1$ and $\varepsilon _{0}^{0}=\varepsilon
_{1}^{1}=0$ due to $\varepsilon _{k}^{j}=\varepsilon ^{ji}\eta _{ik}$), one
arrives at the invariance of $\Gamma _{\mathrm{eff}}\left( B,\Sigma \right) $
under (\ref{Lor_infinite}), (\ref{form_trans}) and (\ref{infinit_loc_tens}).


\begin{thebibliography}{99}
\bibitem{DeW} B.S. DeWitt, Quantum theory of gravity. II. The manifestly
covariant theory, Phys. Rev. \textbf{162}, 1195--1239 (1967).

\bibitem{AFS} I.Ya. Aref'eva, L.D. Faddeev, A.A. Slavnov, Generating
functional for the S matrix in gauge theories, Theor. Math. Phys. \textbf{21}%
, 1165--1172 (1975); in Russian: Teor. Mat. Fiz. \textbf{21}, 311--321
(1974).

\bibitem{Abbott} L.F. Abbott, The background field method beyond one loop,
Nucl. Phys. B \textbf{185}, 189--203 (1981).

\bibitem{17} J.M. Cornwall, R. Jackiw, E. Tomboulis, Effective action for
composite operators, Phys. Rev. D \textbf{10}, 2428--2445 (1974).

\bibitem{18} R.W. Haymaker, Variational methods for composite operators,
Rivista Nuovo Cim. \textbf{14}, 1--89 (1991).

\bibitem{Zwanziger} D. Zwanziger, Action from the Gribov horizon, Nucl.
Phys. B \textbf{321}, 591--604 (1989); Local and renormalizable action from
the Gribov horizon, Nucl. Phys. B \textbf{323}, 513--544 (1989).

\bibitem{Gribov} V.N. Gribov, Quantization of nonabelian gauge theories,
Nucl. Phys. B \textbf{139}, 1--19 (1978).

\bibitem{Abbott:1981ke} L.F. Abbott, Introduction to the background field
method, Acta Phys. Polon. B \textbf{13}, 33--50 (1982).

\bibitem{Weinberg} S. Weinberg, \emph{The Quantum Theory of Fields. Vol. II
Modern Applications} (Cambridge University Press, 1996).

\bibitem{'tH} G. 't Hooft, An algorithm for the poles at dimension four in
the dimensional regularization procedure, Nucl. Phys. B \textbf{62},
444--460 (1973).

\bibitem{K-SZ} H. Kluberg-Stern, J.B. Zuber, Renormalization of non-Abelian
gauge theories in a background-field gauge. I. Green's Functions, Phys. Rev.
D \textbf{12}, 482--488 (1975).

\bibitem{GvanNW} M.T. Grisaru, P. van Nieuwenhuizen, C.C. Wu,
Background-field method versus normal field theory in explicit examples: one
loop divergences in S matrix and Green's functions for Yang--Mills and
gravitational fields, Phys. Rev. D \textbf{12}, 3203--3213 (1975).

\bibitem{CMacL} D.M. Capper, A. MacLean, The background field method at two
loops: a general gauge Yang--Mills calculation, Nucl. Phys. B \textbf{203},
413--422 (1982).

\bibitem{IO} S. Ichinose, M. Omote, Renormalization using the
background-field method, Nucl. Phys. B \textbf{203}, 221--267 (1982).

\bibitem{GS} M.H. Goroff, A. Sagnotti, The ultraviolet behavior of Einstein
gravity, Nucl. Phys. B \textbf{266}, 709--736 (1986).

\bibitem{Ven} A.E.M. van de Ven, Two-loop quantum gravity. Nucl. Phys. B
\textbf{378}, 309--366 (1992).

\bibitem{Gr} P.A. Grassi, Algebraic renormalization of Yang--Mills theory
with background field method, Nucl. Phys. B \textbf{462}, 524--550 (1996).

\bibitem{BC} C. Becchi, R. Collina, Further comments on the background field
method and gauge invariant effective actions, Nucl. Phys. B \textbf{562},
412--430 (1999); [arXiv:hep-th/9907092].

\bibitem{FPQ} R. Ferrari, M. Picariello, A. Quadri, Algebraic aspects of the
background field method, Annals Phys. \textbf{294}, 165--181 (2001);
[arXiv:hep-th/0012090].

\bibitem{BQ} D. Binosi, A. Quadri, The background field method as a
canonical transformation, Phys. Rev. D \textbf{85}, 121702 (2012).

\bibitem{Barv} A.O. Barvinsky, D. Blas, M. Herrero-Valea, S.M. Sibiryakov,
C.F. Steinwachs, Renormalization of gauge theories in the background-field
approach, JHEP \textbf{1807}, 035 (2018); [arXiv:1705.03480 [hep-th]].

\bibitem{FT} J. Frenkel, J.C. Taylor, Background gauge renormalization and
BRST identities, Annals Phys. \textbf{389}, 234--238 (2018).

\bibitem{BLT-YM} I.A. Batalin, P.M. Lavrov, I.V. Tyutin, Multiplicative
renormalization of Yang--Mills theories in the background-field formalism,
Eur. Phys. J. C \textbf{78}, 570 (2018); [arXiv:1806.02552 [hep-th]].

\bibitem{BFMc} F.T. Brandt, J. Frenkel, D.G.C. McKeon, Renormalization of
six-dimensional Yang--Mills theory in a background gauge field, Phys. Rev. D
\textbf{99}, 025003 (2019).

\bibitem{llr1} P. Lavrov, O. Lechtenfeld, A. Reshetnyak, Is soft breaking of
BRST symmetry consistent? JHEP \textbf{1110}, 043 (2011); [arXiv:1108.4820
[hep-th]].

\bibitem{Reshetnyak1} P. Lavrov, O. Radchenko, A. Reshetnyak, Soft breaking
of BRST symmetry and gauge dependence, Mod. Phys. Lett. A \textbf{27},
1250067 (2012); [arXiv:1201.4720 [hep-th]].

\bibitem{Reshetnyak} A. Reshetnyak, \ On gauge independence for gauge models
with soft breaking of BRST symmetry, Int. J. Mod. Phys. A \textbf{29},
1450184 (2014); [arXiv:1312.2092 [hep-th]].

\bibitem{LL2} P. Lavrov, O. Lechtenfeld, Gribov horizon beyond the Landau
gauge, Phys. Lett. B \textbf{725}, 386--388 (2013); [arXiv:1305.2931
[hep-th]].

\bibitem{Reshetnyak2} A.\ Reshetnyak, On composite fields approach to Gribov
copies elimination in Yang--Mills theories, Phys. Part. Nucl. Lett. \textbf{%
11}, 964--967 (2014); [arXiv:1402.3060 [hep-th]].

\bibitem{Capri0} M.A.L. Capri, D. Fiorentini, M.S. Guimaraes, B.W. Mintz,
L.F. Palhares, S.P. Sorella, A local and renormalizable framework for the
gauge-invariant operator $A_{\min }^{2}$ in Euclidean Yang--Mills theories
in linear covariant gauges, Phys. Rev. D \textbf{94}, 065009 (2016);
[arXiv:1606.06601 [hep-th]].

\bibitem{Capri2} M.A.L. Capri, S.P. Sorella, R.C. Terin, Study of a gauge
invariant local composite fermionic field, Annals Phys. \textbf{414}, 168077
(2020); [arXiv:1909.07927 [hep-th]].

\bibitem{Canfora} F.E. Canfora, D. Hidalgo, P. Pais, The Gribov problem in
presence of background field for $SU(2)$ Yang--Mills theory, Phys. Lett. B
\textbf{763}, 94--101 (2016); [arXiv:1610.08067].

\bibitem{ExamComp} Y. Nambu, G. Jona-Lasinio, Dynamical model of elementary
particles based on an analogy with superconductivity, Phys. Rev. \textbf{122}%
, 345--358 (1961); D. Gross, A. Neveu, Dynamical symmetry breaking in
asymptotically free field theories, Phys. Rev. D \textbf{10}, 3235--3253
(1974).

\bibitem{HockMossT} S.W. Hawking, I.G. Moss, Fluctuations in the
inflationary Universe, Nucl. Phys. B \textbf{224}, 180--192 (1983); A.A.
Tseytlin, Conformal anomaly in two-dimensional $\sigma $-model on curved
background and strings, Phys. Lett. B \textbf{178}, 34--40 (1986).

\bibitem{DSCB} R. Casalbuoni, S. DeCurtis, R. Gatto, Composite operator
calculation of chiral symmetry breaking in color gauge theory, Phys. Lett. B
\textbf{140}, 357--362 (1984).

\bibitem{20} W. Bardeen, C. Hill, M. Lindner, Minimal dynamical symmetry
breaking of the Standard Model, Phys. Rev. D \textbf{41}, 1647--1660 (1990).

\bibitem{GT} R.P. Grigorian, I.V. Tyutin, Renormalization group equation for
composite fields, Sov. J. Nucl. Phys. \textbf{26}, 593 (1977); in Russian:
Yad. Fiz. \textbf{26}, 1121--1129 (1977); V.P. Gusynin, V.A. Miransky,
Nonperturbative scale anomaly and composite operators in gauge field
theories, Phys. Lett. B \textbf{198}, 362--366 (1987).

\bibitem{21} N. Seiberg, The power of duality -- exact results in 4D SUSY
field theory, Int. J. Mod. Phys. A \textbf{16}, 4365--4376 (2001);
[arXiv:hep-th/9506077].

\bibitem{Stepanyantz} M. Shifman, K. Stepanyantz, Exact Adler function in
supersymmetric QCD, Phys. Rev. Lett. \textbf{114}, 051601 (2015);
[arXiv:1412.3382 [hep-th]].

\bibitem{Shifman:2015doa} M. Shifman, K.V. Stepanyantz, Derivation of the
exact expression for the D function in $N=1$ SQCD, Phys. Rev. D \textbf{91},
105008 (2015); [arXiv:1502.06655 [hep-th]].

\bibitem{Capri1} M.A.L. Capri, D.M. van Egmond, M.S. Guimar\~{a}es, O.
Holanda, S.P. Sorella, R.C. Terin, H.C. Toledo, Renormalizability of $N=1$
super Yang--Mills theory in Landau gauge with a Stueckelberg-like field,
Eur. Phys. J. C \textbf{78}, 797 (2018); [arXiv:1803.03077 [hep-th]].

\bibitem{FRG1} C. Wetterich, Average action and the renormalization group
equations, Nucl. Phys. B \textbf{352}, 529--584 (1991).

\bibitem{FRG2} M. Reuter, C. Wetterich, Average action for the Higgs model
with Abelian gauge symmetry, Nucl. Phys. B \textbf{391}, 147--175 (1993).

\bibitem{FRG3} M. Reuter, C.Wetterich, Effective average action for gauge
theories and exact evolution equations, Nucl. Phys. B \textbf{417}, 181--214
(1994).

\bibitem{FRG4} P. Lavrov, I. Shapiro, On the functional renormalization
group approach for Yang--Mills Fields, JHEP \textbf{1306}, 086 (2013);
[arXiv:1212.2577 [hep-th]].

\bibitem{FRG5} V.F. Barra, P.M. Lavrov, E.A. dos Reis, T. de Paula Netto,
I.L. Shapiro, Functional renormalization group approach and gauge dependence
in gravity theories, Phys. Rev. D \textbf{101}, 065001 (2020);
[arXiv:1910.06068[hep-th]].

\bibitem{FP} L.D. Faddeev, V.N. Popov, Feynman diagrams for the Yang--Mills
field, Phys. Lett. B \textbf{25}, 29--30 (1967).

\bibitem{LL1} P. Lavrov, O. Lechtenfeld, Field-dependent BRST
transformations in Yang--Mills theory. Phys. Lett. B \textbf{725}, 382--385
(2013); [arXiv:1305.0712[hep-th]].

\bibitem{LM} P.M. Lavrov, P.Yu. Moshin, Quantization of two-dimensional
gravity with dynamical torsion, Class. Quant. Grav. \textbf{16}, 2247--2258
(1999); [arXiv:hep-th/9710100].

\bibitem{BLTfin} I.A. Batalin, P.M. Lavrov, I.V. Tyutin, A systematic study
of finite BRST--BV transformations in field-antifield formalism, Int. J.
Mod. Phys. A \textbf{29}, 1450166 (2014); [arXiv:1405.2621 [hep-th]].

\bibitem{MR1} P.Yu. Moshin, A.A. Reshetnyak, Field-dependent BRST-antiBRST
transformations in Yang--Mills and Gribov--Zwanziger theories, {Nucl. Phys.}
B \textbf{888}, 92--128 (2014); [arXiv:1405.0790[hep-th]].

\bibitem{MR2} P.Yu. Moshin, A.A. Reshetnyak, Finite BRST-antiBRST
transformations in Lagrangian formalism, {Phys. Lett.} B \textbf{739},
110--116 (2014); [arXiv:1406.0179[hep-th]];\textbf{\ }Field-dependent
BRST-antiBRST Lagrangian transformations. Int. J. Mod. Phys. A \textbf{30},
1550021 (2015); [arXiv:1406.5086[hep-th]].

\bibitem{SMMR} P.Yu. Moshin, A.A. Reshetnyak, Finite field-dependent
BRST-antiBRST transformations: Jacobians and application to the Standard
Model. {Int. J. Mod. Phys.} A \textbf{31}, 1650111 (2016);
[arXiv:1506.04660[hep-th]]

\bibitem{UMR} S. Upadhyay, B.P. Mandal, A.A. Reshetnyak, Comments on
interactions in the SUSY models, \ {Eur. Phys. J.} C \textbf{6}, 391 (2016);
[arXiv:1605.02973[physics.gen-ph]].

\bibitem{N34} A.A. Reshetnyak, Generalization of Faddeev--Popov rules in
Yang--Mills theories: $N=3,4$ BRST symmetries, Int. J. Mod. Phys. A \textbf{%
33}, 1850006 (2018); [arXiv:1701.00009 [hep-th]].

\bibitem{JM} S.D. Joglekar, B.P. Mandal, Finite field dependent BRS
transformations. {Phys. Rev.} D \textbf{51}, 1919--1927 (1995).

\bibitem{Upadhyay3} S. Upadhyay, B.P. Mandal, Field dependent nilpotent
symmetry for gauge theories {Eur. Phys. J.} C \textbf{72}, 2065 (2012);
[arXiv:1201.0084[hep-th]].

\bibitem{VK} M.O. Katanaev, I.V. Volovich, Two-dimensional gravity with
dynamical torsion and strings, Ann. Phys. \textbf{197}, 1--32 (1990).

\bibitem{DeWcond} B.S. DeWitt, \emph{Dynamical Theory of Groups and Fields}
(Gordon and Breach, New York, 1965).

\bibitem{BRST1} C. Becchi, A. Rouet, R. Stora, The Abelian Higgs--Kibble
model, unitarity of the S-operator, Phys. Lett. B \textbf{52}, 344--346
(1974); Renormalization of Gauge Theories, Ann. Phys. (N.Y.) \textbf{98},
287--321 (1976).

\bibitem{BRST2} I.V. Tyutin, Gauge invariance in field theory and
statistical mechanics, Lebedev Inst. preprint No. 39 (1975);
[arXiv:0812.0580 [hep-th]].

\bibitem{6} P.M. Lavrov, S.D. Odintsov, The gauge dependence of the
effective action of composite fields in general gauge theories, Int. J. Mod.
Phys. A \textbf{4}, 5205--5212 (1989).

\bibitem{LOR} P.M. Lavrov, S.D. Odintsov, A.A. Reshetnyak, Effective action
of composite fields for general gauge theories in BLT-covariant formalism,%
\emph{\ }J. Math. Phys. \textbf{38}, 3466--3477 (1997)\textbf{;} Physics of
Atomic Nuclei \textbf{60}, 1020--1027 (1997); [arXiv:hep-th/9604061].

\bibitem{lavrovmerzlikin} P.M. Lavrov, B.S. Merzlikin, Loop expansion of the
average effective action in the functional renormalization group approach,
Phys. Rev. D \textbf{92}, 085038 (2015) [arXiv:1506.04941[hep-th]].

\bibitem{lavrovmerzl2} P.M. Lavrov, B.S. Merzlikin, Legendre transformations
and Clairaut-type equations, Phys. Lett. B \textbf{756}, 188--193 (2016);
[arXiv:1602.04911[hep-th]].

\bibitem{Kondo} K.-I. Kondo, The nilpotent \textquotedblleft
BRST\textquotedblright\ symmetry for the Gribov--Zwanziger theory, Report
number: CHIBA-EP-176; [arXiv:0905.1899 [hep-th]].

\bibitem{Capri-1} M.A.L. Capri, D. Fiorentini, M.S. Guimaraes, B.W. Mintz,
L.F. Palhares, S.P. Sorella, D. Dudal, I.F. Justo, A.D. Pereira,
R.\thinspace F. Sobreiro, An exact nilpotent nonperturbative BRST symmetry
for the Gribov--Zwanziger action in the linear covariant gauge, Phys. Rev. D
\textbf{92}, 045039 (2015).

\bibitem{SemenovTyanshan} M. Semenov-Tyan-Shanskii, V. Franke, Zap. Nauch.
Sem. LOMI im. V.A. Steklov, AN SSSR 120 (1982) 159 (English translation: New
York: Plenum Press 1986).

\bibitem{MoshReshIzv} A.A. Reshetnyak, P.Yu. Moshin, On the finite BRST
transformations: the Jacobians and the Standard Model with the
gauge-invariant Gribov horizon, Russian Physics Journal, \textbf{59} (2017)
1921, [arXiv:1607.07253[hep-th]].

\bibitem{localtransGZ} M.A.L. Capri, D. Dudal, D. Fiorentini, M.S.
Guimaraes, et al, A local and BRST-invariant Yang--Mills theory within the
Gribov horizon, Phys. Rev. D 94, no. 2, 025035 (2016), [arXiv:1605.02610
[hep-th]].

\bibitem{1708.01543} M.A.L. Capri, D. Fiorentini, A.D. Pereira, S.P.
Sorella, Renormalizability of the refined Gribov--Zwanziger action in the
linear covariant gauges, Phys. Rev. D \textbf{96}, 054022 (2017);
[arXiv:1708.01543[hep-th]].
\end{thebibliography}
\end{document}